\documentclass[aps,preprint,floatfix,nofootinbib,showpacs]{revtex4-1}
\pdfoutput=1
\usepackage{graphicx,color}
\usepackage{hyperref}

\newcommand{\imag}{\Im {\rm m}}
\newcommand{\real}{\Re {\rm e}}
\newcommand{\lsim}{\raisebox{-0.13cm}{~\shortstack{$<$ \\[-0.07cm] $\sim$}}~}
\newcommand{\gsim}{\raisebox{-0.13cm}{~\shortstack{$>$ \\[-0.07cm] $\sim$}}~}

\begin{document}

{\small
\begin{flushright}
CNU-HEP-13-02
\end{flushright} }

\title{Higgcision in the Two-Higgs Doublet Models}

\renewcommand{\thefootnote}{\arabic{footnote}}

\author{
Kingman Cheung$^{1,2}$, Jae Sik Lee$^3$, and
Po-Yan Tseng$^1$}
\affiliation{
$^1$ Department of Physics, National Tsing Hua University,
Hsinchu 300, Taiwan \\
$^2$ Division of Quantum Phases and Devices, School of Physics, 
Konkuk University, Seoul 143-701, Republic of Korea \\
$^3$ Department of Physics, Chonnam National University, \\
300 Yongbong-dong, Buk-gu, Gwangju, 500-757, Republic of Korea
}
\date{October, 2013}

\begin{abstract}
  We perform global fits to general two-Higgs doublet models (2HDMs)
  with generalized couplings
using the most updated data from ATLAS, CMS, and Tevatron.  
  We include both scenarios with CP-conserving and CP-violating
  couplings. 
By relaxing the requirement on the
discrete symmetries that are often imposed on the Yukawa couplings,
we try to see which of the 2HDMs is preferred.
We found that (i) Higgcision in 2HDMs can be performed
  efficiently by using only 4 parameters including the charged
  Higgs contributions to the Higgs couplings to two photons, (ii) the
  differences among various types of 2HDMs are very small with respect
  to the chi-square fits, (iii) $\tan\beta$ is constrained to be small,
(iv) the $p$-values for various fits in 2HDMs are worse
  than that of the standard model. Finally, we put emphasis on our findings
  that future precision measurements of 
the Higgs coupling to the scalar top-quark bilinear ($C_u^S$)
and $\tan\beta$ may
  endow us with the discriminating power among various types of 2HDMs
  especially when $C_u^S$ deviates from its SM value $1$.
\end{abstract}

\maketitle

\section{Introduction}
Upon the observation of a new boson 
at a mass around 125 GeV at the Large Hadron Collider (LHC) 
\cite{atlas,cms}, 
the Higgs precision (Higgcision) era has just begun.
A study based on a generic framework for the deviations of the couplings from 
their standard model (SM) values shows~\cite{Cheung:2013kla} that 
the SM Higgs boson~\cite{higgs} provides the best fit to all the
most updated Higgs data from 
ATLAS~\cite{atlas_h_aa_2013,atlas_com_2013},
CMS~\cite{cms_aa_2013,cms_zz_2013,cms_ww_2013,cms_tau_2013}, 
and Tevatron~\cite{tev,tev_h_bb}.
%

In addition to a number of more or less model-independent studies~ \cite{r1,
r2,r3,r4,r5,r6,r7,r8,r9,r10,r11,r12,r13,r14,r15,r16,r17,r18,r19,r20,r21,r22,anom1,anom2,anom3,anom4,anom5,anom6},
there are also studies done in the
2HDM~\cite{2hdm0,2hdm1,2hdm2,2hdm3,2hdm4,2hdm5,2hdm6,2hdm7,2hdm8,2hdm9,2hdm10,2hdm11,
2hdm12,2hdm13,2hdm14,2hdm15,2hdm16,2hdm17} and
supersymmetric~\cite{susy1,susy2,susy3,susy4,susy5} frameworks.
In this work,
we perform global fits to the general 2HDMs ({\it Higgcision in 2HDMs})
closely following the generic framework suggested in Ref.~\cite{Cheung:2013kla}.
We use the most updated data from the ATLAS, CMS, and the
Tevatron and include the scenarios with CP-conserving (CPC) and 
CP-violating (CPV) couplings. 
We find that Higgcision in 2HDMs can be performed very efficiently
by using {\it only} 3 parameters ($C_u^S$, $C_u^P$, and $\tan\beta$, as 
shown later), if one can neglect the charged-Higgs contribution to
the Higgs couplings to two photons. To consider the case when the charged-Higgs 
contribution to the $H\gamma\gamma$ couplings is significant,
one may need only one additional parameter.

Furthermore, we relax the requirement on the   
discrete symmetries, which are often imposed on the Yukawa couplings
to guarantees the absence of tree-level Flavor Changing Neutral Current 
(FCNC)~\cite{Glashow:1976nt}, to see which of the 2HDMs is preferred.
We find that the differences in the chi-squares among various types of 2HDMs
are very small  and one cannot see any preferences in both the CP-conserving
and CP-violating cases.

A number of important findings in this work are:
\begin{enumerate}
\item the SM provides the best fit in terms of $p$-values. The general
2HDM fits at most improve marginally in the total $\chi^2$ at the
expense of additional parameters though, and so the $p$-values
do not improve at all;

\item the differences among various types of 2HDMs are negligible in
fitting the Higgs data;

\item the gauge boson coupling $C_v$ is constrained to be close to 1,
which means that the observed Higgs boson is responsible for the
most part of the electroweak symmetry breaking; and 

\item the $\tan\beta$ is constrained to a small value. 
\end{enumerate}
Finally, we emphasize that future precision
measurements of $C_u^S$ and $\tan\beta$ can provide us with the
discriminating power among various types of 2HDMs especially when
$C_u^S$ deviates from its SM value $1$.

The organization of the paper is as follows. 
In the next section, 
we describe the interactions
of the Higgs bosons, including deviations in the Yukawa couplings and
deviations in the loop functions of $H\gamma\gamma$, $Hgg$, and $HZ\gamma$
vertices, as well as the notation used in the analysis. 
In Sec. III, we fix the Higgs potential and Yukawa couplings of the
general 2HDMs under consideration and describe how to 
perform  Higgcision in 2HDMs. We
articulate that only 4 fitting parameters are needed if we concentrate on the
couplings of the candidate for the 125 GeV Higgs boson.
We present the results of various fits in Sec. V
and conclude in Sec. VI.

\section{Formalism}
For the Higgs couplings to the SM particles assuming the Higgs boson is
a generic CP-mixed state without carrying any definite CP--parity,
we follow the conventions and notations of
{\tt CPsuperH}~\cite{Lee:2003nta,Lee:2007gn,Lee:2012wa} in which
the Higgs couplings to fermions are given as 
\begin{equation}
\label{eq:hff}
{\cal L}_{H\bar{f}f}\ =\ - \sum_{f=u,d,l}\,\frac{g m_f}{2 M_W}\,
 H\, \bar{f}\,\Big( g^S_{H\bar{f}f}\, +\,
ig^P_{H\bar{f}f}\gamma_5 \Big)\, f\,,
\end{equation}
where $f=u,d,l$ stands for the up- and down-type  quarks and charged leptons,
respectively,
and those to the massive vector bosons are
\begin{equation}
\label{eq:hvv}
{\cal L}_{HVV}  =  g\,M_W \, \left(
g_{_{HWW}} W^+_\mu W^{- \mu}\ + \
g_{_{HZZ}} \frac{1}{2c_W^2}\,Z_\mu Z^\mu\right) \, H\,.
\end{equation}
In the SM, $g^S_{H\bar{f}f}=1$, $g^P_{H\bar{f}f}=0$, and
$g_{_{HWW}}=g_{_{HZZ}}\equiv g_{_{HVV}}=1$. For the loop-induced Higgs couplings
to two photons, two gluons and $Z\gamma$, and their relevance to
the couplings $g^{S,P}_{H\bar{f}f}$ and $g_{_{HVV}}$, we refer to 
Refs.~\cite{Lee:2003nta,Lee:2007gn,Lee:2012wa,Cheung:2013kla}.
Without loss of generality, we use the following notation for the
parameters in the fits:
\begin{eqnarray}
\label{eq:c}
&&
C_u^S=g^S_{H\bar uu}\,, \ \
C_d^S=g^S_{H\bar dd}\,, \ \
C_\ell^S=g^S_{H\bar ll}\,; \ \
C_v=g_{_{HVV}}\,; \nonumber \\
&&
C_u^P=g^P_{H\bar uu}\,, \ \
C_d^P=g^P_{H\bar dd}\,, \ \
C_\ell^P=g^P_{H\bar ll}\, ; \nonumber \\
&&
 \Delta S^\gamma\,, \ \ 
 \Delta S^g \,, \ \
 \Delta P^\gamma \,, \ \
 \Delta P^g \,
 ; \nonumber \\
&& 
 \Delta \Gamma_{\rm tot} \,,
\end{eqnarray}
where $\Delta S^{\gamma}$ and  $\Delta P^\gamma$ denote additional loop
contributions to the loop factor $S^\gamma$ and $P^\gamma$, respectively;
and similarly for $\Delta S^g$ and $\Delta P^g$. The
$\Delta \Gamma_{\rm tot}$ represents an additional nonstandard decay width
of the Higgs boson (e.g., decay into the lighter Higgses).
Here we assume generation independence and also custodial
symmetry between the $W$ and $Z$ bosons. 

Our analysis is based on the theoretical signal strength 
which may be approximated  as the product
\begin{equation}
\widehat\mu({\cal P},{\cal D}) \simeq
\widehat\mu({\cal P})\ \widehat\mu({\cal D}) 
\end{equation}
where ${\cal P}={\rm ggF}, {\rm VBF}, {\rm VH}, {\rm ttH}$ denote 
the production mechanisms
and ${\cal D}=\gamma\gamma , ZZ, WW, b\bar{b}, \tau\bar\tau$
the decay channels.
For explicit expressions of
$\widehat\mu({\cal P})$ and $\widehat\mu({\cal D})$, we again refer to
Ref.~\cite{Cheung:2013kla}, but by noting they are basically given by the
ratios of the Higgs couplings to the corresponding SM ones.

\section{2HDMs}
The general 2HDM potential may be given by~\cite{Battye:2011jj}
\begin{eqnarray}
\label{V2HDM}
\mathrm{V} &=& -\mu_1^2 (\Phi_1^{\dagger} \Phi_1) - \mu_2^2
(\Phi_2^{\dagger} \Phi_2) - m_{12}^2 (\Phi_1^{\dagger} \Phi_2) -
m_{12}^{*2}(\Phi_2^{\dagger} \Phi_1) \nonumber \\
&&+ \lambda_1 (\Phi_1^{\dagger} \Phi_1)^2 + \lambda_2
(\Phi_2^{\dagger} \Phi_2)^2 + \lambda_3 (\Phi_1^{\dagger}
\Phi_1)(\Phi_2^{\dagger} \Phi_2) + \lambda_4 (\Phi_1^{\dagger}
\Phi_2)(\Phi_2^{\dagger} \Phi_1) \nonumber \\
&&+ \frac{\lambda_5}{2} (\Phi_1^{\dagger} \Phi_2)^2 +
\frac{\lambda_5^{*}}{2} (\Phi_2^{\dagger} \Phi_1)^2 + \lambda_6
(\Phi_1^{\dagger} \Phi_1) (\Phi_1^{\dagger} \Phi_2) + \lambda_6^{*}
(\Phi_1^{\dagger} \Phi_1)(\Phi_2^{\dagger} \Phi_1) \nonumber \\
&& + \lambda_7 (\Phi_2^{\dagger} \Phi_2) (\Phi_1^{\dagger} \Phi_2) +
\lambda_7^{*} (\Phi_2^{\dagger} \Phi_2) (\Phi_2^{\dagger} \Phi_1)\; .
\end{eqnarray}
With the parameterization
\begin{equation}
\Phi_1=\left(\begin{array}{c}
\phi_1^+ \\ \frac{1}{\sqrt{2}}\,(v_1+\phi_1^0+ia_1)
\end{array}\right)\,; \ \ \
\Phi_2={\rm e}^{i\xi}\,\left(\begin{array}{c}
\phi_2^+ \\ \frac{1}{\sqrt{2}}\,(v_2+\phi_2^0+ia_2)
\end{array}\right)
\end{equation}
and denoting $v_1=v \cos\beta=vc_\beta$
and $v_2=v \sin\beta=vs_\beta$, one may remove $\mu_1^2$, $\mu_2^2$,
and $\imag(m_{12}^2{\rm e}^{i\xi})$ from the 2HDM potential using three tadpole
conditions.  
Then, including the vacuum expectation value $v$,
  one may need the following 13 parameters plus one sign:
\begin{eqnarray}
\label{eq:2hdmpara}
&& v \,, t_\beta\,, |m_{12}| \,;  \nonumber \\
&& \lambda_1\,, \lambda_2\,,\lambda_3\,,\lambda_4\,,
|\lambda_5|\,,|\lambda_6|\,,|\lambda_7|\,; \nonumber \\
&&
\phi_5+2\xi\,,\phi_6+\xi\,,\phi_7+\xi\,,{\rm sign}[\cos(\phi_{12}+\xi)]\,.
\end{eqnarray}
to fully specify the general 2HDM potential.
Here $m_{12}^2=|m_{12}|^2e^{i\phi_{12}}$ and
$\lambda_{5,6,7}=|\lambda_{5,6,7}|e^{i\phi_{5,6,7}}$ and
we note that $\sin(\phi_{12}+\xi)$ is fixed by the CP-odd tadpole condition 
when the CP phases $\phi_5+2\xi\,,\phi_6+\xi$ and $\phi_7+\xi$ are given
and, accordingly,
$\cos(\phi_{12}+\xi)$ is determined up to the two-fold ambiguity.
One may take the convention with $\xi=0$ without loss of generality.

On the other hand, the Yukawa couplings are given in the interactions
\begin{eqnarray}
-{\cal L}_Y&=& h_u\, \overline{u_R}\, Q^T\,(i\tau_2)\,\Phi_2
+h_d\, \overline{d_R}\, Q^T\,(i\tau_2)\,
\left(-\eta_1^d\,\widetilde\Phi_1 -\eta_2^d\,\widetilde\Phi_2\right)
\nonumber \\[2mm] &+&
h_l\, \overline{l_R}\, L^T\,(i\tau_2)\,
\left(-\eta_1^l\,\widetilde\Phi_1 -\eta_2^l\,\widetilde\Phi_2\right)
\ + \ {\rm h.c.}
\end{eqnarray}
where $Q^T=(u_L\,,d_L)$, $L^T=(\nu_L\,,l_L)$, and
$\widetilde\Phi_i=i\tau_2 \Phi_i^*$ with
\begin{equation}
i\tau_2=\left(\begin{array}{cc}
0&1 \\ -1 & 0
\end{array}\right)\,.
\end{equation}
We note that there is a freedom to
redefine the two linear combinations of $\Phi_2$ and $\Phi_1$ to eliminate
the coupling of the up-type quarks to $\Phi_1$~\cite{Davidson:2005cw}. The 2HDMs
are classified according to the values of $\eta_{1,2}^l$ and $\eta_{1,2}^d$
as in Table~\ref{tab:2hdtype}.
%
\begin{table}[!hbt]
\caption{\label{tab:2hdtype}
Classification of 2HDMs satisfying the Glashow-Weinberg condition
\cite{Glashow:1976nt}
which guarantees the absence of tree-level FCNC.
}
\begin{center}
\begin{tabular}{|l|cccc|}
\hline
\hline
         & \hspace{0.5cm} 2HDM I\hspace{0.5cm} & 2HDM II\hspace{0.5cm} 
& 2HDM III\hspace{0.5cm} & 2HDM IV\hspace{0.5cm}  \\
\hline
$\eta_1^d$  & $0$ & $1$ & $0$ & $1$  \\
$\eta_2^d$  & $1$ & $0$ & $1$ & $0$  \\
\hline
\hline
$\eta_1^l$  & $0$ & $1$ & $1$ & $0$  \\
$\eta_2^l$  & $1$ & $0$ & $0$ & $1$  \\
\hline
\hline
\end{tabular}
\end{center}
\end{table}

By identifying the couplings
\begin{equation}
h_u = \frac{\sqrt{2}m_u}{v}\,\frac{1}{s_\beta}\,; \ \
h_d = \frac{\sqrt{2}m_d}{v}\,\frac{1}{\eta_1^dc_\beta+\eta_2^d s_\beta}\,; \ \
h_l = \frac{\sqrt{2}m_l}{v}\,\frac{1}{\eta_1^lc_\beta+\eta_2^l s_\beta}\,,
\end{equation}
we have obtained the following Higgs-fermion-fermion interactions
\begin{eqnarray}
\label{eq:nhff.2hdm}
-{\cal L}_{H_i\bar{f}f} &=&
\frac{m_u}{v}\left[
\bar{u}\, \left(
\frac{O_{\phi_2 i}}{s_\beta} -i\,\frac{c_\beta}{s_\beta}O_{ai}\,\gamma_5\,
\right)\,u \right]\,H_i
\nonumber \\[2mm] &+&
\frac{m_d}{v}\left[
\bar{d}\, \left(
\frac{\eta_1^dO_{\phi_1 i}+\eta_2^dO_{\phi_2
i}}{\eta_1^dc_\beta+\eta_2^ds_\beta}
-i\,\frac{\eta_1^ds_\beta-\eta_2^dc_\beta}{\eta_1^dc_\beta+\eta_2^ds_\beta}
O_{ai}\,\gamma_5\,
\right)\,d \right]\,H_i
\nonumber \\[2mm] &+&
\frac{m_l}{v}\left[
\bar{l}\, \left(
\frac{\eta_1^lO_{\phi_1 i}+\eta_2^lO_{\phi_2
i}}{\eta_1^lc_\beta+\eta_2^ls_\beta}
-i\,\frac{\eta_1^ls_\beta-\eta_2^lc_\beta}{\eta_1^lc_\beta+\eta_2^ls_\beta}
O_{ai}\,\gamma_5\,
\right)\,l \right]\,H_i
\end{eqnarray}
and 
\begin{eqnarray}
\label{eq:chff.2hdm}
-{\cal L}_{H^\pm\bar{u}d} &=&
-\frac{\sqrt{2}m_u}{v}
\left(\frac{c_\beta}{s_\beta}\right) \,\bar{u}\,P_L\,d\,H^+
- \frac{\sqrt{2}m_d}{v}\left(\,
\frac{\eta_1^ds_\beta-\eta_2^dc_\beta}{\eta_1^dc_\beta+\eta_2^ds_\beta}
\right)\,\bar{u}\,P_R\,d\,H^+
\nonumber \\ &&
- \frac{\sqrt{2}m_l}{v}\left(\,
\frac{\eta_1^ls_\beta-\eta_2^lc_\beta}{\eta_1^lc_\beta+\eta_2^ls_\beta}
\right)\,\bar{\nu}\,P_R\,l\,H^+
\ + \ {\rm h.c.}
\end{eqnarray}
Here we take the convention with $\xi=0$ and the couplings $h_{u,d,l}$ are
supposed to be real. The $3\times 3$ mixing matrix $O$ is defined through
\begin{eqnarray}
(\phi_1^0,\phi_2^0,a)^T_\alpha&=&O_{\alpha i} (H_1,H_2,H_3)^T_i
\end{eqnarray}
such that $O^T {\cal M}_0^2 O={\rm diag}(M_{H_1}^2,M_{H_2}^2,M_{H_3}^2)$
with the ordering of $M_{H_1}\leq M_{H_2}\leq M_{H_3}$.
Here the $3\times 3$ mass matrix of the neutral Higgs bosons 
${\cal M}_0^2$ is given by
\begin{equation}
{\cal M}^2_0 = M_A^2 \left(\begin{array}{ccc}
s_\beta^2 & -s_\beta c_\beta & 0 \\
-s_\beta c_\beta & c_\beta^2 & 0 \\
0 & 0 & 1 \end{array}\right) \ + \ {\cal M}^2_\lambda
\end{equation}
with (reinstating the relative phase $\xi$)
\begin{eqnarray}
M_A^2&=&M_{H^\pm}^2+\frac{1}{2}\lambda_4 v^2 -\frac{1}{2}\real(\lambda_5{\rm
e}^{2i\xi})v^2\,, \\
M_{H^\pm}^2&=&\frac{\real(m_{12}^2{\rm e}^{i\xi})}{c_\beta s_\beta}
-\frac{v^2}{2c_\beta s_\beta}\left[\lambda_4 c_\beta s_\beta+
c_\beta s_\beta\real(\lambda_5{\rm e}^{2i\xi})+
c_\beta^2\real(\lambda_6{\rm e}^{i\xi})+
s_\beta^2\real(\lambda_7{\rm e}^{i\xi}) \right]\,,\nonumber
\end{eqnarray}
and
\begin{equation}
\frac{{\cal M}^2_\lambda}{v^2} = \left(\begin{array}{lll}
2\lambda_1 c_\beta^2 +\real(\lambda_5{\rm e}^{2i\xi})s_\beta^2 &
\lambda_{34}c_\beta s_\beta + \real(\lambda_6{\rm e}^{i\xi}) c_\beta^2 &
-\frac{1}{2}\imag(\lambda_5{\rm e}^{2i\xi})s_\beta \\
+2\real(\lambda_6{\rm e}^{i\xi}) s_\beta c_\beta &
+\real(\lambda_7{\rm e}^{i\xi}) s_\beta^2 &
-\imag(\lambda_6{\rm e}^{i\xi}) c_\beta \\[5mm]
\lambda_{34}c_\beta s_\beta + \real(\lambda_6{\rm e}^{i\xi}) c_\beta^2 &
2\lambda_2 s_\beta^2 +\real(\lambda_5{\rm e}^{2i\xi})c_\beta^2 &
-\frac{1}{2}\imag(\lambda_5{\rm e}^{2i\xi})c_\beta \\
+\real(\lambda_7{\rm e}^{i\xi}) s_\beta^2 &
+2\real(\lambda_7{\rm e}^{i\xi}) s_\beta c_\beta &
-\imag(\lambda_7{\rm e}^{i\xi}) s_\beta \\[3mm]
-\frac{1}{2}\imag(\lambda_5{\rm e}^{2i\xi})s_\beta &
-\frac{1}{2}\imag(\lambda_5{\rm e}^{2i\xi})c_\beta &
~~~~~~~~~0 \\
-\imag(\lambda_6{\rm e}^{i\xi}) c_\beta &
-\imag(\lambda_6{\rm e}^{i\xi}) s_\beta &
\end{array}\right)
\end{equation}
where $\lambda_{34}=\lambda_3+\lambda_4$ and, in passing,
we note $v=gM_W/2$, $a=-s_\beta a_1+c_\beta a_2$ and
$H^+=-s_\beta \phi_1^++c_\beta \phi_2^+$.
We need to specify, therefore, the
13 parameters plus one sign listed in Eq.~(\ref{eq:2hdmpara})
to fix all the Higgs-fermion-fermion couplings.

Nevertheless, in order to calculate the signal strengths on which our chi-square
analysis is based, we need to know only
the couplings of the 125 GeV Higgs boson.
Regarding the $i$-th Higgs boson $H_i$ as the 
candidate for the 125 GeV Higgs boson, and
by looking into Eqs.~(\ref{eq:nhff.2hdm}) and (\ref{eq:chff.2hdm}), 
the relevant Higgs 
couplings can be fully determined by knowing the components
$O_{\phi_1i}$, $O_{\phi_2i}$, and $O_{ai}$ of the mixing matrix and $t_\beta$
in each 2HDM.
Comparing Eqs.~(\ref{eq:nhff.2hdm}) and (\ref{eq:hff})
we find
\begin{eqnarray}
\label{eq:oai}
&& 
O_{\phi_2 i}=s_\beta\,C_u^S\,,  \ \ \
O_{ai}=-t_\beta\,C_u^P\,; \nonumber \\
&&
O_{\phi_1 i} = \pm \left[1-s_\beta^2(C_u^S)^2-t_\beta^2(C_u^P)^2\right]^{1/2}\,,
\end{eqnarray}
where $C_u^S=g^S_{H_i\bar uu}$ and $C_u^P=g^P_{H_i\bar uu}$ and
the orthogonality relation $(O_{\phi_1 i})^2+(O_{\phi_2 i})^2+(O_{a
i})^2=1$ is used
\footnote{ 
Depending on the values of $\tan\beta$, $C_u^S$, and $C_u^P$,
one may take one or both of the two signs for $O_{\phi_1 i}$
by fixing the relative sign between the Yukawa and
$g_{H_iVV}$ couplings.  Without loss of generality
we take the convention of $g_{H_iVV}=C_v>0$ in this work.
}.
Therefore, by specifying only the 3 parameters of $C_u^S$, $C_u^P$, and 
$t_\beta$,
the couplings of the 125 GeV Higgs to all the SM fermions can be determined
in each 2HDM as summarized in Table~\ref{tab:cdcl}. In addition, 
the Higgs coupling to the massive vector bosons 
is determined by
\begin{equation}
\label{eq:cv}
C_v= c_\beta O_{\phi_1 i} + s_\beta O_{\phi_2 i}
= \pm c_\beta \left[1-s_\beta^2(C_u^S)^2-t_\beta^2(C_u^P)^2\right]^{1/2}
+s_\beta^2 \,C_u^S\,.
\end{equation}

\begin{table}[!hbt]
\caption{\label{tab:cdcl}
The couplings $C_{d,l}^{S,P}$ as functions of $C_u^{S,P}$ and
$\tan\beta$ in each 2HDM.
}
\begin{center}
\begin{tabular}{||l||c|c||c|c||}
\hline
2HDM I & $C_d^S = C_u^S$ & $C_l^S = C_u^S$ &
$C_d^P=-C_u^P$ & $C_l^P=-C_u^P$ \\
\hline
2HDM II & $C_d^S = \pm\frac{\left[1-s_\beta^2 (C_u^S)^2
-t_\beta^2(C_u^P)^2\right]^{1/2}}{c_\beta}$ &
$C_l^S = \pm\frac{\left[1-s_\beta^2 (C_u^S)^2
-t_\beta^2(C_u^P)^2\right]^{1/2}}{c_\beta}$ &
$C_d^P=t_\beta^2C_u^P$ & $C_l^P=t_\beta^2C_u^P$ \\
\hline
2HDM III & $C_d^S = C_u^S$ & $C_l^S = \pm\frac{\left[1-s_\beta^2 (C_u^S)^2
-t_\beta^2(C_u^P)^2\right]^{1/2}}{c_\beta}$  &
$C_d^P=-C_u^P$ & $C_l^P=t_\beta^2C_u^P$ \\
\hline
2HDM IV & $C_d^S = \pm\frac{\left[1-s_\beta^2 (C_u^S)^2
-t_\beta^2(C_u^P)^2\right]^{1/2}}{c_\beta}$ & $C_l^S = C_u^S$ &
$C_d^P=t_\beta^2C_u^P$ & $C_l^P=-C_u^P$ \\
\hline
\end{tabular}
\end{center}
\end{table}

To recapitulate, we need 13 parameters (plus one sign) to fix all the
Higgs couplings to the SM particles and the Higgs boson spectrum fully
in general 2HDMs.
In contrast,
only 3 parameters are needed for the couplings of the
125 GeV Higgs candidate to the SM fermions and massive vector bosons.
These 3 parameters are the two couplings of the 125 GeV Higgs 
candidate to the scalar and pseudoscalar top-quark bilinears 
($C_u^S$ and $C_u^P$, respectively) and $\tan\beta$. 
One may use $C_v$ instead of $\tan\beta$ as shown later.
In this work, we take advantage of the avenue with
the smaller number of parameters to analyze the Higgs data. 

With $C_u^S$, $C_u^P$, and $C_v$ (or $\tan\beta$) given, 
we also need to know the charged Higgs 
contribution to the Higgs coupling to two photons 
in order to calculate the signal strengths.
The charged Higgs contribution to the Higgs coupling to two photons is given by
\begin{equation}
\label{eq:dsgamma}
\left(\Delta S^\gamma_i\right)^{H^\pm} =
-g_{H_iH^+H^-}\frac{v^2}{2M_{H^\pm}^2}F_0(\tau_{iH^\pm})\;,
\end{equation}
where $\tau_{iH^\pm}=M_{H_i}^2/4M_{H^\pm}^2$ and
$F_0(\tau)=\tau^{-1}\,[-1+\tau^{-1}f(\tau)]$ with
\begin{eqnarray}
f(\tau)=-\frac{1}{2}\int_0^1\frac{{\rm d}y}{y}\ln\left[1-4\tau y(1-y)\right]
       =\left\{\begin{array}{cl}
           {\rm arcsin}^2(\sqrt{\tau}) \,:   & \qquad \tau\leq 1\,, \\
   -\frac{1}{4}\left[\ln \left(\frac{\sqrt{\tau}+\sqrt{\tau-1}}{
                                     \sqrt{\tau}-\sqrt{\tau-1}}\right)
                    -i\pi\right]^2\,: & \qquad \tau\geq 1\,.
\end{array}\right.
\end{eqnarray}
The $g_{H_iH^+H^-}$ coupling is defined in the interaction
\begin{equation}
{\cal L}_{3H}  = v\,\sum_{i=1}^3 g_{_{H_iH^+H^-}}\,H_iH^+H^-\,,\\[3mm]
\end{equation}
with
$g_{_{H_iH^+H^-}}\ =\ \sum_{\alpha =\phi_1,\phi_2,a} O_{\alpha i}\,
g_{_{\alpha H^+H^-}}$.
The  effective couplings  $g_{_{\alpha H^+H^-}}$
indeed involve all of the Higgs quartic couplings again
and read~\cite{CHL}
\footnote{Note the convention difference for $\lambda_5$
by a factor $2$.}:
\begin{eqnarray}
g_{_{\phi_1H^+H^-}}  \!\!&=& 2s^2_\beta c_\beta\lambda_1\: +\:
c^3_\beta\lambda_3\: -\: s^2_\beta c_\beta \lambda_4\: -\:
s^2_\beta c_\beta\, \real {\lambda_5}\:
+\: s_\beta (s^2_\beta - 2c^2_\beta)\, \real\lambda_6\nonumber\\
&&+\: s_\beta c^2_\beta \real \lambda_7\, ,\nonumber\\
g_{_{\phi_2 H^+H^-}} \!\!&=& 2s_\beta c^2_\beta \lambda_2\: +\:
s^3_\beta\lambda_3\: -\: s_\beta c^2_\beta \lambda_4\: -\:
s_\beta c^2_\beta \, \real {\lambda_5}\:
+\: s^2_\beta c_\beta \, \real\lambda_6\nonumber\\
&&+\: c_\beta (c^2_\beta - 2s^2_\beta)\, \real \lambda_7\, ,\nonumber\\
g_{_{aH^+H^-}} \!\!&=& s_\beta c_\beta\, \imag{\lambda_5}\: -\:
s^2_\beta\, \imag\lambda_6\: -\: c^2_\beta\, \imag\lambda_7\, .
\end{eqnarray}
Therefore, in order to include $\left(\Delta S^\gamma_i\right)^{H^\pm}$ 
one may specify all the
quartic couplings and the charged Higgs mass in principle, 
but, then, the situation goes back to
the original case with 13 parameters plus one sign. 
Nevertheless, even in this case
one can still keep the spirit of efficiency and simplicity
by treating
$\left(\Delta S^\gamma_i\right)^{H^\pm}$ itself as another free parameter
in addition to the other three ones $C_u^{S}\,,C_u^{P}$ and $C_v$.
And then, the results on
$\left(\Delta S^\gamma_i\right)^{H^\pm}$
could be directly interpreted in terms of 
the coupling $g_{_{H_i H^+H^-}}$ of the 125 GeV Higgs boson to the 
charged Higgses and the charged Higgs boson mass $M_{H^\pm}$,
as shown in Eq.~(\ref{eq:dsgamma}).

One caveat of our
approach to analyze the Higgs data with only 3 or 4
parameters is that one cannot say much about the other two neutral Higgs bosons
and the charged one which, in principle, can be either heavier or lighter than the
candidate for the 125 GeV Higgs.
Before moving to the next section to present the results of various $2$-, $3$- and
$4$-parameter fits,
we would like to briefly comment on the status of
experimental searches for the additional Higgs bosons.

At the LHC, both the ATLAS and CMS
collaborations have searched for the additional neutral Higgses bosons up to $1$
TeV through their decays into two massive vector bosons,
$H_i\rightarrow ZZ$ or $WW$ \cite{heavyH_atlas,heavyH_cms}. 
Without observing any positive signal, they put an
upper bound on the relevant cross section $\sigma(pp\rightarrow H_i\rightarrow VV)$
\footnote{We note that,if $CP$ is conserved, the constraints provided by
these search channels cannot be
applied to the CP-odd state.}.
The ATLAS collaboration performed the neutral Higgs-boson searches 
through the tau-lepton channel, $H_i \rightarrow \tau \tau$~\cite{heavyH_atlas_tau}.
While this applies for both the CP-even and CP-odd neutral Higgses up to 500 GeV,
it was reported that the constraint for the additional CP-even Higgs from this
channel is weaker than that from $H_i \rightarrow ZZ$~\cite{1310.7941}.

For the charged Higgs boson with mass around a few hundred GeV, the strongest
constrain may come from ${\rm BR}(\bar{B}\rightarrow X_s \gamma)$
through the additional loop contributions from the charged-Higgs bosons 
to the process $b\to s \gamma$~\cite{1310.7941}.
When the charged Higgs boson is lighter than the top quark, it can be searched at
the LHC through the top-quark decay
channel $t\rightarrow H^+ b$ with the charged Higgs boson 
subsequently decaying
into $c\bar{b}$, $c\bar{s}$, and $\tau^+ \nu_{\tau}$.
The direct searches of the charged Higgs boson at the LHC also set
limits on the interactions of charged Higgs boson, but 
their constraints are still weaker than those from
$\bar{B}\rightarrow X_s \gamma$~\cite{1310.7941}. 

The current direct experimental searches for the
additional Higgs bosons and their indirect effects on some flavor
observables such as ${\rm BR}(\bar{B}\rightarrow X_s \gamma)$
should provide more stringent restrictions on the model parameters in addition to
those obtained by fitting the 125-GeV Higgs data only.
This may deserve an independent study and we will discuss
these crucial issues in detail in a future publication.

\section{Fits}

As shown in the previous section, the whole analysis of the couplings of 
the observed Higgs boson (denoted by $H_i$) in 2HDMs, including the
CP-conserving and CP-violating cases, 
can be performed with only 4 parameters:
$C_u^{S}\,,C_u^{P}\,,C_v\,,  \left(\Delta S^\gamma_i\right)^{H^\pm}$.
In particular, we consider the following cases with respect to 
CP-conserving or CP-violating, and with/without charged Higgs contributions:
\begin{itemize}
\item{CP-conserving (CPC) cases}
\begin{itemize}
\item\underline{\bf CPC2}: $C_u^S$, $C_v$
\item\underline{\bf CPC3}: $C_u^S$, $C_v$, $(\Delta S^\gamma)^{H^\pm}$
\end{itemize}
\item{CP-violating (CPV) cases}
\begin{itemize}
\item\underline{\bf CPV3}: $C_u^S$, $C_u^P$, $C_v$
\item\underline{\bf CPV4}: $C_u^S$, $C_u^P$, $C_v$, $(\Delta S^\gamma)^{H^\pm}$
\end{itemize}
\end{itemize}
Here
{\bf CPC} and {\bf CPV} represent CP-conserving and CP-violating fits,
respectively, and the number denotes the number of varying parameters in each
fit. In {\bf CPC2} and {\bf CPV3}, the charged Higgs contribution
$(\Delta S^\gamma)^{H^\pm}=0$. 
Note that 
the varying parameters should satisfy the following relations due to the
unitarity of the mixing matrix:
\begin{eqnarray}
\label{eq:cons}
s_\beta^2 (C_u^S)^2 \leq 1\,, \ \ \
t_\beta^2(C_u^P)^2 \leq 1\,, \ \ \
s_\beta^2 (C_u^S)^2+t_\beta^2(C_u^P)^2 \leq 1\,.
\end{eqnarray}
One can use $\tan\beta$ in place of $C_v$ in the analysis 
by exploiting the relation derived from Eq.~(\ref{eq:cv}):
\begin{equation}
\label{eq:sbsq}
s_\beta^2=\frac{1-C_v^2}{1+(C_u^S)^2+(C_u^P)^2-2C_vC_u^S}\,,
\end{equation}
which is independent of ${\rm sign}\,[O_{\phi_1 i}]$.
When $C_u^S=1$ and $C_u^P=0$, the above relation becomes $s_\beta^2=(1+C_v)/2$,
which leads to $\tan\beta=\infty$ in the SM limit of $C_v=1$.
On the other hand, as in many models beyond the SM,
if $C_u^S$ and/or $C_u^P$ deviate from its SM values $1$ and $0$, respectively,
one may end up in the opposite limit, $\tan\beta=0$,
when the dynamics of the fit pushes $C_v$ to its maximally allowed value
or $1$. 
In practice, one may wish to avoid the regions with small or (very) 
large $\tan\beta$ to maintain the perturbativity of the top and bottom
Yukawa couplings $h_t$ and $h_b$, respectively.
We therefore restrict the range of $\tan\beta$ between
$10^{-4}$ and $10^2$.

Before presenting our numerical results, we briefly review the current Higgs
data.
Current Higgs data focus on a few decay channels of the Higgs boson:
(i) $h\to \gamma\gamma$, (ii) $h\to Z Z^* \to \ell^+ \ell^- \ell^+ \ell^-$,
(iii) $h\to W W^* \to \ell^+ \bar \nu \ell^- \nu$,
(iv) $h \to b\bar b$, and (v) $h \to \tau^+ \tau^-$.
We have used 22 data points in our analysis as in Ref.~\cite{Cheung:2013kla}.
To briefly summarize, the chi-square of all these 22 data points relative
to the SM is
\begin{eqnarray}
18.94=7.89 (\gamma\gamma:6)+1.65 (ZZ^*:2)+3.70 (WW^*:5)+3.55 (b\bar{b}:4)
+2.15 (\tau^+\tau^-:5) \nonumber \,,
\end{eqnarray}
where the numbers in parentheses denote the number of data points in each
decay mode.
The chi-square per degree of freedom
(dof) is about $18.94/22=0.86$ and the $p$-value is about $p_{\rm SM}=0.65$.
We note the chi-square is dominated by the diphoton
data with $\mu_{ggH+ttH}^{\rm ATLAS}=1.6\pm 0.4$ and
$\mu_{\rm untagged}^{\rm CMS}=0.78^{+0.28}_{-0.26}$.
Since the ATLAS data is
about $1.5\sigma$ larger than the SM while the CMS one is about $1\sigma$
smaller, the dynamics of the
fit cannot force the parameters to go into either direction.

\subsection{\bf CP conserving fits}
In this subsection, we study the CP-conserving case  with $C_u^P=0$.
In our numerical study, we find that $\tan\beta$ is bounded from above when
$C_u^S$ deviates from its SM value $1$. Before presenting numerical results,
we look into the correlation among the 
varying parameters $C_u^S$, $C_v$, and $\tan\beta$.

\begin{figure}[t!]
\centering
\includegraphics[width=3.2in]{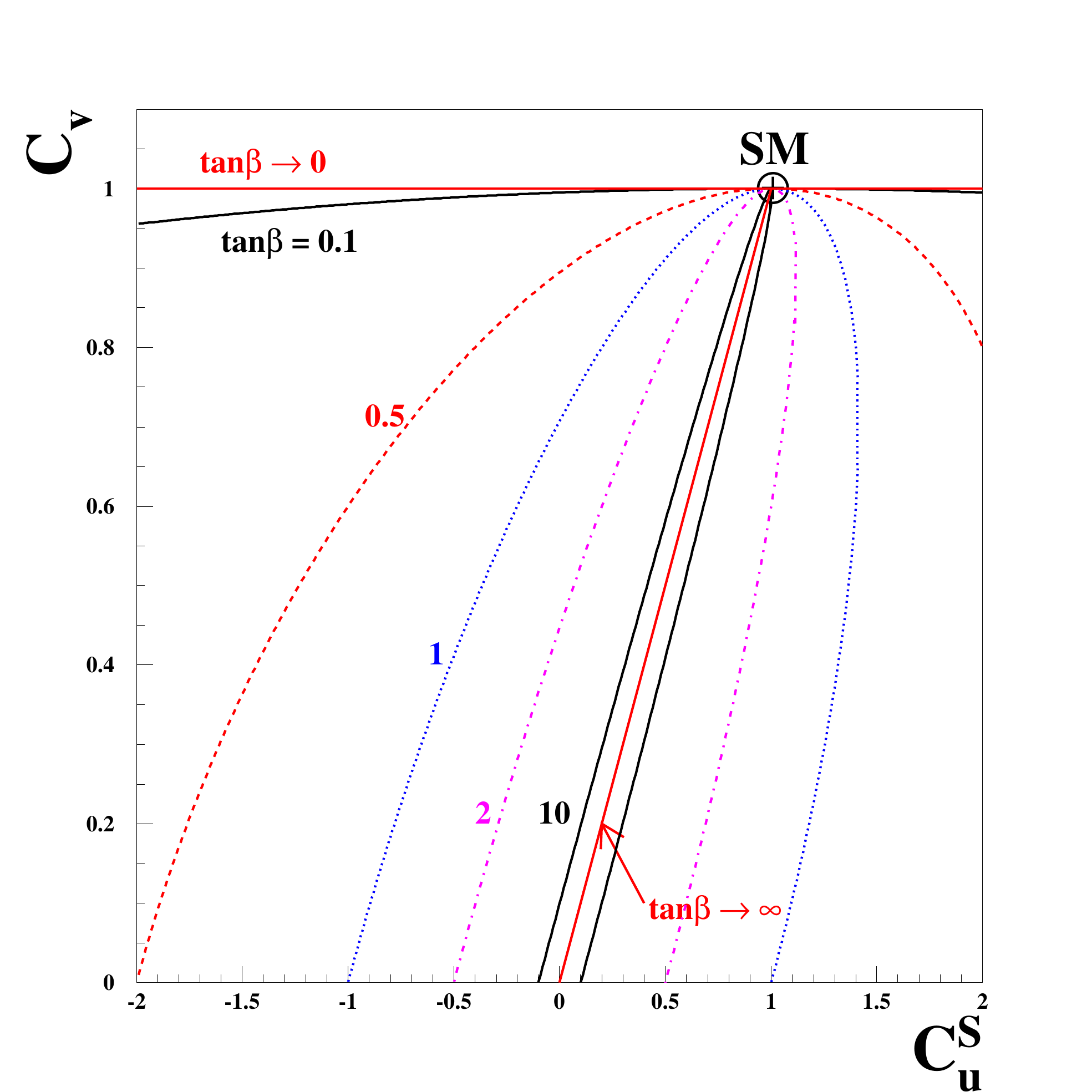}
\includegraphics[width=3.2in]{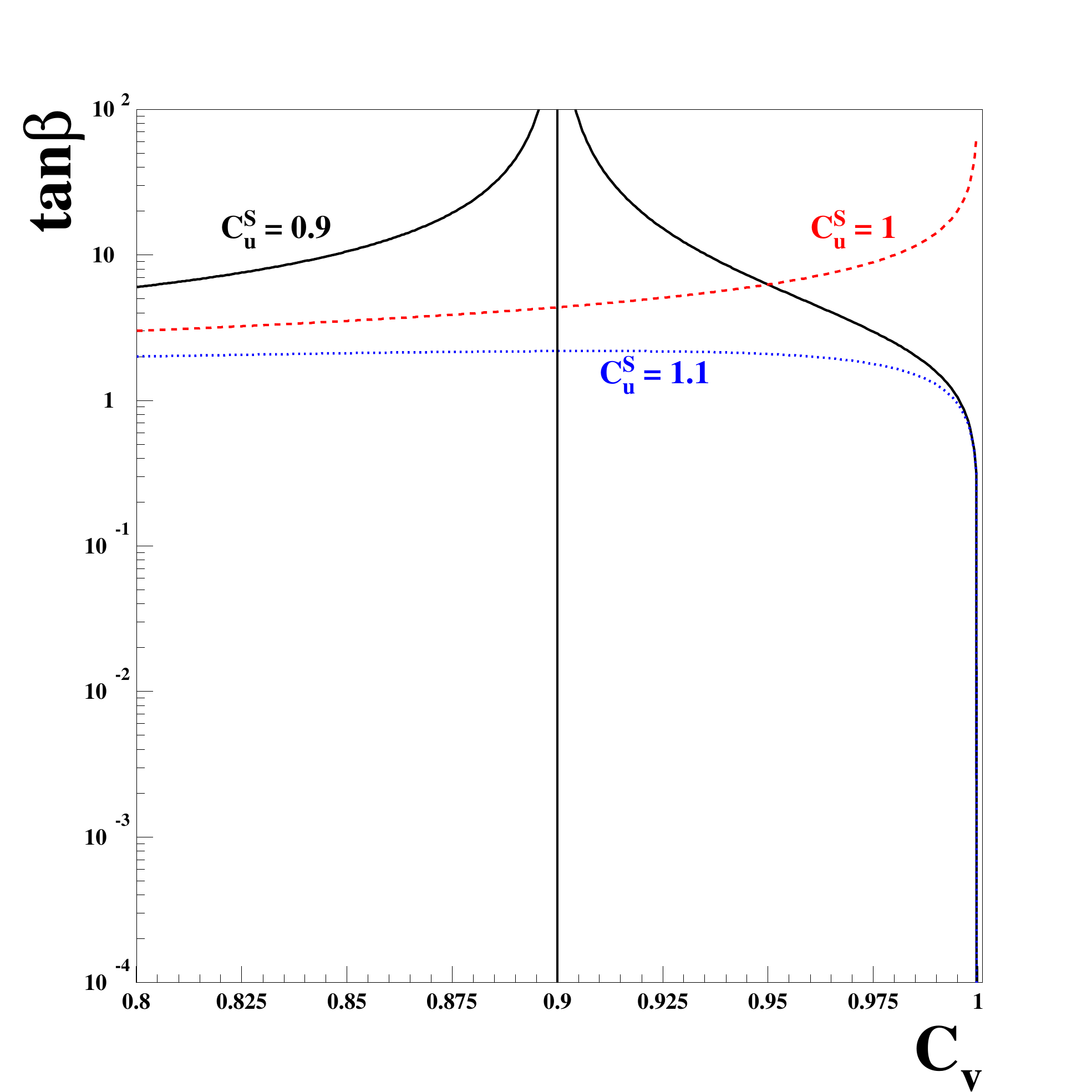}
\caption{\small \label{cpc-cuscvtb}
(Left)
$C_v$ as functions of $C_u^S$ for several values of $\tan\beta=0.1$ (black)
$0.5$ (red), $1$ (blue), $2$ (magenta), and $10$ (black). The horizontal
red line is for the limit $\tan\beta\to 0$ and the straight red line with
$C_v=C_u^S$ represents the limits $\tan\beta\to \infty$. The SM point with
$C_v=C_u^S=1$ is denoted by $\oplus$. (Right) $\tan\beta$ as functions
$C_v$ for three values of $C_u^S=0.9$ (black), $1$ (red), and $1.1$ (blue).
The vertical line shows the location $C_v=0.9$.
}
\end{figure}

In the CP-conserving case, Eq.~(\ref{eq:cv}) simplifies into
\begin{eqnarray}
C_v&=& \pm c_\beta \left[1-s_\beta^2(C_u^S)^2\right]^{1/2}
+s_\beta^2 \,C_u^S\,, \nonumber 
\end{eqnarray}
with the constraint $|s_\beta C_u^S | \leq 1$, which can be 
recast into the form
\begin{equation}
\label{eq:25}
-\frac{1}{t_\beta} \leq C_u^S \leq\sqrt{1+\frac{1}{t_\beta^2}}\,,
\end{equation}
taking into account our convention of $C_v>0$.
For a given value of $\tan\beta$,
we find that $C_v$ takes the plus($+$) sign as $C_u^S$
increases from $-1/t_\beta$ (where $C_v=0$) to $\sqrt{1+1/t_\beta^2}$. While
it takes the minus($-$) sign when $C_u^S$ goes from the maximum value 
$\sqrt{1+1/t_\beta^2}$ back to $1/t_\beta$ where again $C_v=0$. 
Therefore, $C_v$ has two positive
solutions if $C_u^S$ lies between $1/t_\beta$ and  $\sqrt{1+1/t_\beta^2}$.
This behavior is shown in the left frame of Fig.~\ref{cpc-cuscvtb}.
{}From Eq.~(\ref{eq:sbsq}) which now can be rearranged into the form
\begin{eqnarray}
s_\beta^2&=&\frac{(1-C_v^2)}{(1-C_v^2)+(C_u^S-C_v)^2}\nonumber\,,
\end{eqnarray}
we can see that $\sin\beta=1$ or $\tan\beta=\infty$ along the line
$C_v=C_u^S$. 
Also, the larger $\tan\beta$ the smaller $C_v$ will be.
Therefore,  
$\tan\beta$ will be bounded from above when $C_v$ is pushed to be close to $1$, 
unless $C_u^S=1$.

To be more precise, we consider the situation in which
$C_v$ is constrained as $C_v>(C_v)_{\rm min}$.
As illustrated in the right frame of Fig.~\ref{cpc-cuscvtb}
with three values of $C_u^S=0.9$ (black), $1$ (red), and $1.1$ (blue),
we have found that
$\tan\beta$ has an upper bound when $C_u^S<(C_v)_{\rm min}$ 
for $C_u^S<1$.
We observe that the upper bound on $\tan\beta$ is stronger when
$(C_v)_{\rm min}$ is closer to $1$ but it disappears when $(C_v)_{\rm min}<C_u^S$ 
or $C_u^S=1$.
On the other hand, when
$C_u^S >1$, $\tan\beta$ is always bounded by
$\tan\beta \leq 1/\sqrt{(C_u^S)^2-1}$, see Eq.~(\ref{eq:25}).
Requiring $C_v>0.95$, for example, we find $\tan\beta\lsim 6$ for
$C_u^S=0.9$ and $\tan\beta\lsim 1/\sqrt{(C_u^S)^2-1} \simeq 2$ for
$C_u^S=1.1$. 

In the following sub-subsections,  we illustrate that
the precise and independent measurements of $C_u^S$ and $\tan\beta$ 
can tell us the phenomenological viability of 2HDMs and/or
enable us to make discrimination among them.

The results for various fits ({\bf CPC2}, {\bf CPC3}, {\bf CPV3}, and 
{\bf CPV4}) are tabulated
in Tables~\ref{tab:best} and \ref{tab:best2}, 
and confidence regions are shown in 
Figs.~\ref{cpc1-cucv} -- \ref{cpv3-cl-clp}.

\subsubsection{\bf CPC2}

The fit {\bf CPC2} analyzes the Higgs data by varying $C_u^S$ and    
$C_v$ (or equivalently $\log_{10}\tan\beta$). The total $\chi^2$,
$\chi^2/{\rm dof}$, $p$-value and the best-fit values  of
$C_u^S$, $C_v$, and $\tan\beta$ for the types I -- IV of 2HDMs
are shown at the top of Table~\ref{tab:best}.
We have found that the type-I model gives the smallest $\chi^2$ 
but the variation of
total $\chi^2$ among the 4 types is very small, within $0.29$. 
Statistically, there is no preference among any type I to IV of 2HDMs.
We note that the $p$-values of the fits are all worse than the SM one
$p_{\rm SM} = 0.65$.
The best-fit values
for $C_u^S$ are about $0.9$ for type I and III, and about 
$0.96$ for type II and IV.
The fitted $C_v$'s are very close to the theoretically allowed maximum value
$1$ independent of the type.
In the actual implementation, we used $\log_{10}\tan\beta$ as the scanning
variable with $-4 <  \log_{10}\tan\beta < 2$, instead of $C_v$.
Again, independent of the type,
$\chi^2$ continues to decrease as $\tan\beta$ falls below its lower
limit $\tan\beta=10^{-4}$, though extremely slowly.  
The best fitted values for $\tan\beta$ are denoted by
{\it limit} in Table~\ref{tab:best}.

\begin{figure}[th!]
\centering
\includegraphics[width=3.2in,height=2.4in]{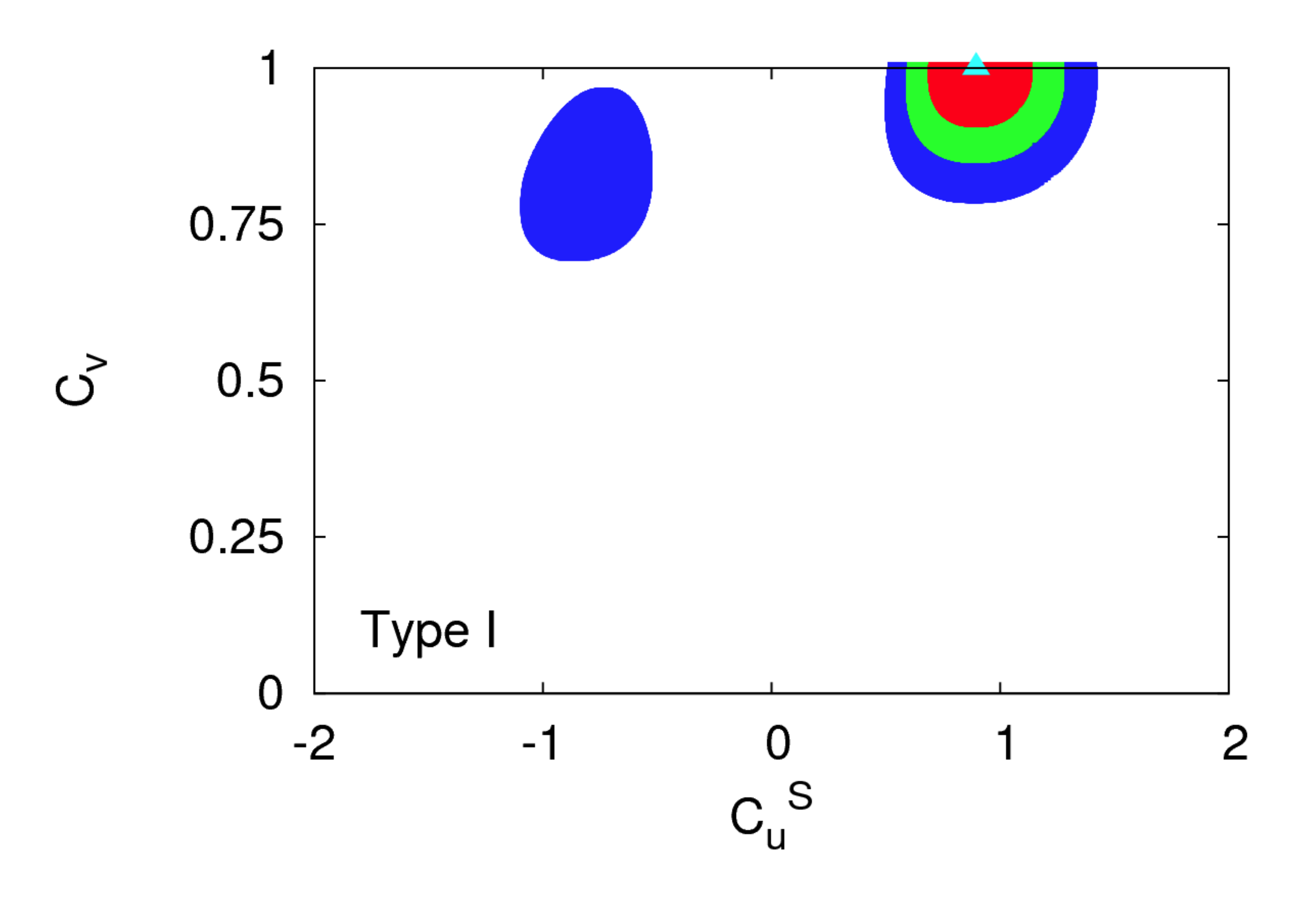}
\includegraphics[width=3.2in]{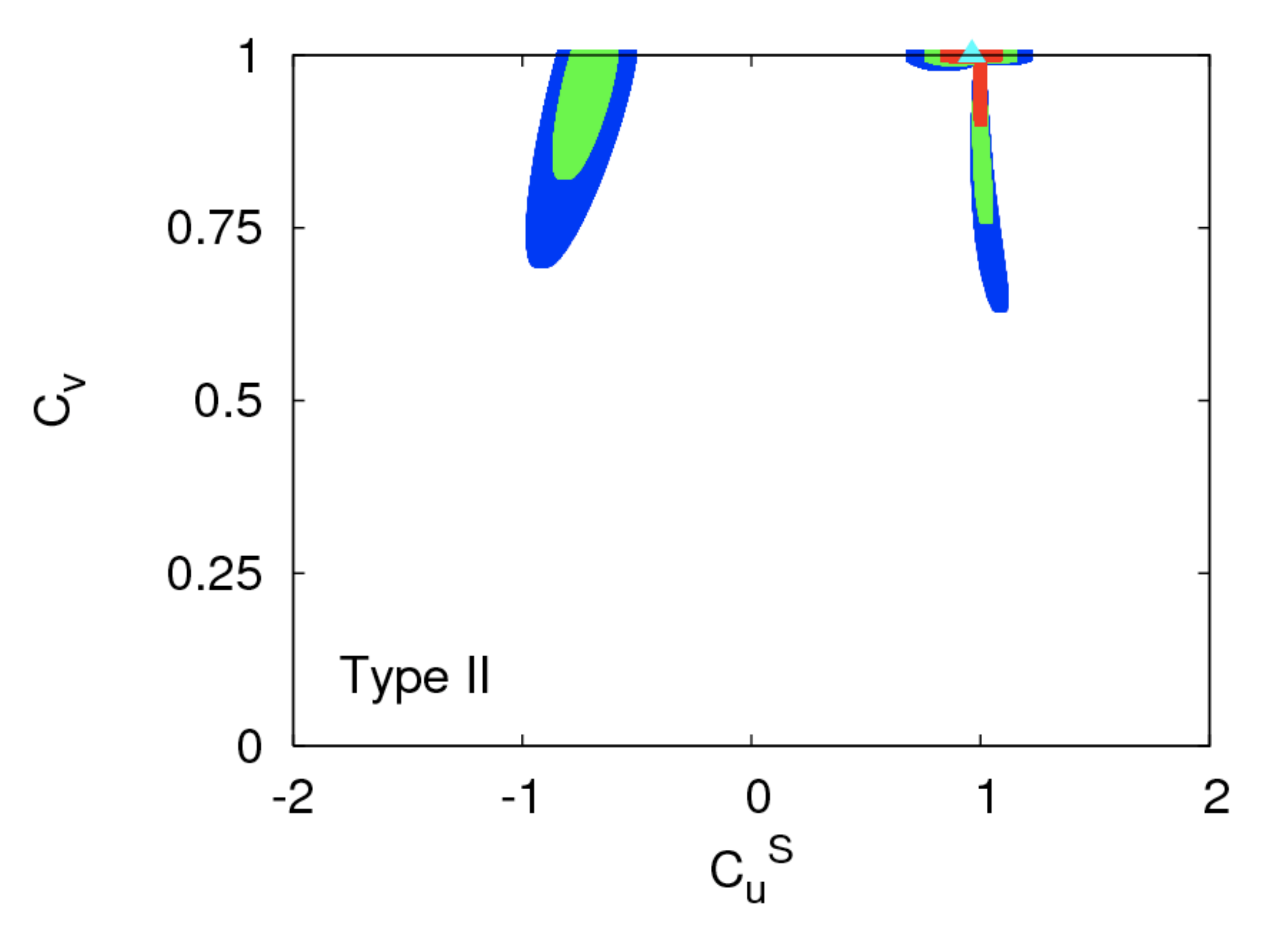}
\includegraphics[width=3.2in]{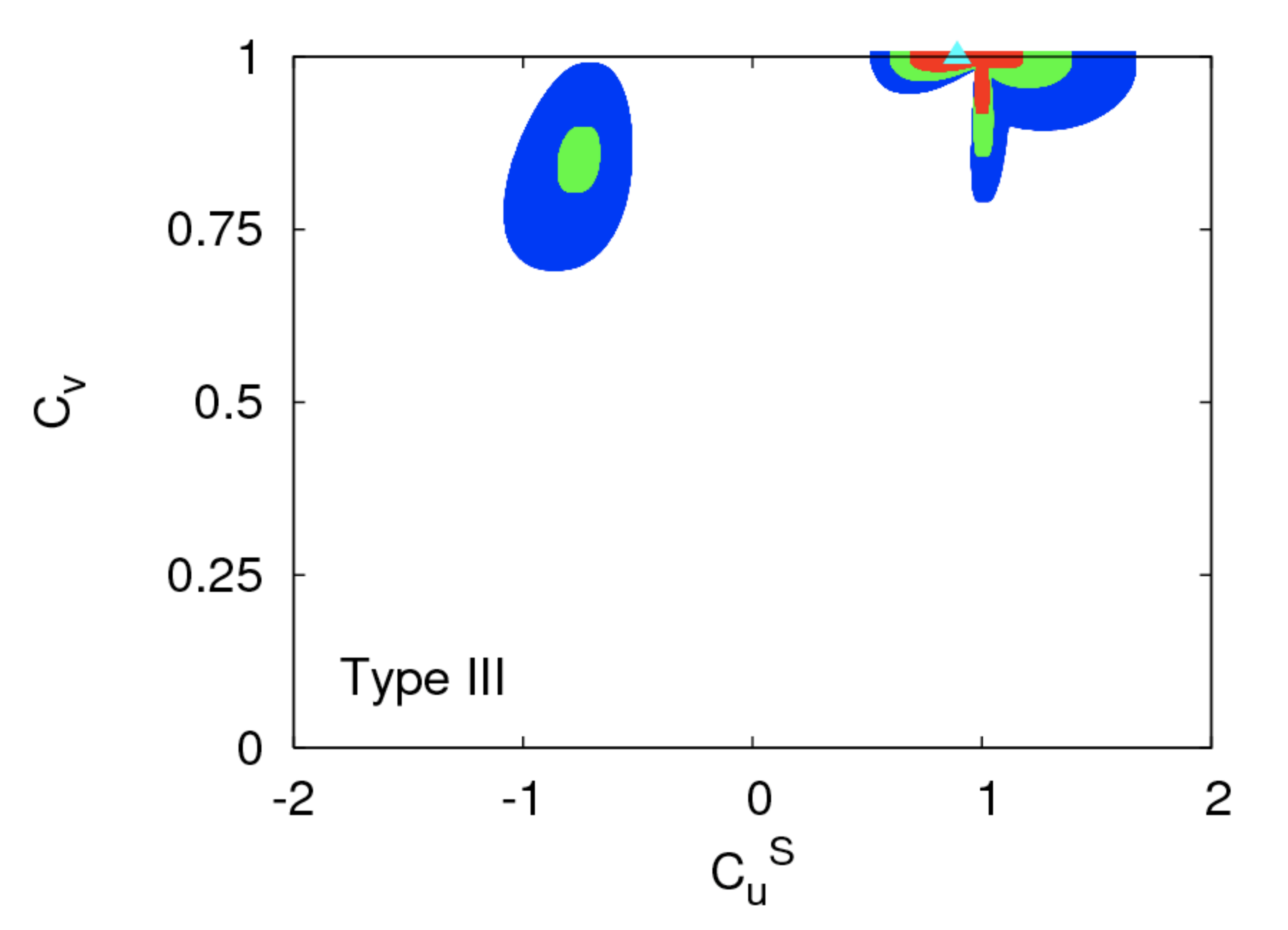}
\includegraphics[width=3.2in]{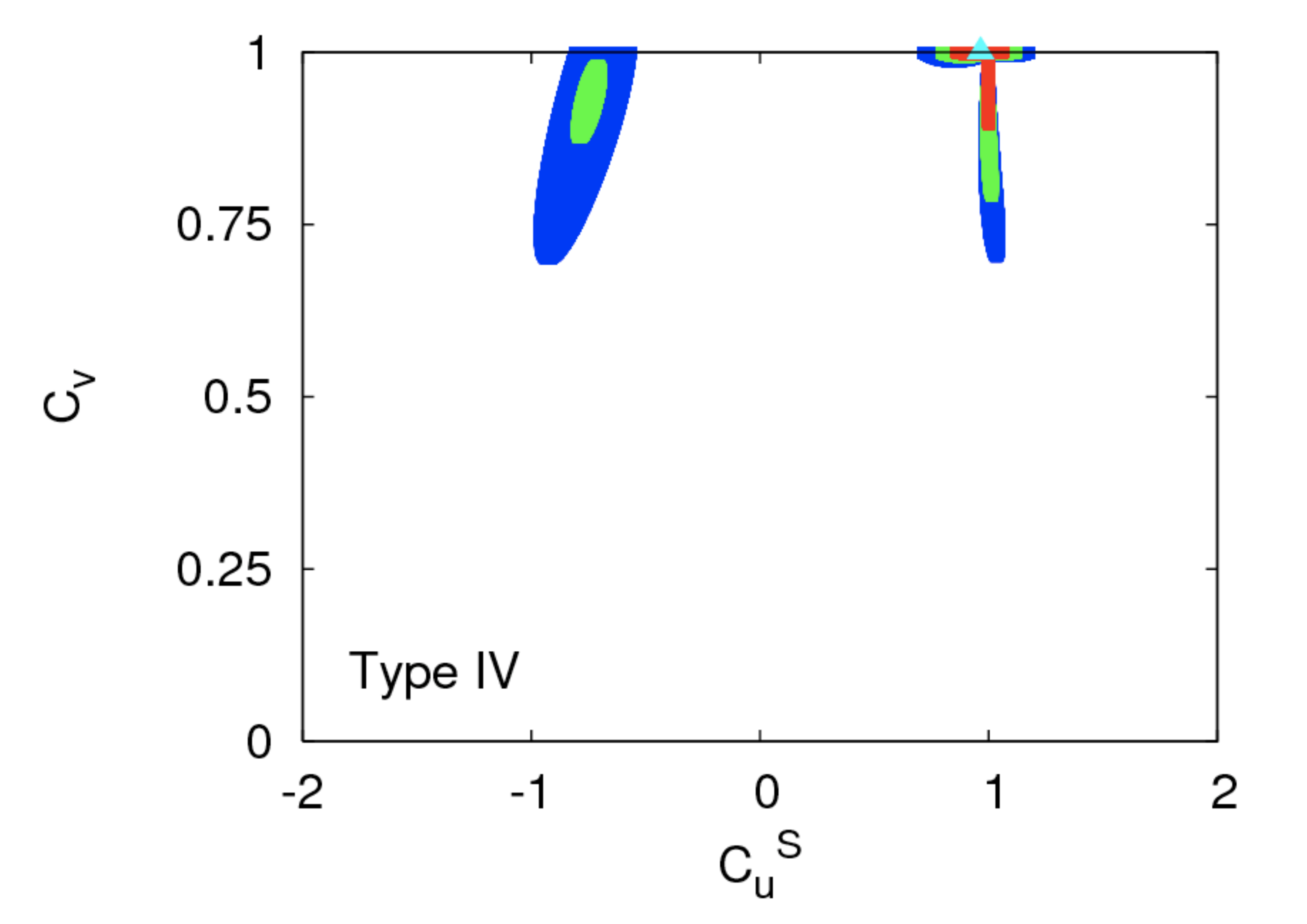}
\caption{\small \label{cpc1-cucv}
The confidence-level regions of the fit by varying $C_u^S$ and 
$C_v$ (or equivalently $\log_{10}\tan\beta$) only ({\bf CPC2} case) in the
plane of $C_u^S$ vs $C_v$ for Type I -- IV.
The contour regions shown are for 
$\Delta \chi^2 \le 2.3$ (red), $5.99$ (green), and $11.83$ (blue) 
above the minimum, which 
correspond to confidence levels of
$68.3\%$, $95\%$, and $99.7\%$, respectively.
The best-fit points are denoted by the triangle.
}
\end{figure}

\begin{figure}[th!]
\centering
\includegraphics[width=3.2in]{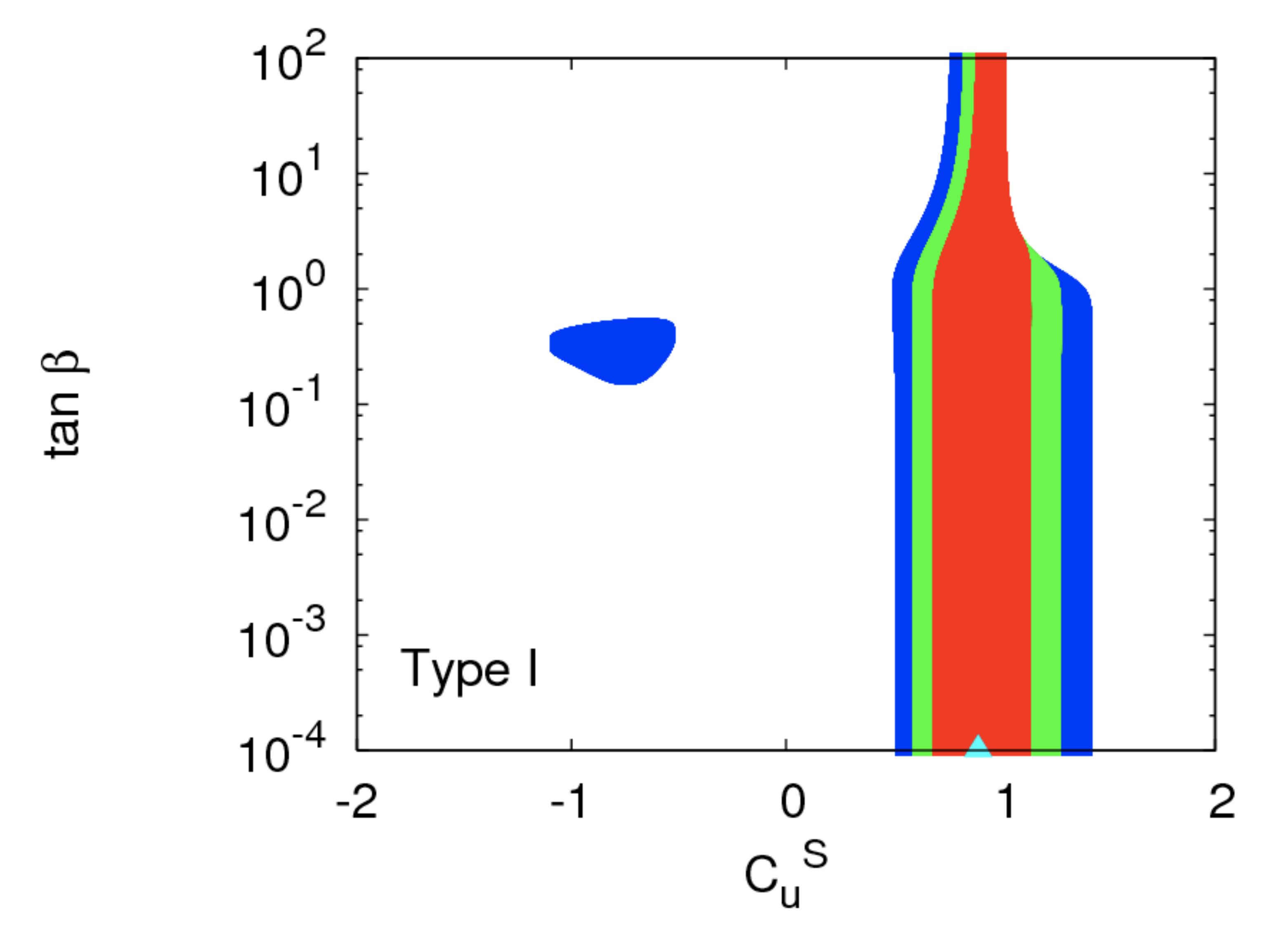}
\includegraphics[width=3.2in]{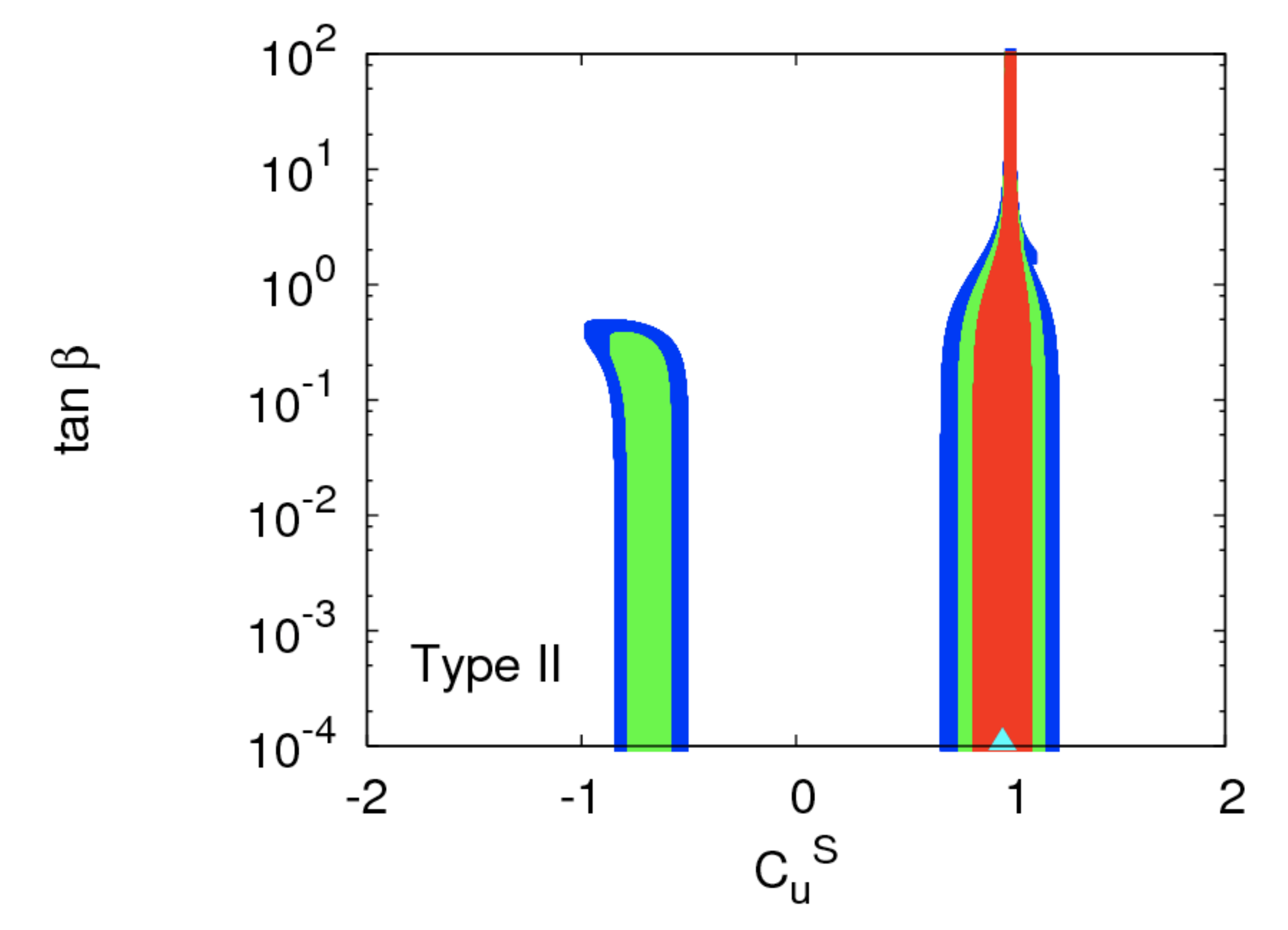}
\includegraphics[width=3.2in]{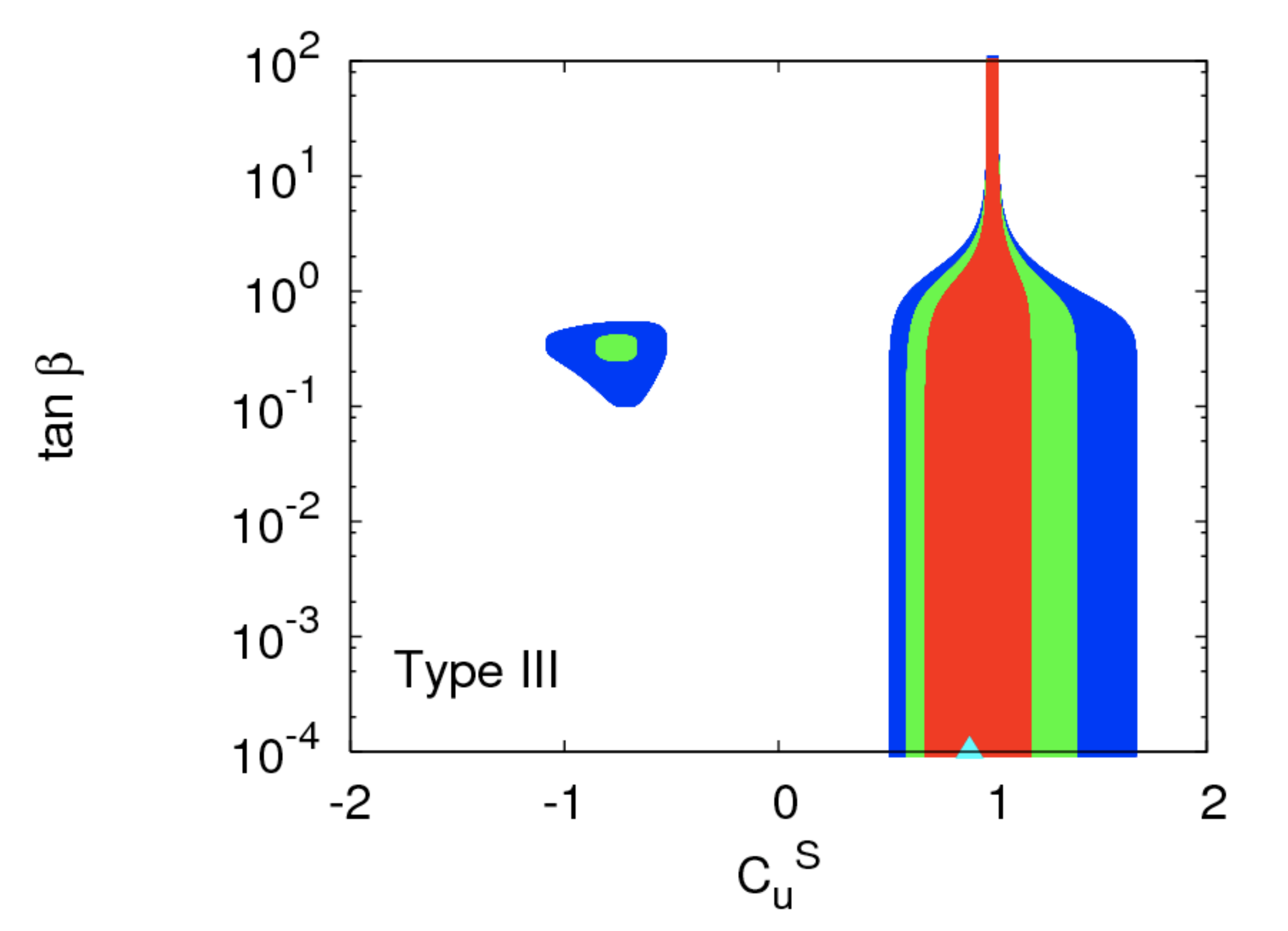}
\includegraphics[width=3.2in]{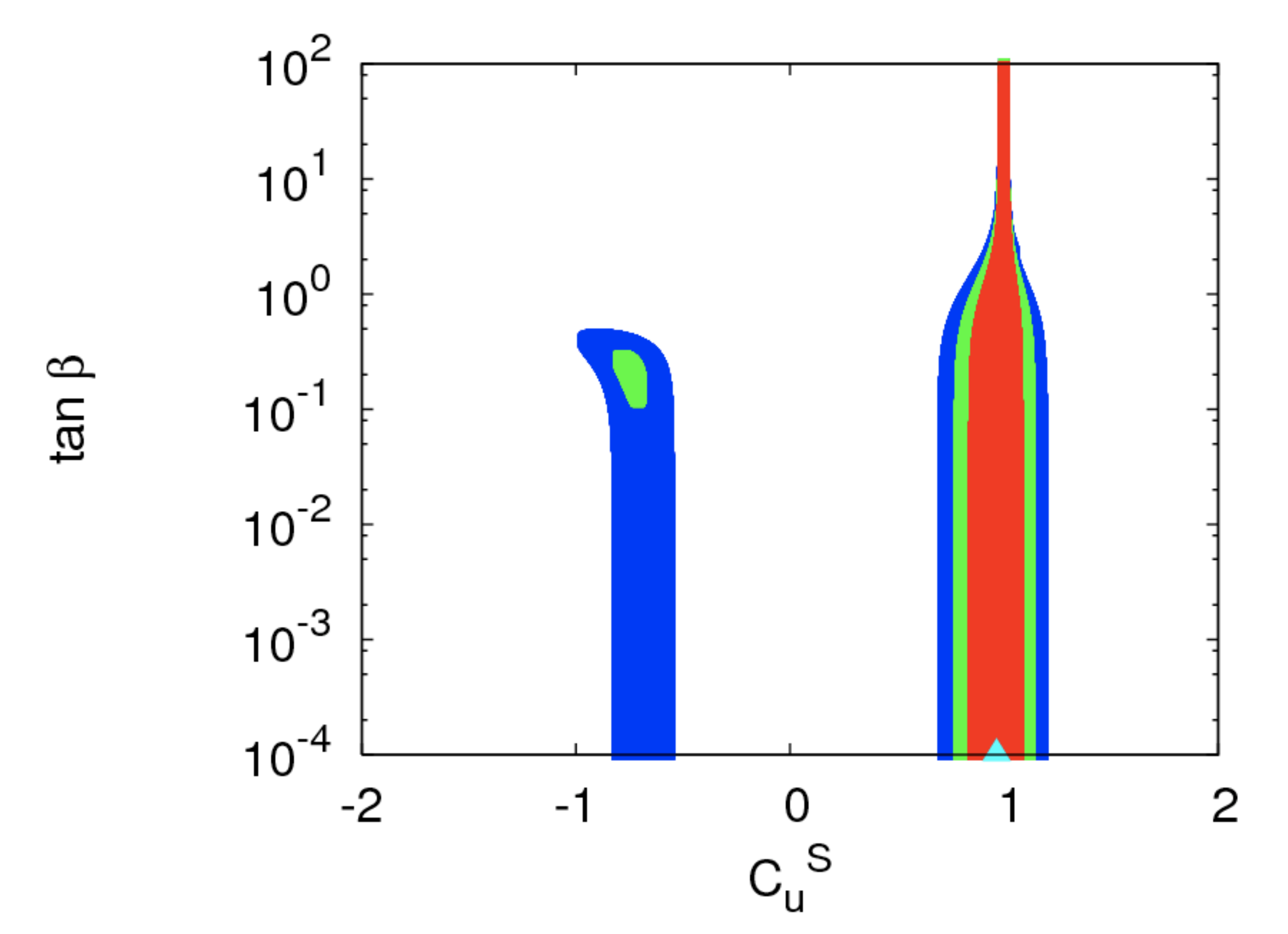}
\caption{\small \label{cpc1-cutan}
The same as Fig.~\ref{cpc1-cucv} but in the plane of
$C_u^S$ vs $\tan\beta$ ({\bf CPC2}).
The description of the confidence regions is the same as Fig.~\ref{cpc1-cucv}.
}
\end{figure}

\begin{figure}[th!]
\centering
\includegraphics[width=3.2in]{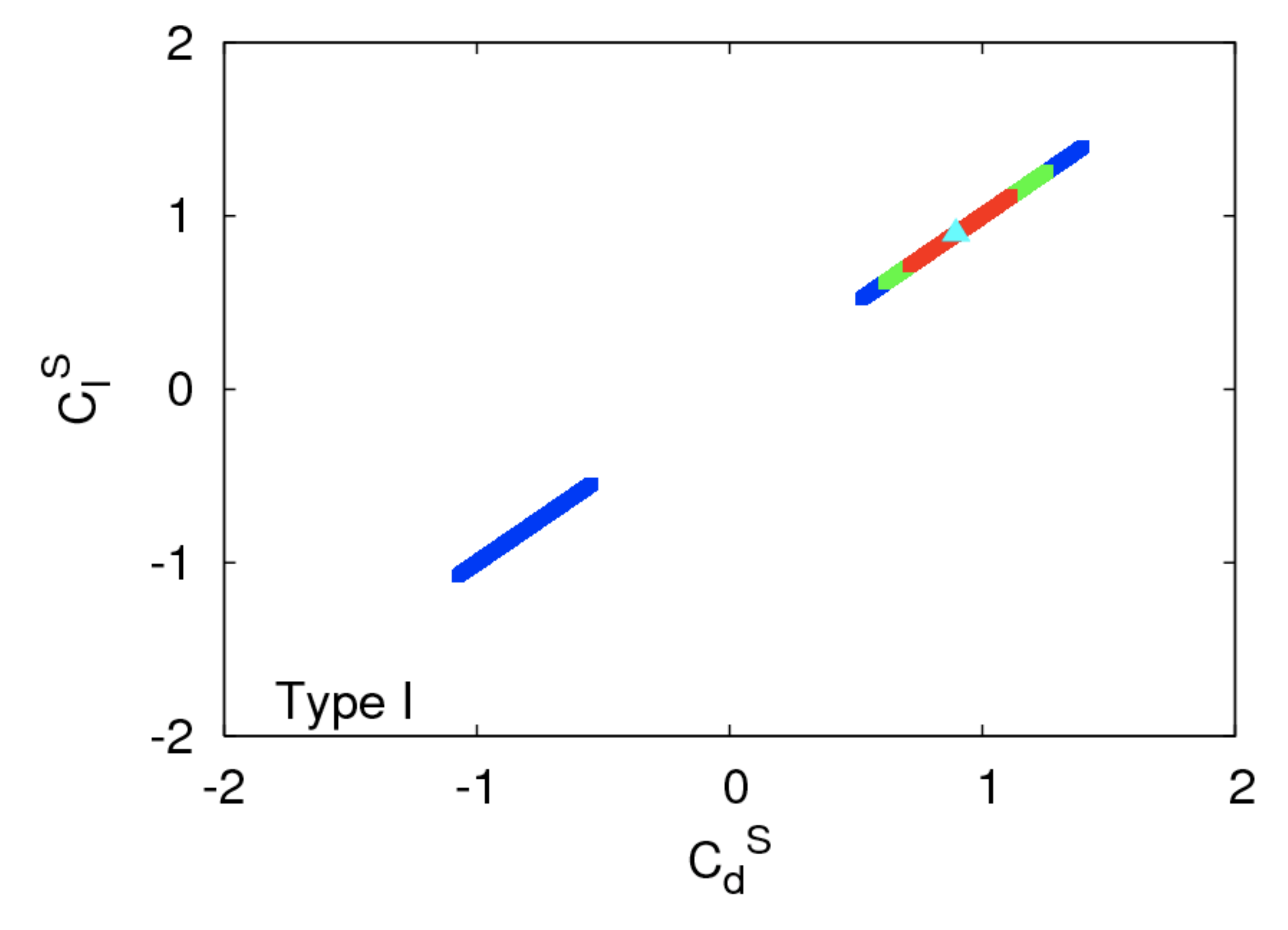}
\includegraphics[width=3.2in]{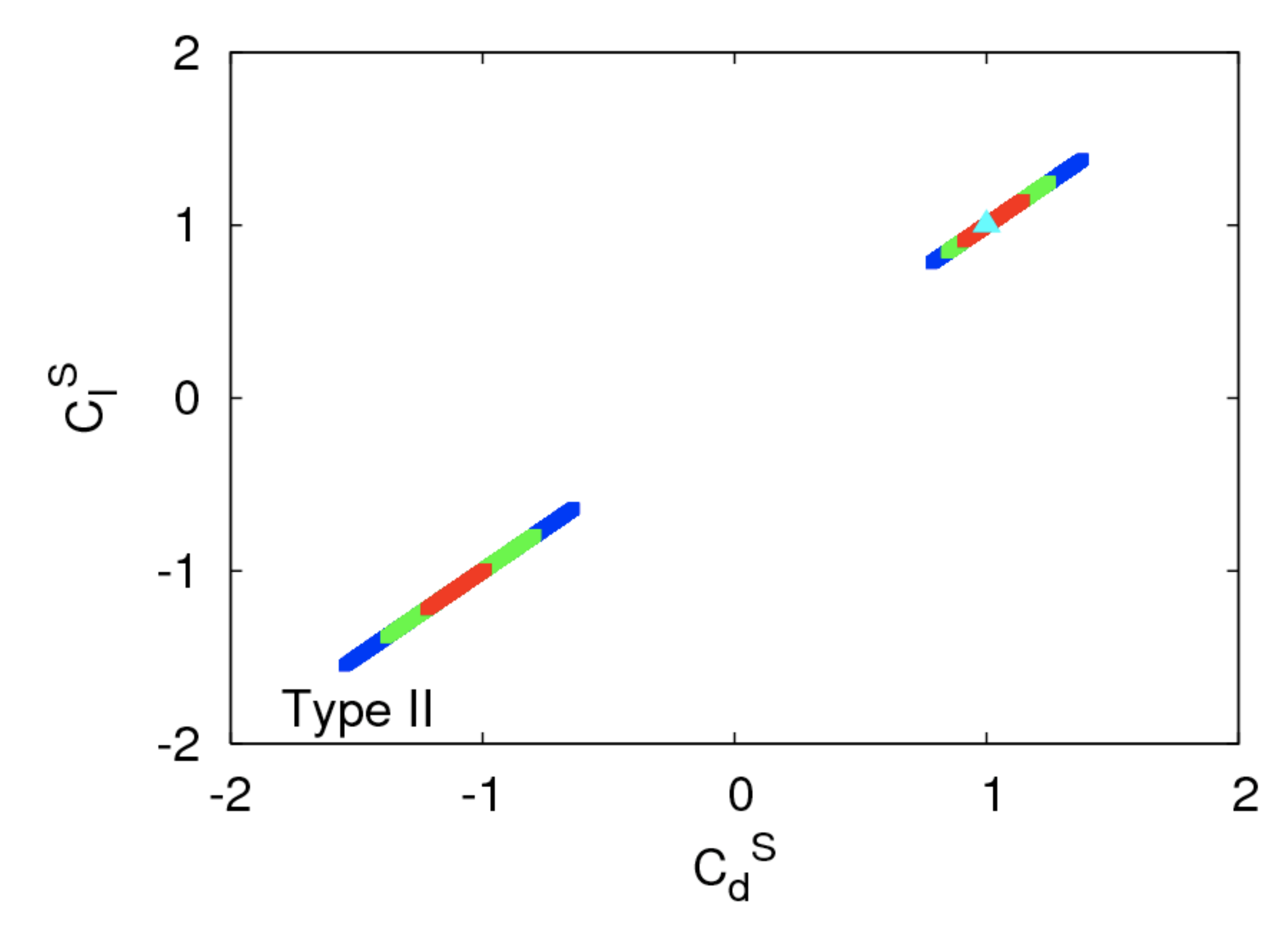}
\includegraphics[width=3.2in]{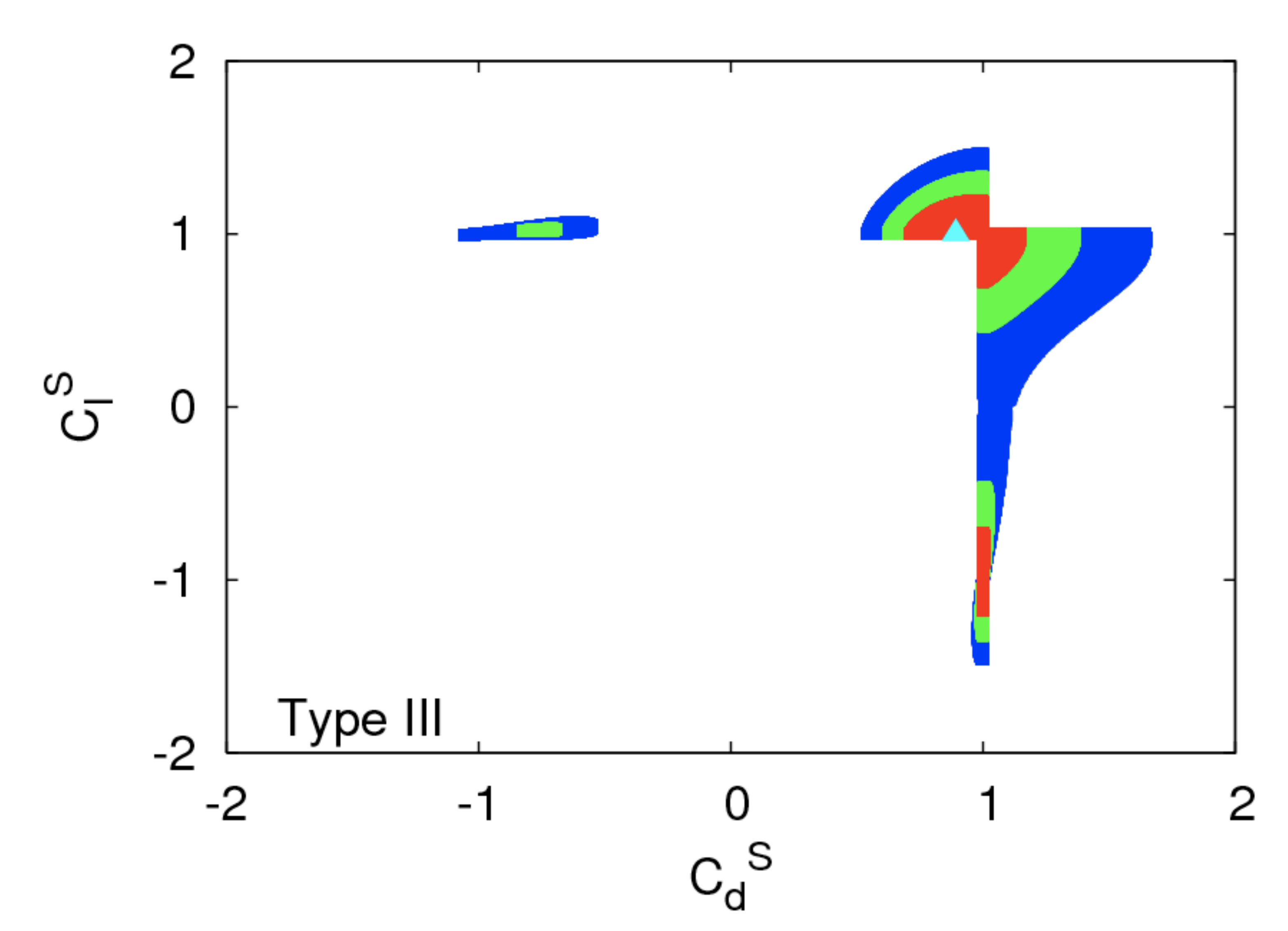}
\includegraphics[width=3.2in]{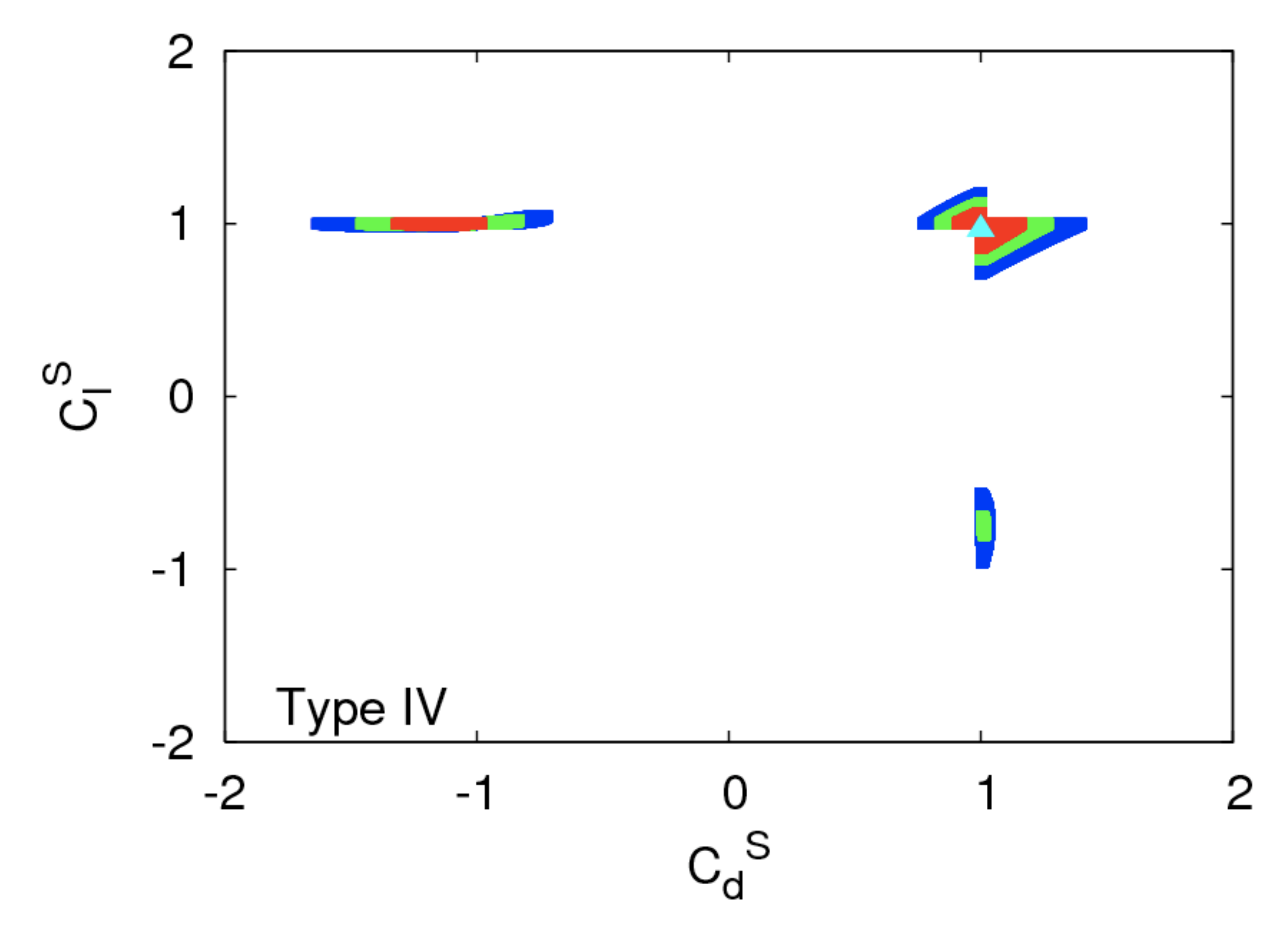}
\caption{\small \label{cpc1-cdcl}
The same as Fig.~\ref{cpc1-cucv} but in the plane of
$C_d^S$ vs $C_\ell^S$ ({\bf CPC2}).
The description of the confidence regions is the same as Fig.~\ref{cpc1-cucv}.
}
\end{figure}

We show the contour plots for confidence-level regions
as functions $C_u^S$ vs $C_v$, $C_u^S$ vs $\tan\beta$, and
$C_d^S$ vs $C_l^S$ in Figs.~\ref{cpc1-cucv} -- \ref{cpc1-cdcl}, 
respectively.
The regions shown are for 
$\Delta \chi^2 \le 2.3$ (red), $5.99$ (green), and $11.83$ (blue) 
above the minimum, which 
correspond to confidence levels of
$68.3\%$, $95\%$, and $99.7\%$, respectively.
The best-fit point is denoted by the triangle.
We note that from Fig.~\ref{cpc1-cucv} there are two islands and 
positive $C_u^S$ is preferred.
At $99.7\%$ confidence level (CL), $C_v\gsim 0.7$.
We also find that $C_u^S$ takes on the values between
$0.71$ and $1.2$ (I),
$0.86$ and $1.1$ (II),
$0.71$ and $1.2$ (III), and
$0.86$ and $1.1$ (IV)
at 68.3 \% CL.
Comparing type I with the other three types, we find that
the preference for $C_u^S=1$ is stronger in type II, III, and IV,
and $C_v$ is more strongly constrained to be close to $1$ unless $C_u^S=1$.
Furthermore, the $\tan\beta=\infty$ line with $C_u^S=C_v$ passes
through the CL regions only in type I.

In Fig.~\ref{cpc1-cutan}, we show the CL regions in the plane of $C_u^S$
and $\tan\beta$. For $\tan\beta \lsim 0.5$, we find $\chi^2$ is 
almost independent
of $\tan\beta$ for a fixed value of $C_u^S$; while for $\tan\beta \gsim 1$, 
the values of $C_u^S$ is constrained 
by $C_u^S \leq \sqrt{1+1/t_\beta^2}$.
For type I, 
as we observed in Fig.~\ref{cpc1-cucv}, the $\tan\beta=\infty$ 
line passes through the CL regions and it explains
why we can have very large $\tan\beta$ in relatively broader range 
of $C_u^S$. For the other three types it is only possible to have very large
$\tan\beta$ in the narrow region around $C_u^S=1$. Thus, in these cases
we find that 
$\tan\beta\lsim 3$ (II), $2$ (III), $3$ (IV) 
at 99.7 \% CL when 
the best-fit value of $C_u^S$ is taken in each of the type II, III,
and IV.
If precise and independent measurements of $C_u^S$ and $\tan\beta$ 
are available in future experiments, one
can tell the phenomenological viability of 2HDMs. For example, 
if $\tan\beta\gsim 10$  and $C_u^S\neq 1$, then one can
rule out the type II, III, and IV models based on Fig.~\ref{cpc1-cutan}.

In Fig.~\ref{cpc1-cdcl}, we show the CL regions in the plane of $C_d^S$
and $C_l^S$. From Table~\ref{tab:cdcl},
the following relations 
$C_d^S=C_l^S=C_u^S$ (I),
$C_d^S=C_l^S$ (II),
$C_d^S=C_u^S$ (III), and
$C_l^S=C_u^S$ (IV) are hold.
In Table~\ref{tab:best2}, 
we can see that the best-fit values of $C_d^S$
and/or $C_l^S$ are $+1$ unless either or both of them
are equal to $C_u^S$.
This can be understood
from the relation, for example in type II,
\begin{equation}
C_d^S=C_l^S
=\frac{\sqrt{1-s_\beta^2(C_u^S)^2}}{c_\beta }
=\sqrt{1+t_\beta^2[1-(C_u^S)^2]}
\end{equation}
with the best-fit values of $C_u^S=0.963$ and $\tan\beta=limit=10^{-4}$. 
Note that the positive sign is selected to
explain the best-fit values of $C_{d,l}^S$. 
Taking into account the negative sign,
we observe that the points around
$(C_d^S,C_l^S)=(-1,-1)$ (II),
$(C_d^S,C_l^S)=(+1,-1)$ (III), and
$(C_d^S,C_l^S)=(-1,+1)$ (IV) are also allowed at
68.3 \% CL even when $C_u^S$ is positive.

\begin{figure}[th!]
\centering
\includegraphics[width=3.2in]{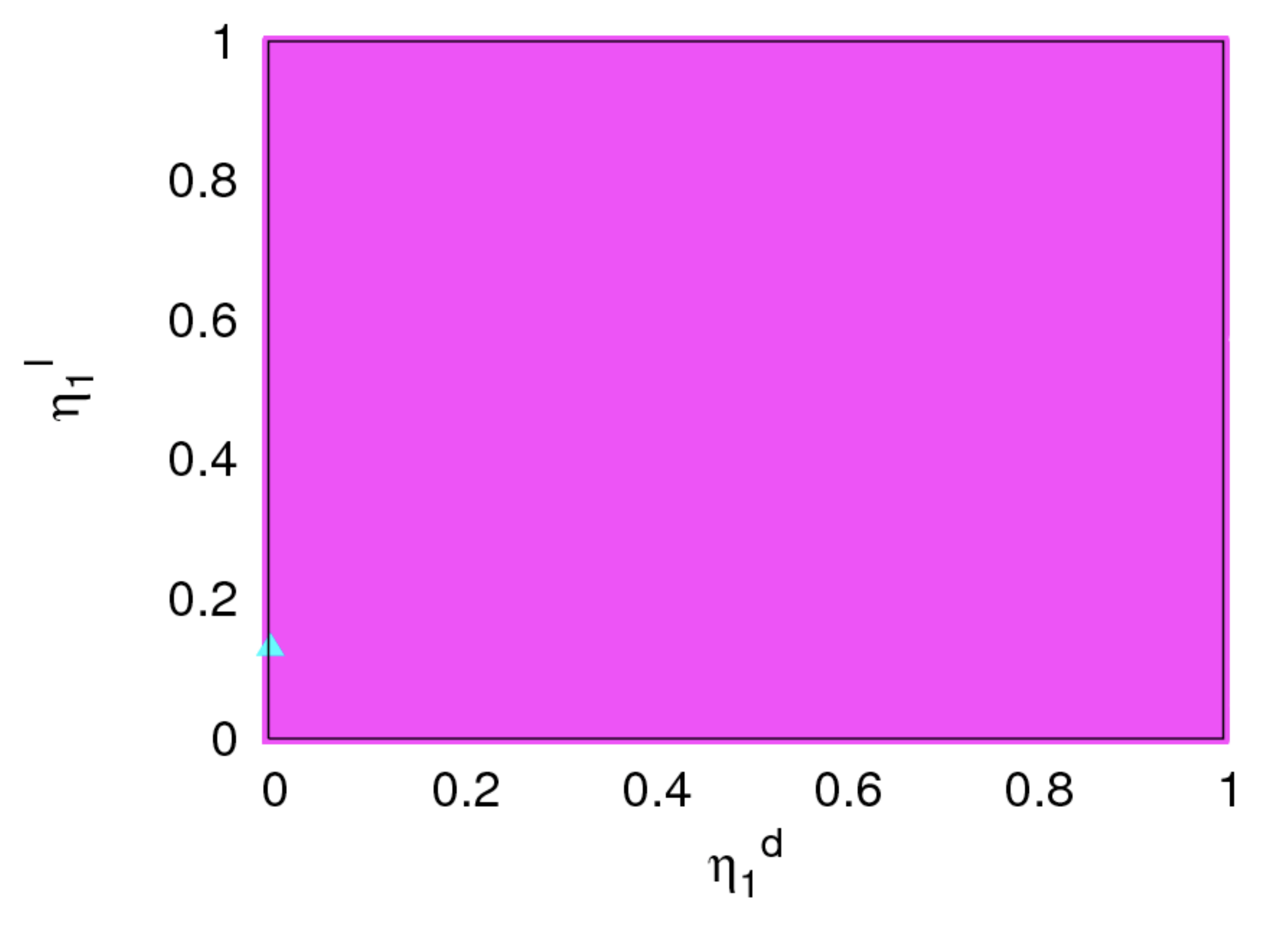}
\caption{\small \label{cpc2}
The confidence-level regions of the fit by varying $C_u^S$,
$C_v$, $\eta_1^d$, and $\eta_1^\ell$ 
in the plane of 
$\eta_1^d$ vs $\eta_1^\ell$.
The best-fit points are denoted by the triangle.
Here the entire region is for $\Delta\chi^2 < 1.0$.
}
\end{figure}

So far in this {\bf CPC2} fit we only found very small $\chi^2$ differences
among the four types. What if the discrete symmetries are relaxed,
do we get a better $\chi^2$ fit?
We relax the requirement on the discrete symmetries, which enforces
$\eta_{1,2}^{d,\ell}$ to be either $0$ or $1$, but still require
$(\eta_1^{d,\ell})^2+(\eta_2^{d,\ell})^2=1$. We therefore have two more
free parameters in our scan, and they are 
$\eta_1^{d,\ell}$, leading to
a four-parameter fit by varying
$C_u^S$, $C_v$, $\eta_1^d$, and $\eta_1^\ell$.
In Fig.~\ref{cpc2}, we show the CL
regions of the fit by varying $C_u^S$,
$C_v$, $\eta_1^d$, and $\eta_1^\ell$ 
in the plane of $\eta_1^d$ and $\eta_1^\ell$
\footnote{We obtain the minimum 
$\chi^2=18.30$ and $\chi^2/{\rm dof}=1.02$ for this fit.}. 
We observe that
$\Delta\chi^2<1$ in the whole $(\eta_1^d,\;\eta_1^\ell)$ plane,
and so conclude that one cannot say any preference
based on the current Higgs data.

\subsubsection{\bf CPC3}

\begin{figure}[th!]
\centering
\includegraphics[width=3.2in]{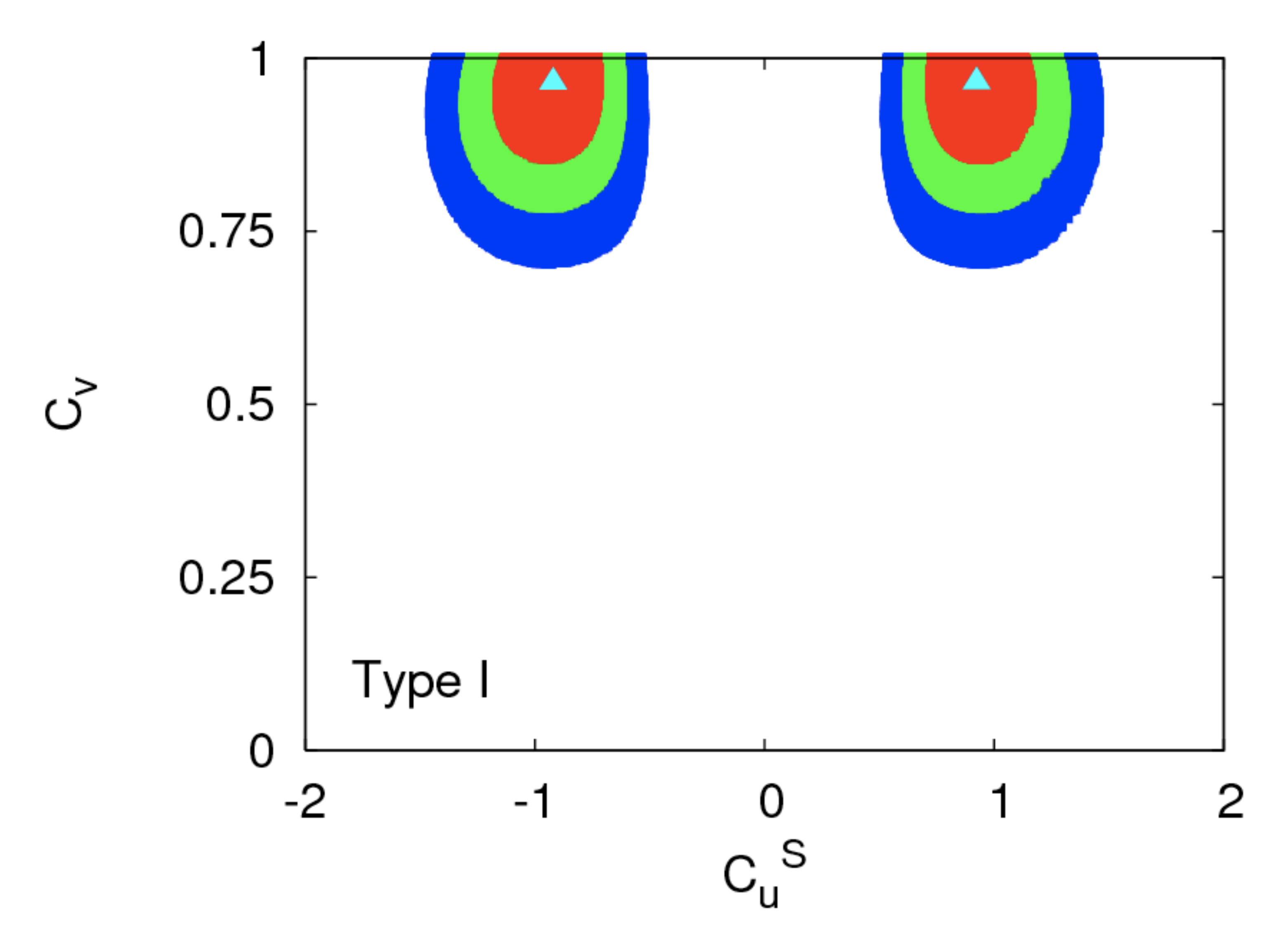}
\includegraphics[width=3.2in]{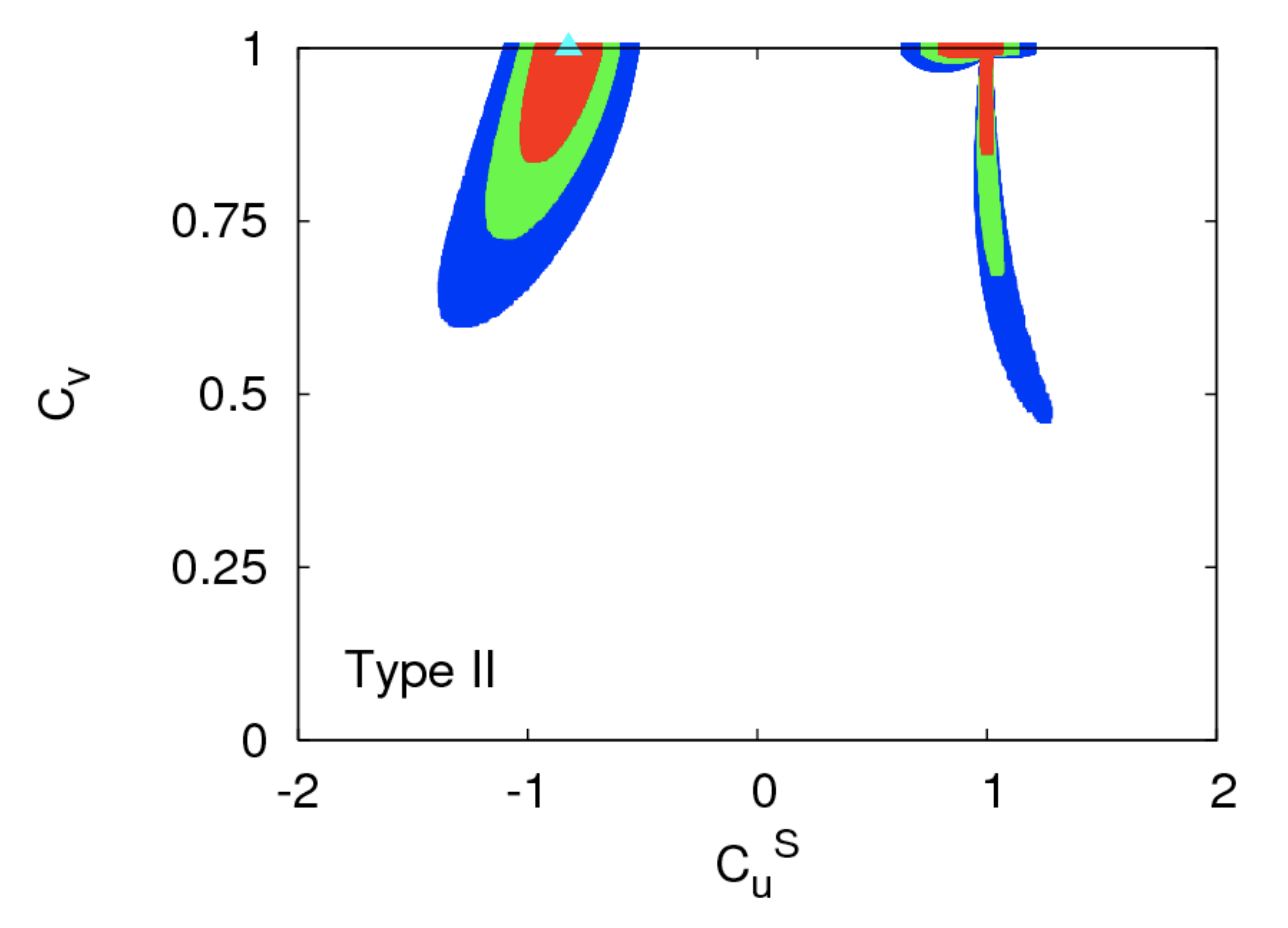}
\includegraphics[width=3.2in]{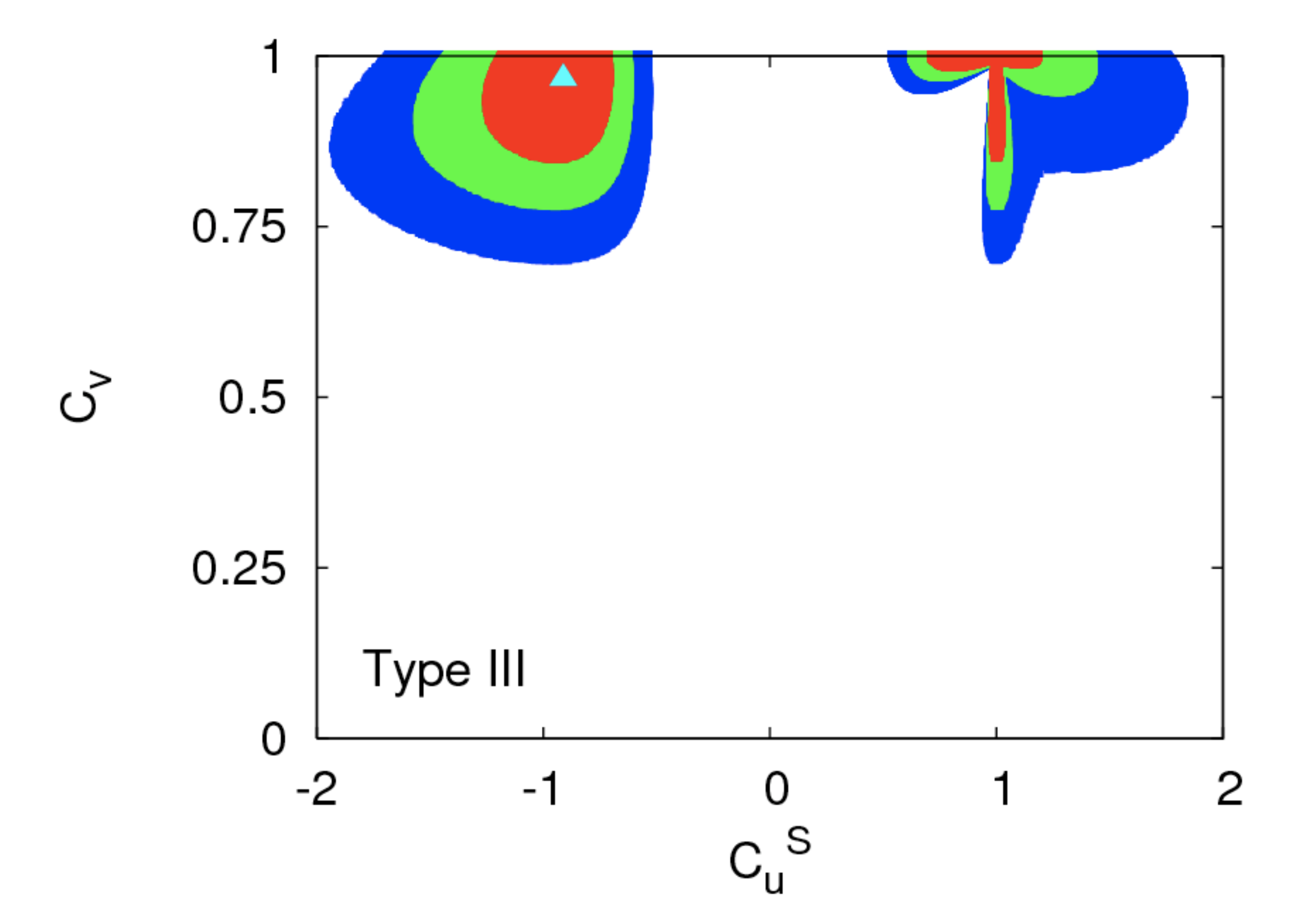}
\includegraphics[width=3.2in]{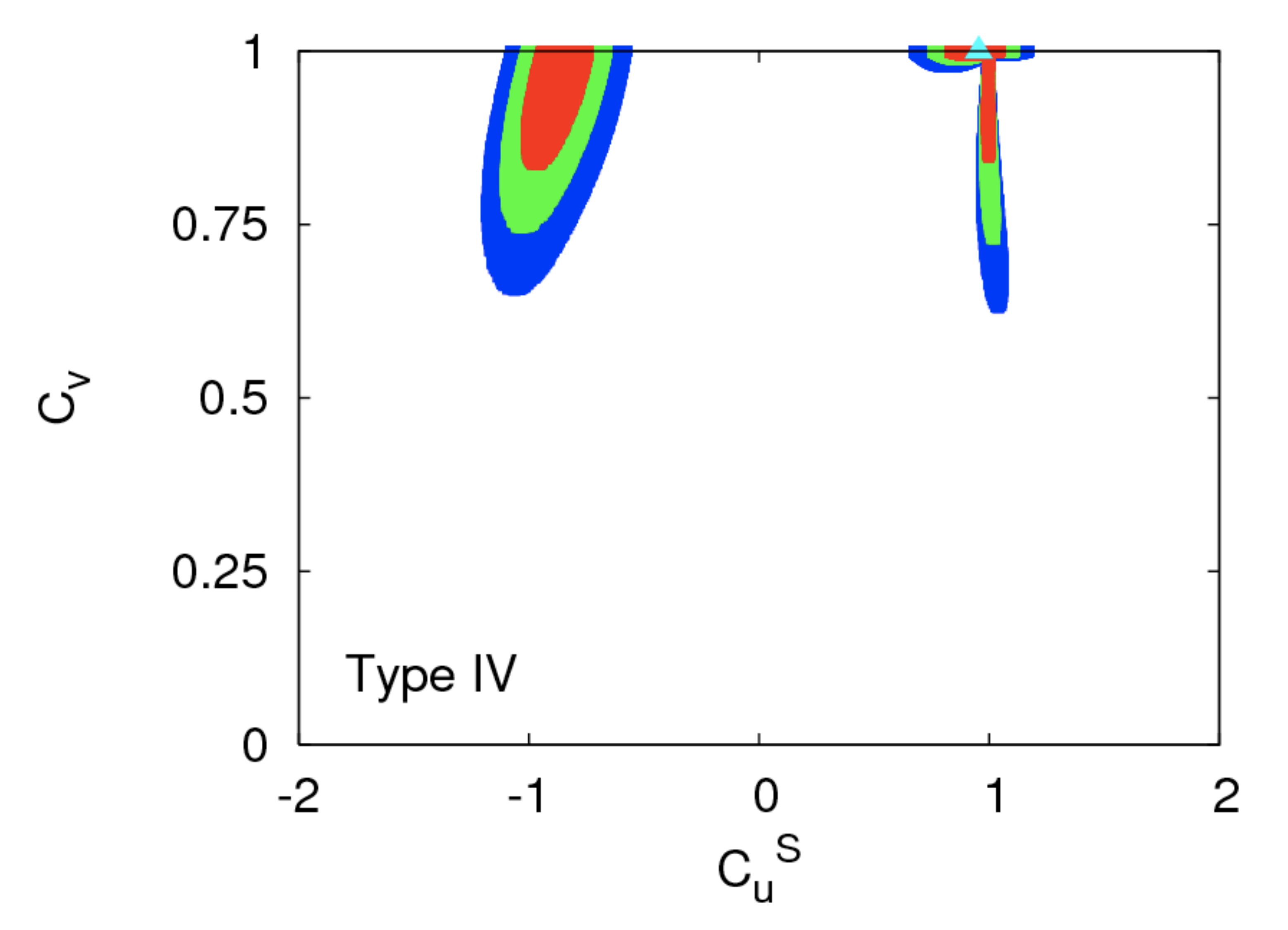}
\caption{\small \label{cpc3-cucv}
The confidence-level regions of the fit by varying $C_u^S$,
$\log_{10} \tan\beta$, and $\Delta S^\gamma $ ({\bf CPC3} case) 
in the plane of $C_u^S$ vs $C_v$ for Type I -- IV.
The contour regions shown are for 
$\Delta \chi^2 \le 2.3$ (red), $5.99$ (green), and $11.83$ (blue) 
above the minimum, which 
correspond to confidence levels of
$68.3\%$, $95\%$, and $99.7\%$, respectively.
The best-fit points are denoted by the triangle.
}
\end{figure}

\begin{figure}[th!]
\centering
\includegraphics[width=3.2in]{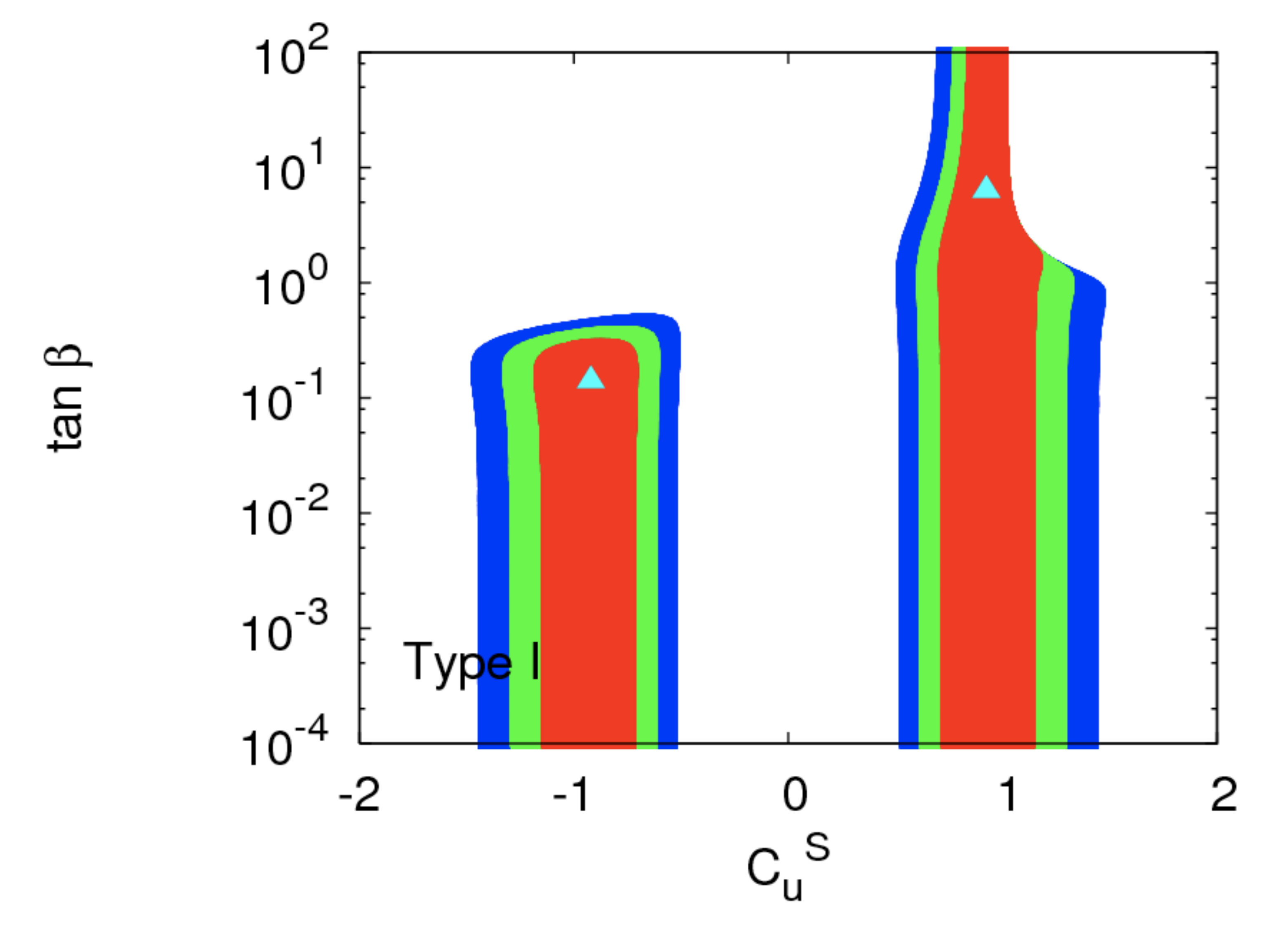}
\includegraphics[width=3.2in]{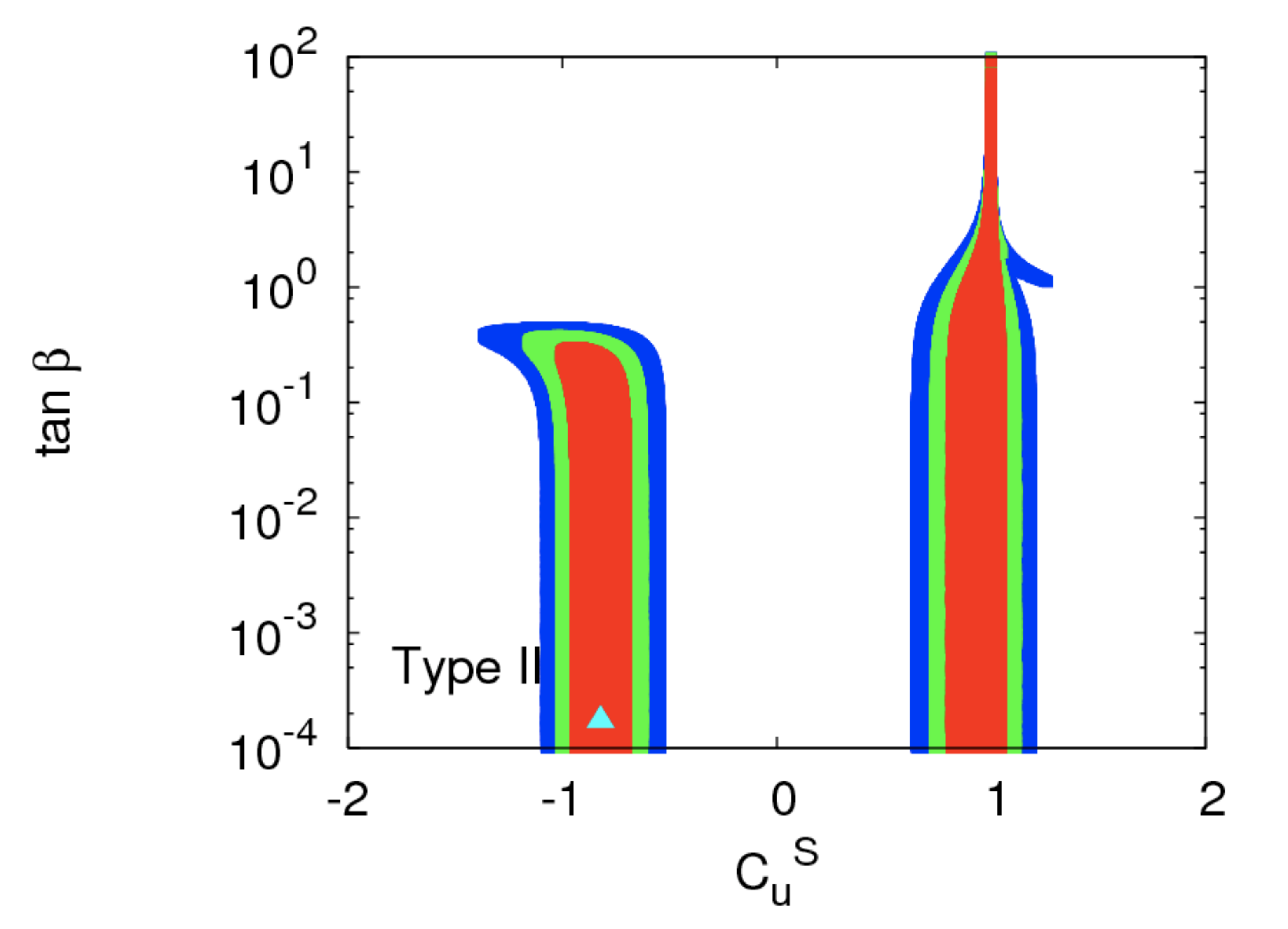}
\includegraphics[width=3.2in]{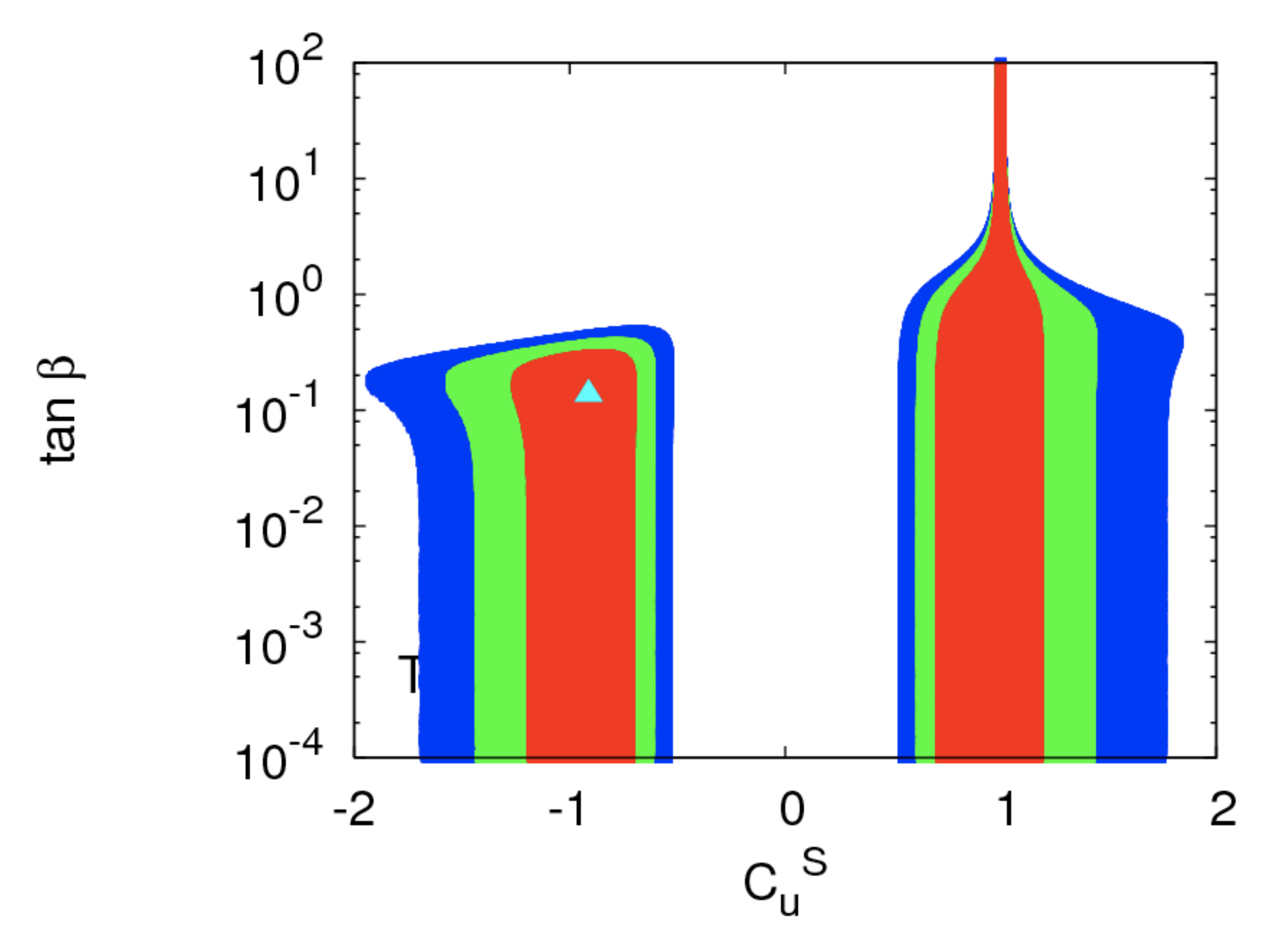}
\includegraphics[width=3.2in]{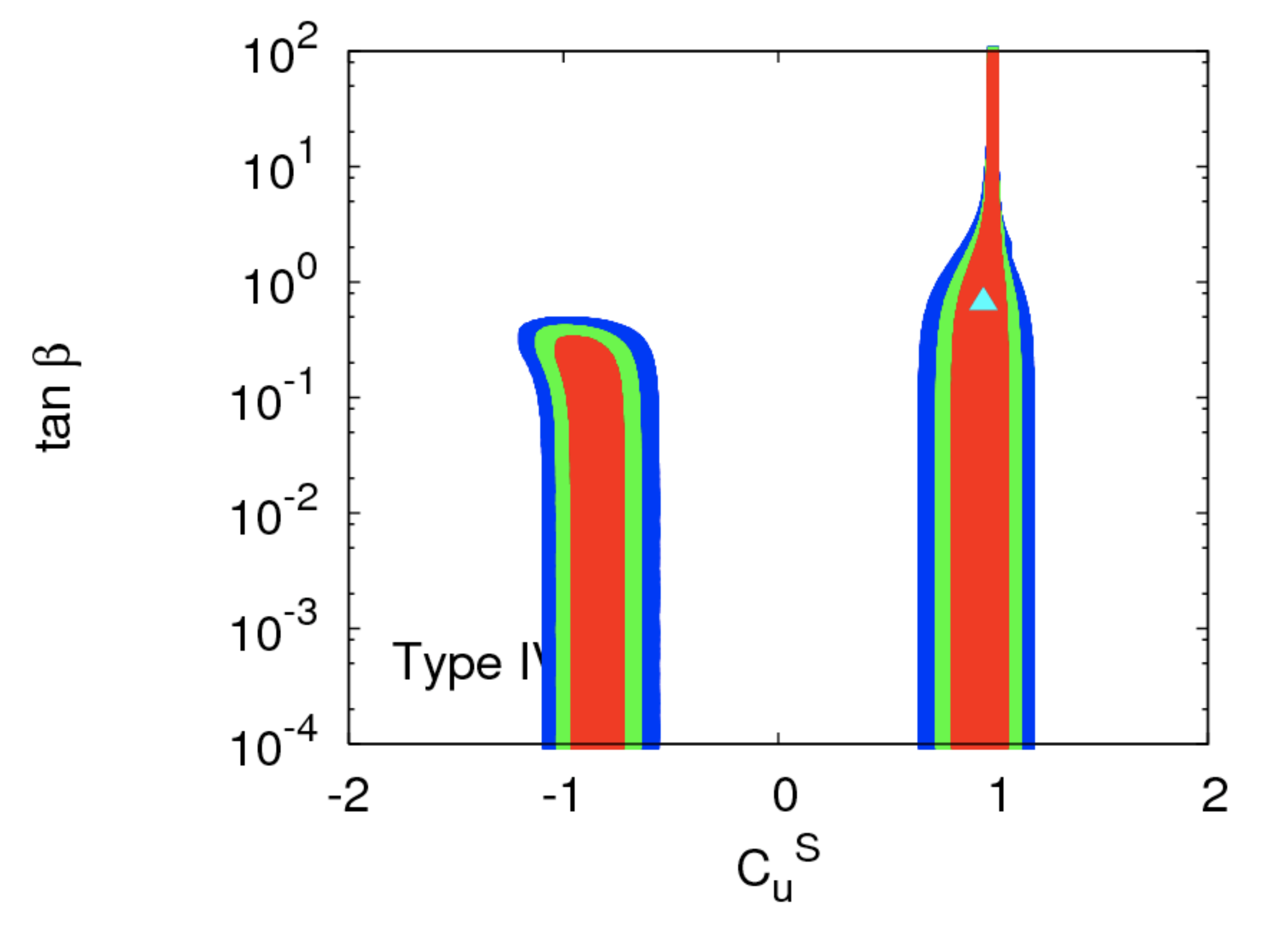}
\caption{\small \label{cpc3-cutanb}
The same as Fig.~\ref{cpc3-cucv} but 
in the plane of $C_u^S$ vs $\tan\beta$ for Type I -- IV ({\bf CPC3}).
The description of the confidence regions is the same as Fig.~\ref{cpc3-cucv}.
}
\end{figure}

\begin{figure}[th!]
\centering
\includegraphics[width=3.2in]{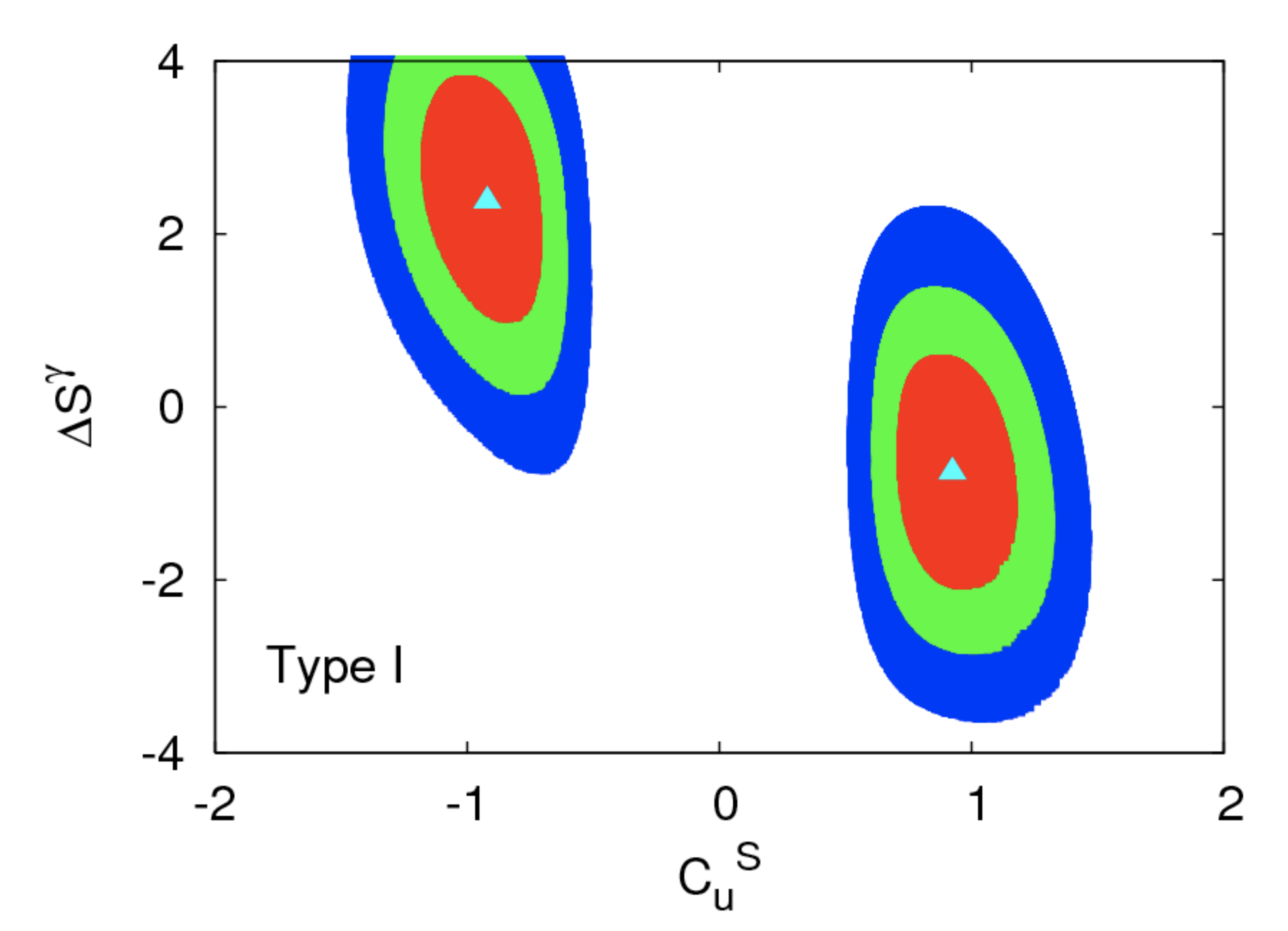}
\includegraphics[width=3.2in]{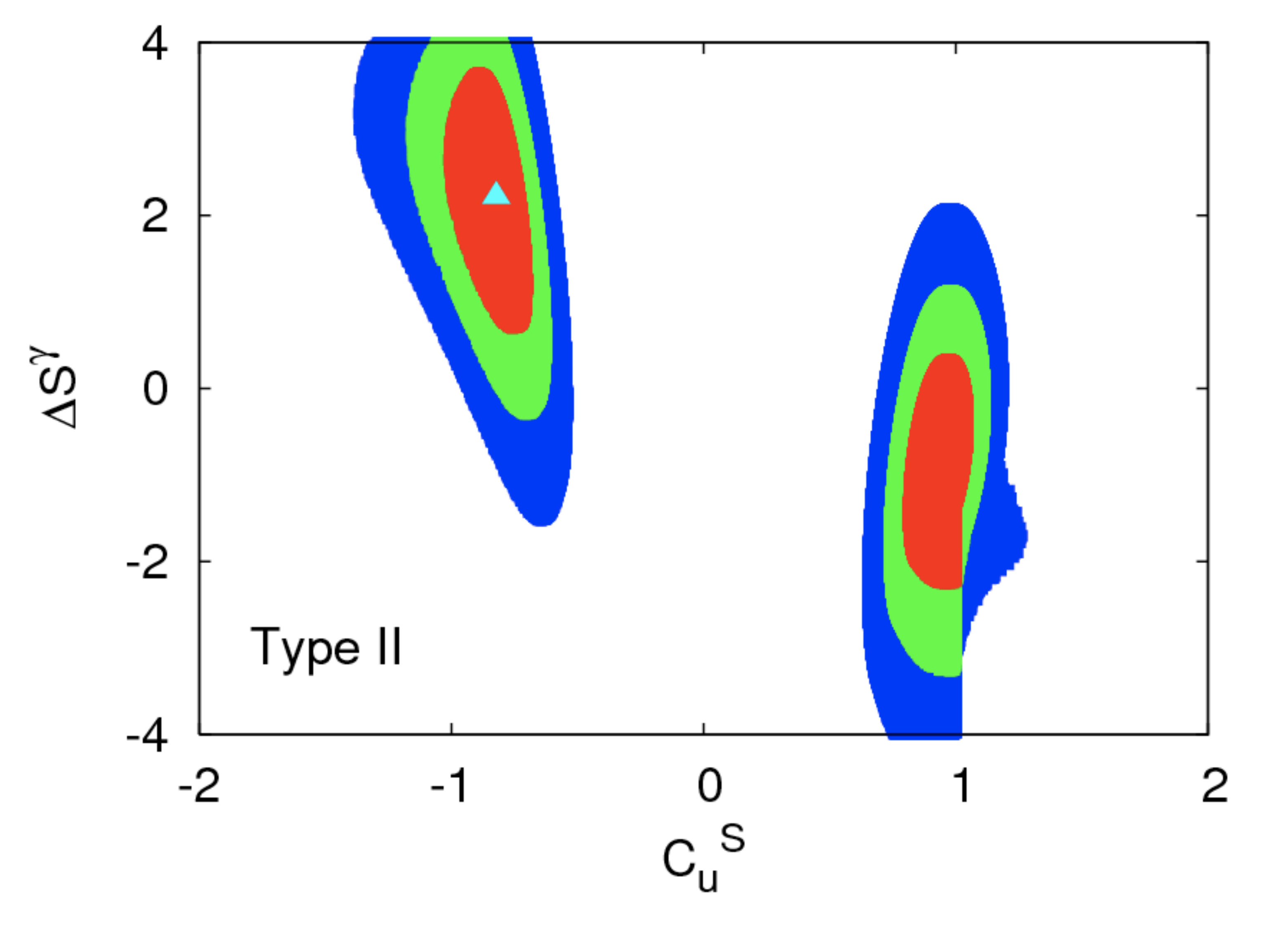}
\includegraphics[width=3.2in]{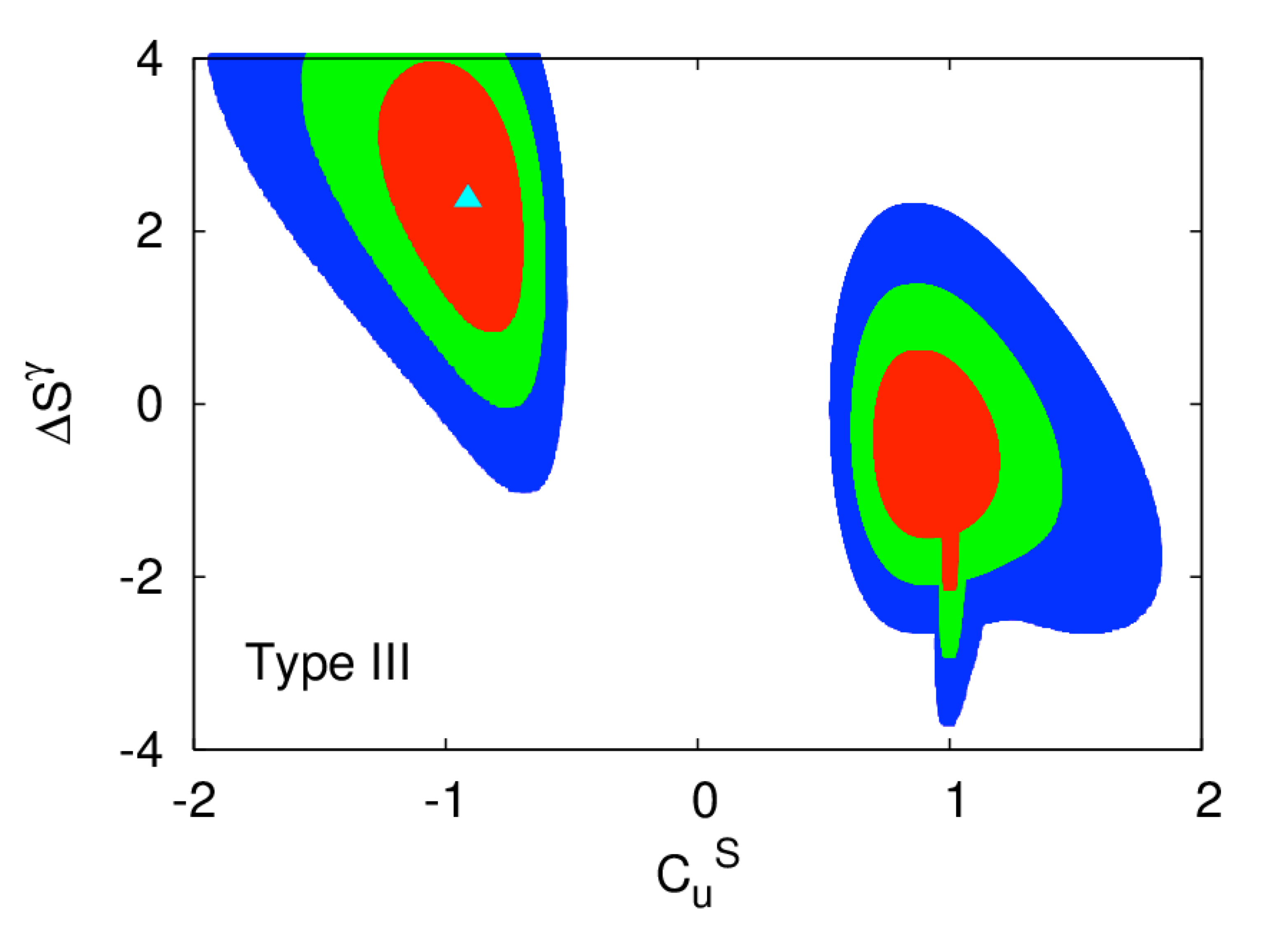}
\includegraphics[width=3.2in]{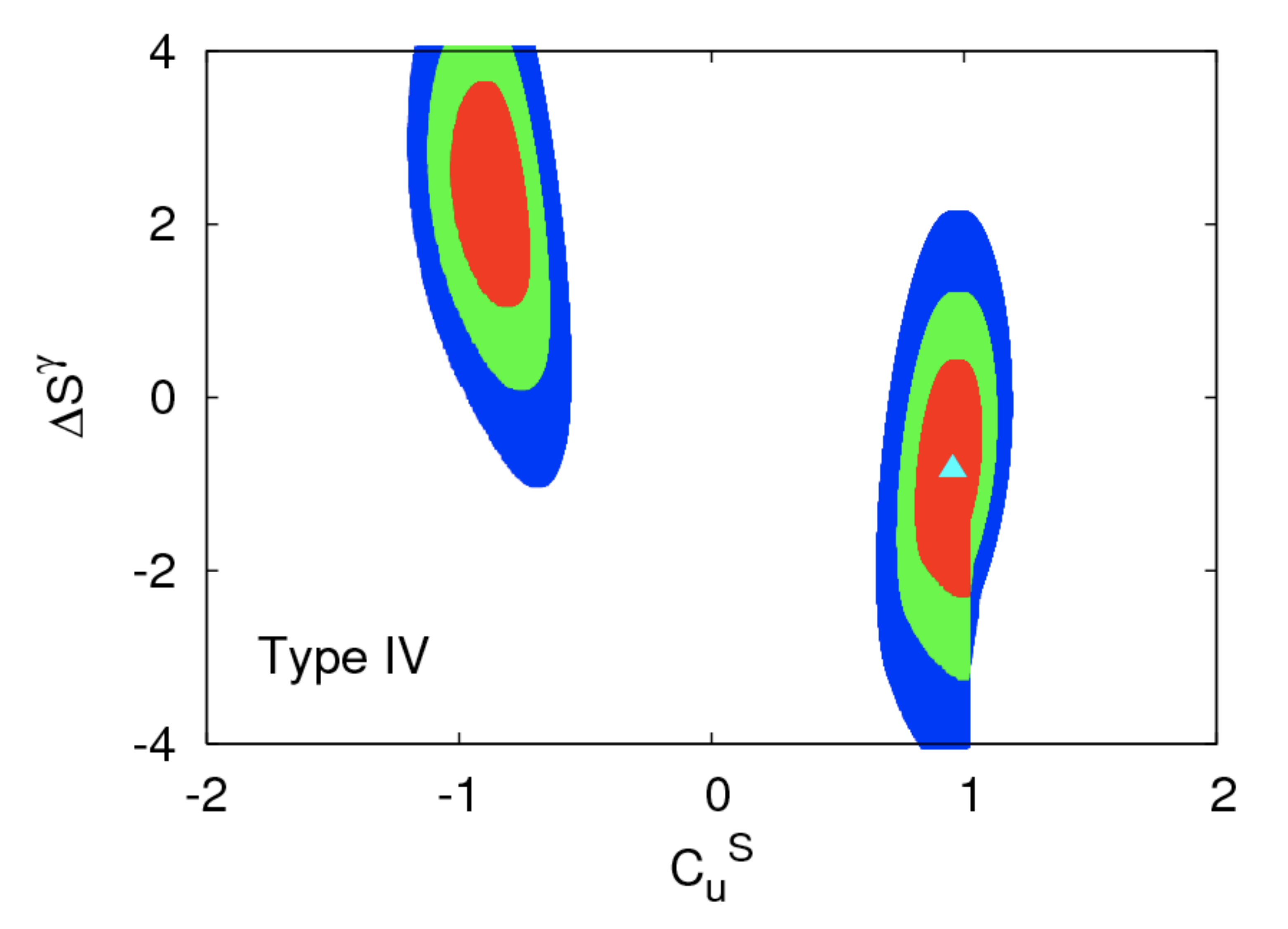}
\caption{\small \label{cpc3-cudsp}
The same as Fig.~\ref{cpc3-cucv} but 
in the plane of $C_u^S$ vs $\left ( \Delta S^{\gamma} \right )^{H^\pm}$
for Type I -- IV ({\bf CPC3}).
The description of the confidence regions is the same as Fig.~\ref{cpc3-cucv}.
}
\end{figure}

\begin{figure}[th!]
\centering
\includegraphics[width=3.2in]{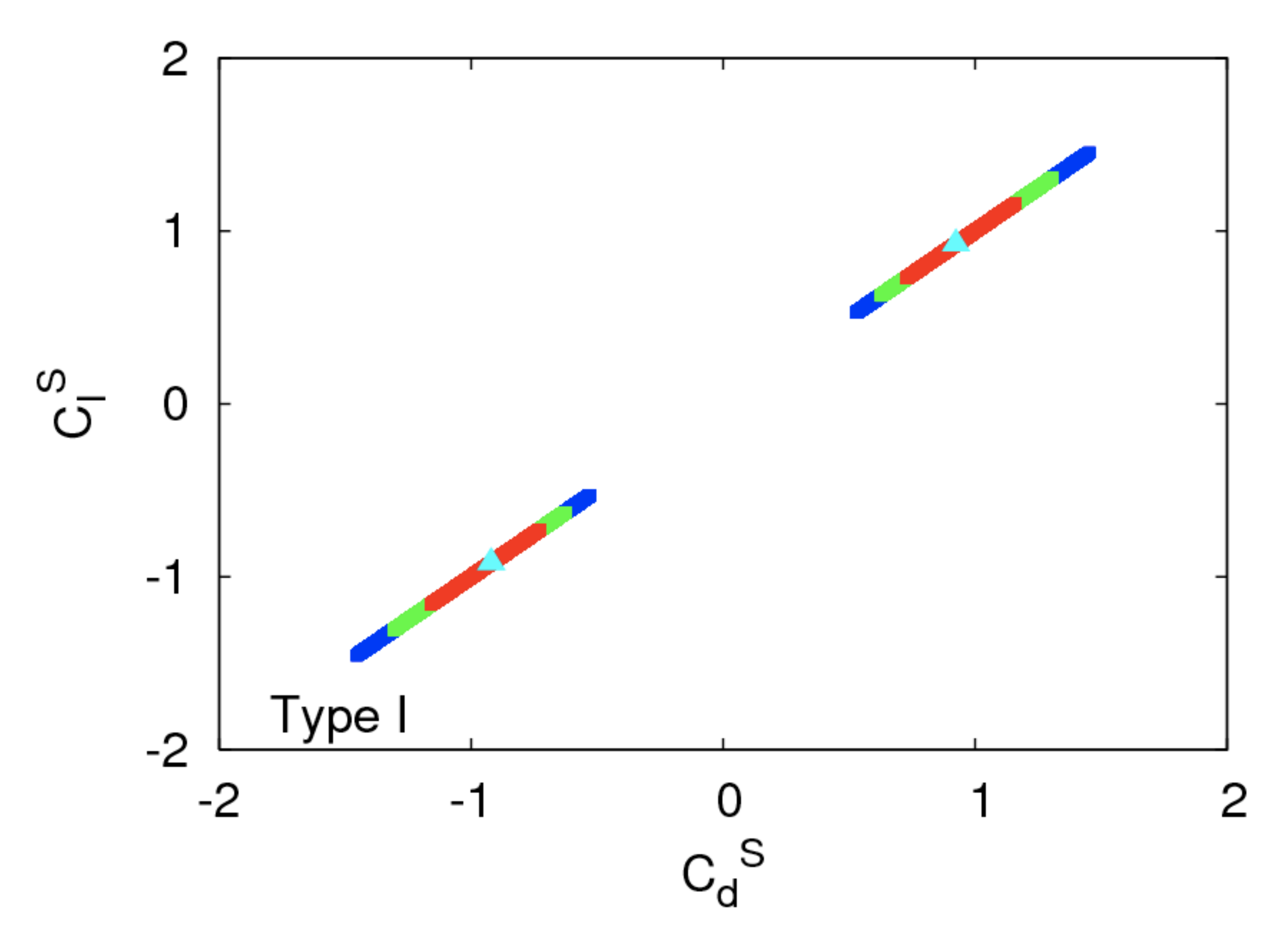}
\includegraphics[width=3.2in]{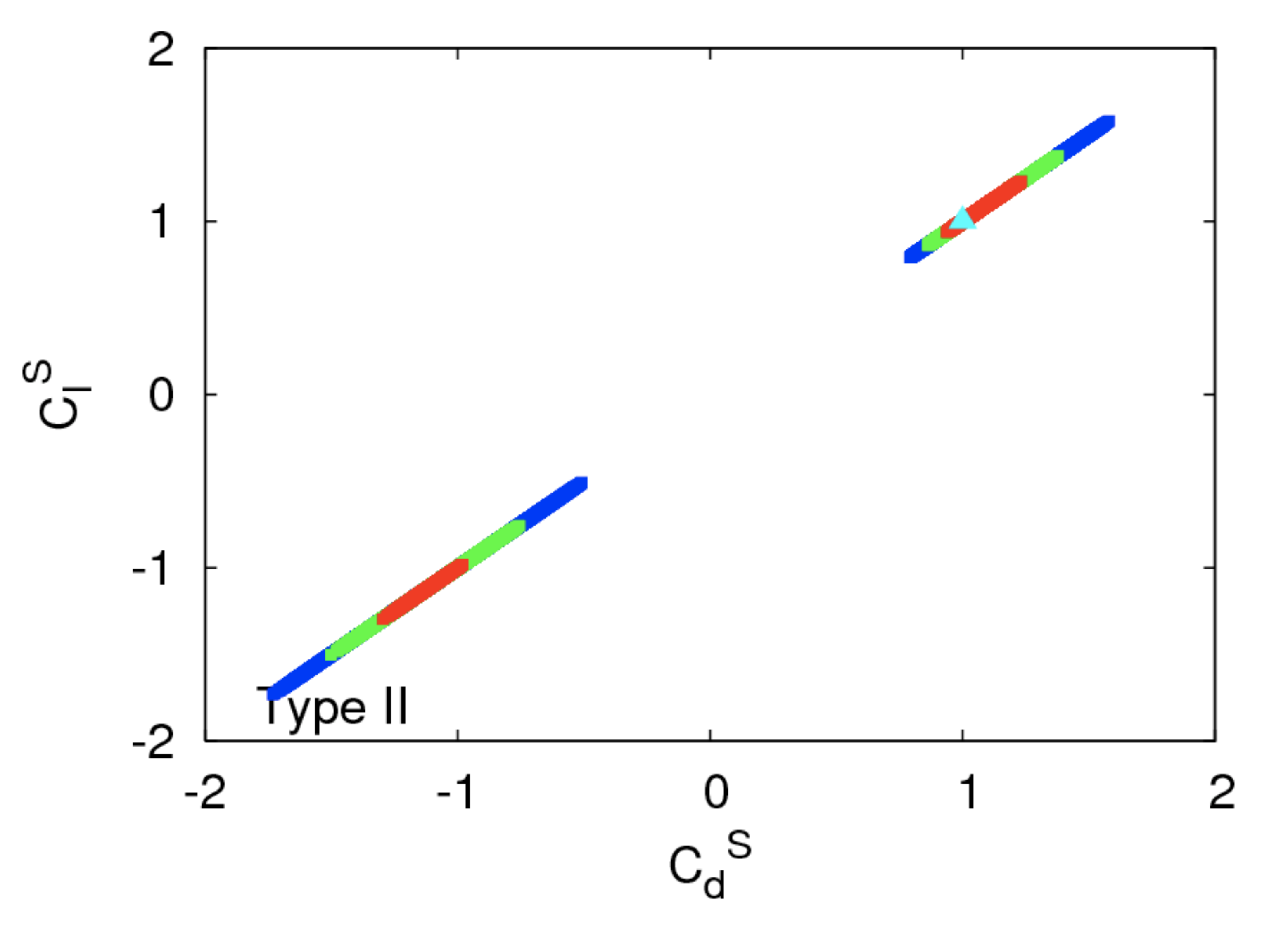}
\includegraphics[width=3.2in]{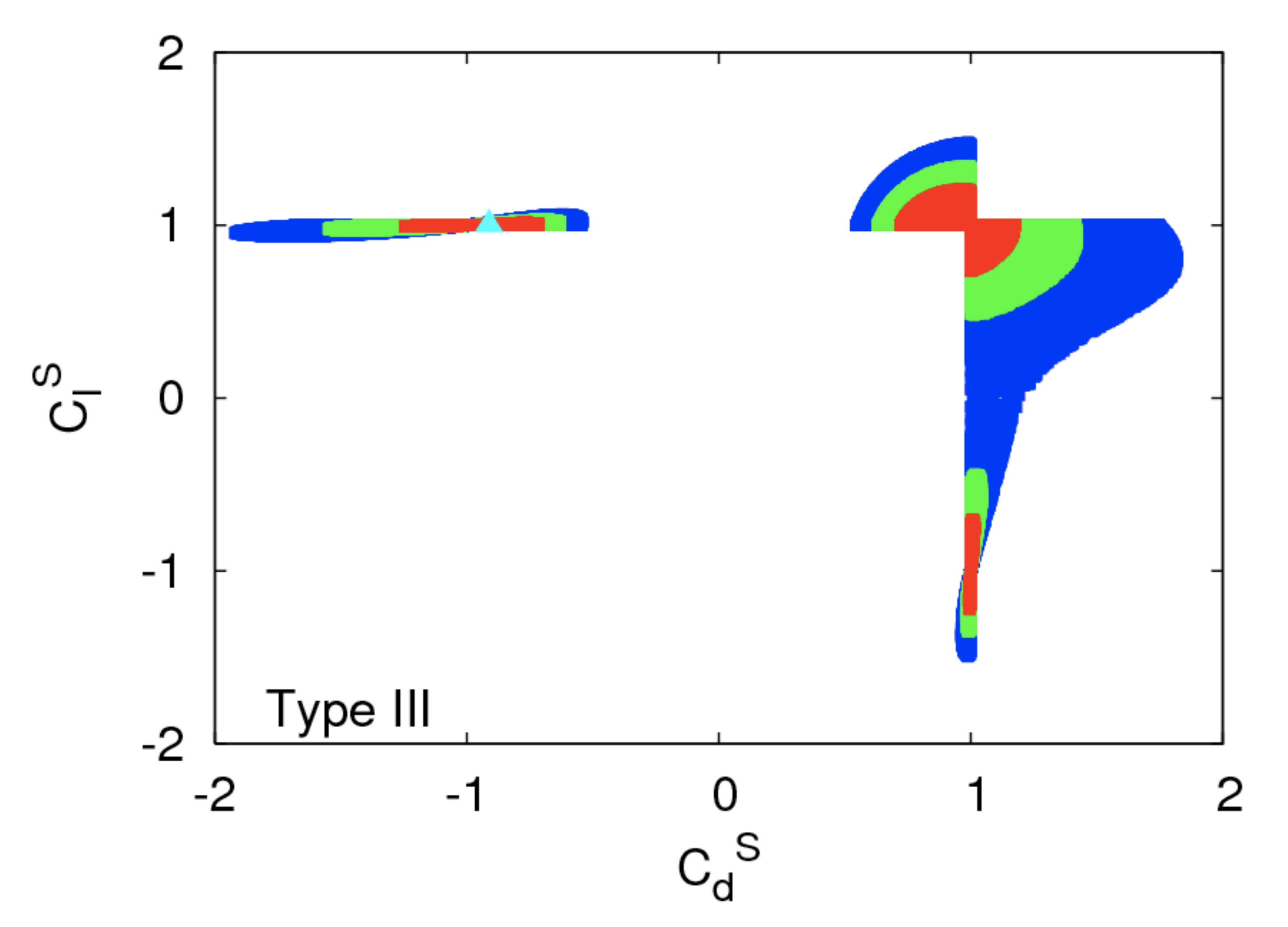}
\includegraphics[width=3.2in]{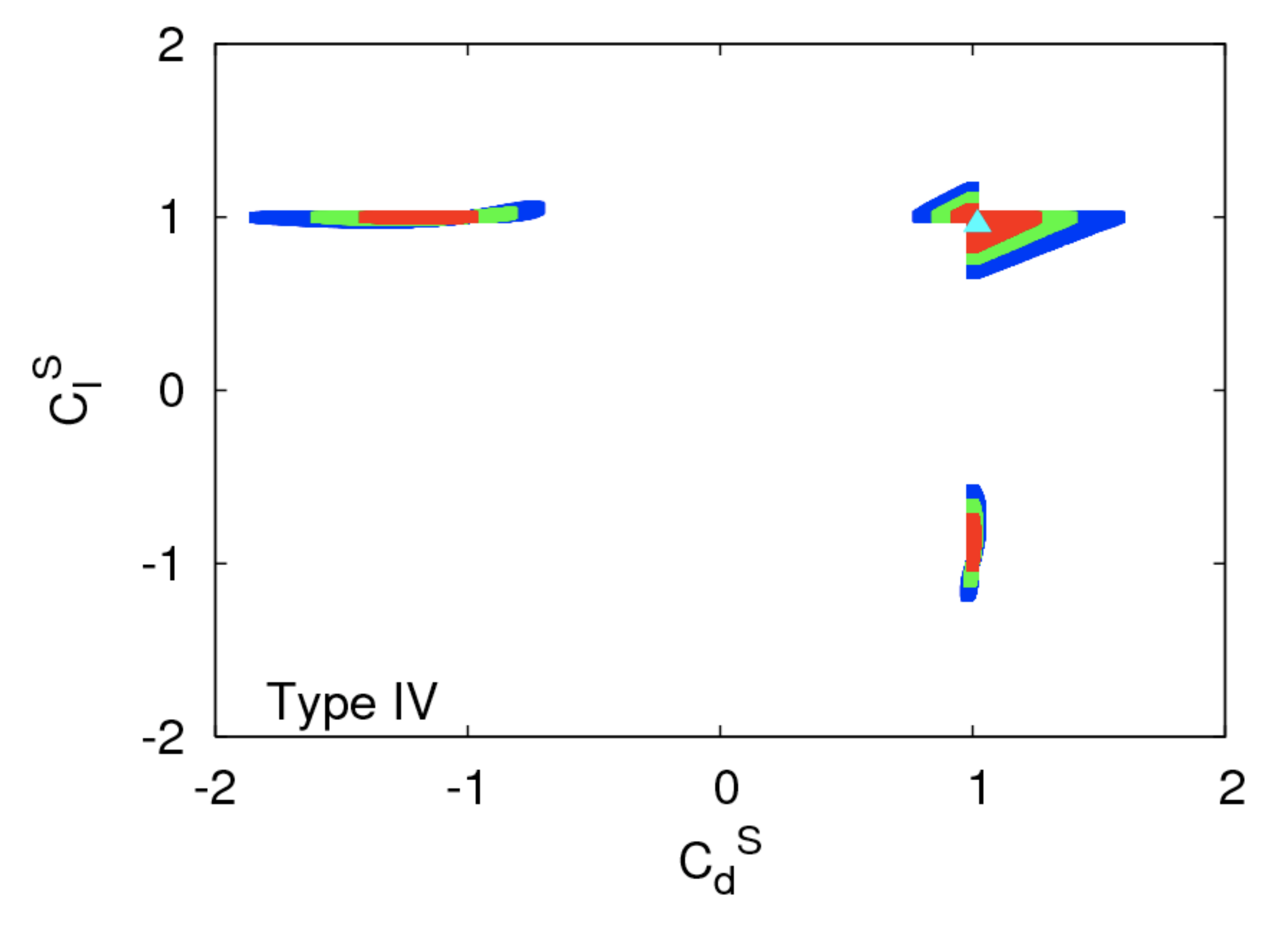}
\caption{\small \label{cpc3-cdcl}
The same as Fig.~\ref{cpc3-cucv} but 
in the plane of $C_d^S$ vs $C_\ell^S$ for Type I -- IV ({\bf CPC3}).
The description of the confidence regions is the same as Fig.~\ref{cpc3-cucv}.
}
\end{figure}

In this {\bf CPC3} fit, we vary three parameters: 
$C_u^S$, $C_v$ (or equivalently $\log_{10}\tan\beta$), and 
$\left(\Delta S^\gamma\right)^{H^\pm}$. 
The total $\chi^2$,
$\chi^2/{\rm dof}$, $p$-value and the best-fit values  of
$C_u^S$, $C_v$ ($\tan\beta$), and $\left(\Delta S^\gamma\right)^{H^\pm}$
for the types I -- IV of 2HDMs  
are shown in the upper half of Table~\ref{tab:best}.
We show the contour plots for confidence-level regions
as functions $C_u^S$ vs $C_v$, $C_u^S$ vs $\tan\beta$, 
$C_u^S$ vs $\left(\Delta S^\gamma\right)^{H^\pm}$, and
$C_d^S$ vs $C_l^S$ in Figs.~\ref{cpc3-cucv} -- \ref{cpc3-cdcl}, 
respectively.

We found that type II gives the smallest $\chi^2$ 
but the variation of total $\chi^2$ among the four types is very small, 
within $0.34$. The {\bf CPC3}
 fit is slightly better than the {\bf CPC2}, as it has
one more parameter in the fit.
However, the $p$-values of the fits are still worse than the SM one
$(p_{\rm SM} = 0.65)$.  
The best-fit values
for $C_u^S$ are about $\pm 0.92$ (I), $-0.82$ (II), $-0.91$ (III),
and $0.96$ (IV)
and those of $C_v$ are $0.97$ for I and III and 1 for II and IV.
We also implement independent fits with 
$\log_{10}\tan\beta$  as
the scanning variable taking
$-4 < \log_{10}\tan\beta < 2$, instead of $C_v$.
The best-fit values
for $\tan\beta$ are either small or very small, except for 
type I with positive $C_u^S$. Again, we note that
$\chi^2$ hardly changes as $\tan\beta$ varies in wide range of parameter space.

For $\left(\Delta S^\gamma\right)^{H^\pm}$, we have obtained
$\left(\Delta S^\gamma\right)^{H^\pm} \simeq -0.8$ or $2.3$ when
$C_u^S \sim +0.9$ or $-0.9$, respectively. 
This can be understood from the numerical
expression for $S^\gamma$~\cite{Cheung:2013kla}
\begin{equation}
\label{eq:sgamma}
S^\gamma \simeq
-8.35\,C_v+1.76\,C_u^S +\left(\Delta S^\gamma\right)^{H^\pm}\,.
\end{equation}
When $C_u^S$ changes from $+0.9$ to $-0.9$, 
$\left(\Delta S^\gamma\right)^{H^\pm}$ 
changes from $-0.8$ to $+2.3$ 
so that the sum $1.76 C_u^S   + \left(\Delta S^\gamma\right)^{H^\pm} \approx 0.7$.

The contour plots for the CL regions in the plane of $C_u^S$ vs $C_v$ 
for type I -- IV are shown in Fig.~\ref{cpc3-cucv},
which can be directly compared to Fig.~\ref{cpc1-cucv}. 
In contrast, the negative $C_u^S$ is now equally as good as the positive one.
We show the CL regions in the plane of $C_u^S$ and $\tan\beta$
in Fig.~\ref{cpc3-cutanb}. 
For the negative $C_u^S$ case, we find that $\tan\beta$ is 
smaller than $\sim 0.6$ at 99.7 \% CL.
In Fig.~\ref{cpc3-cudsp}, we show the CL regions in the plane of $C_u^S$
and $\left(\Delta S^\gamma\right)^{H^\pm}$. For positive $C_u^S$,
it lies between $2$ and $-4$ while 
$\left(\Delta S^\gamma\right)^{H^\pm} > -0.7 \sim -1.6$
for negative $C_u^S$ at 99.7 \% CL.
The CL regions for $C_l^S$ and $C_d^S$ 
are similar to the {\bf CPC2} case as 
shown in Fig.~\ref{cpc3-cdcl} but with the larger regions 
allowed at 68.5 \% CL around 
the negative values of couplings.

\begin{figure}[th!]
\centering
\includegraphics[width=3.2in]{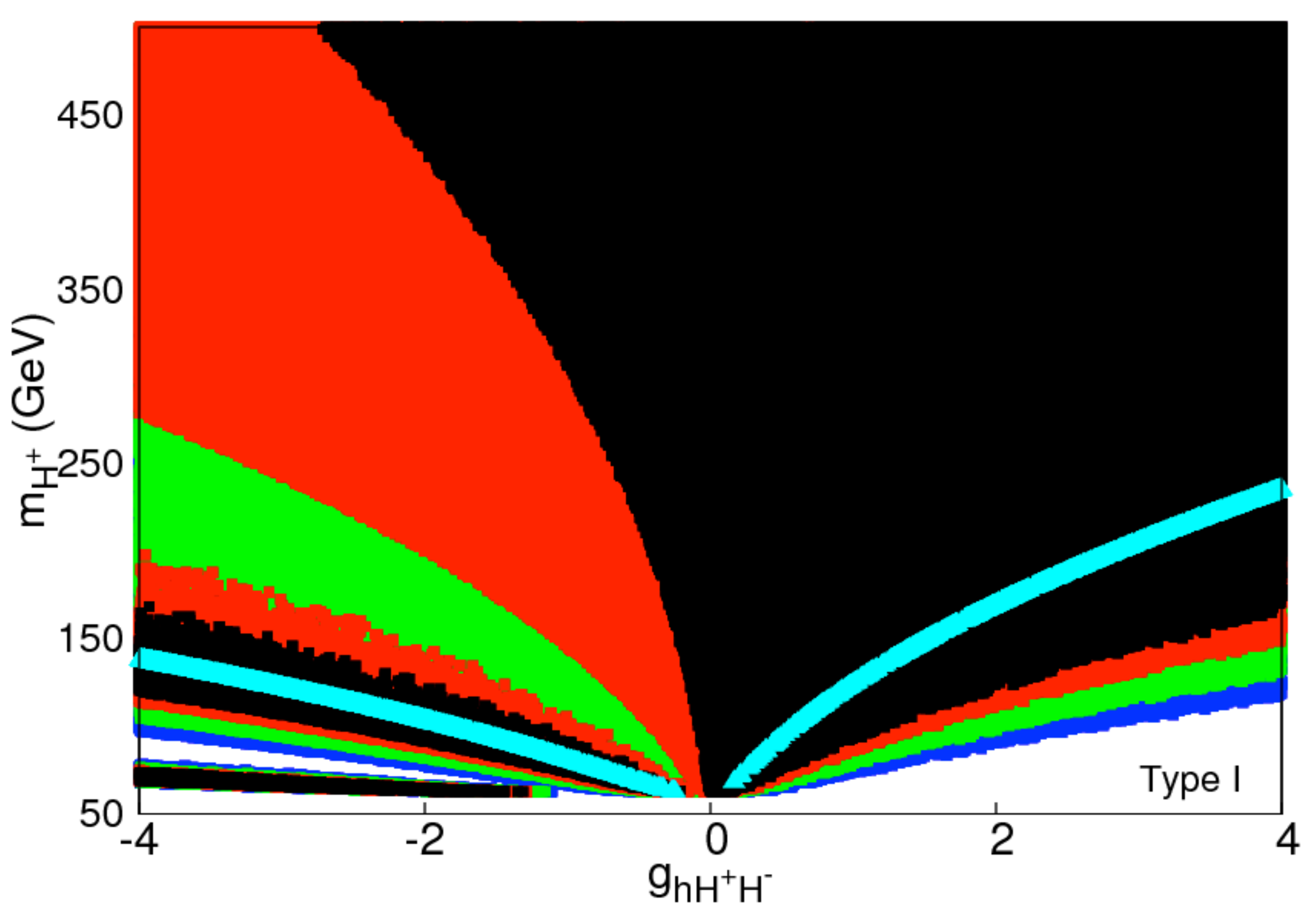}
\includegraphics[width=3.2in]{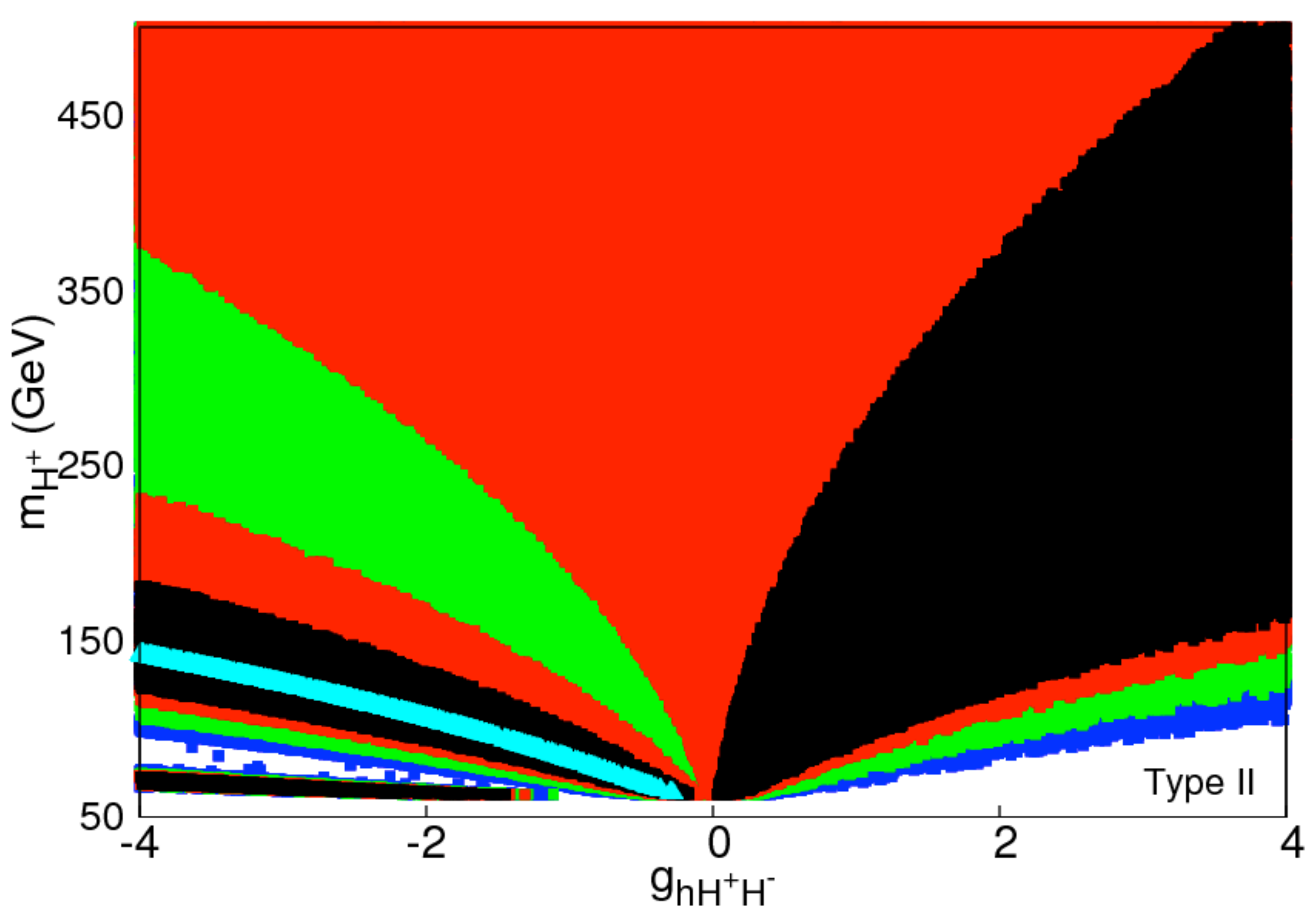}
\includegraphics[width=3.2in]{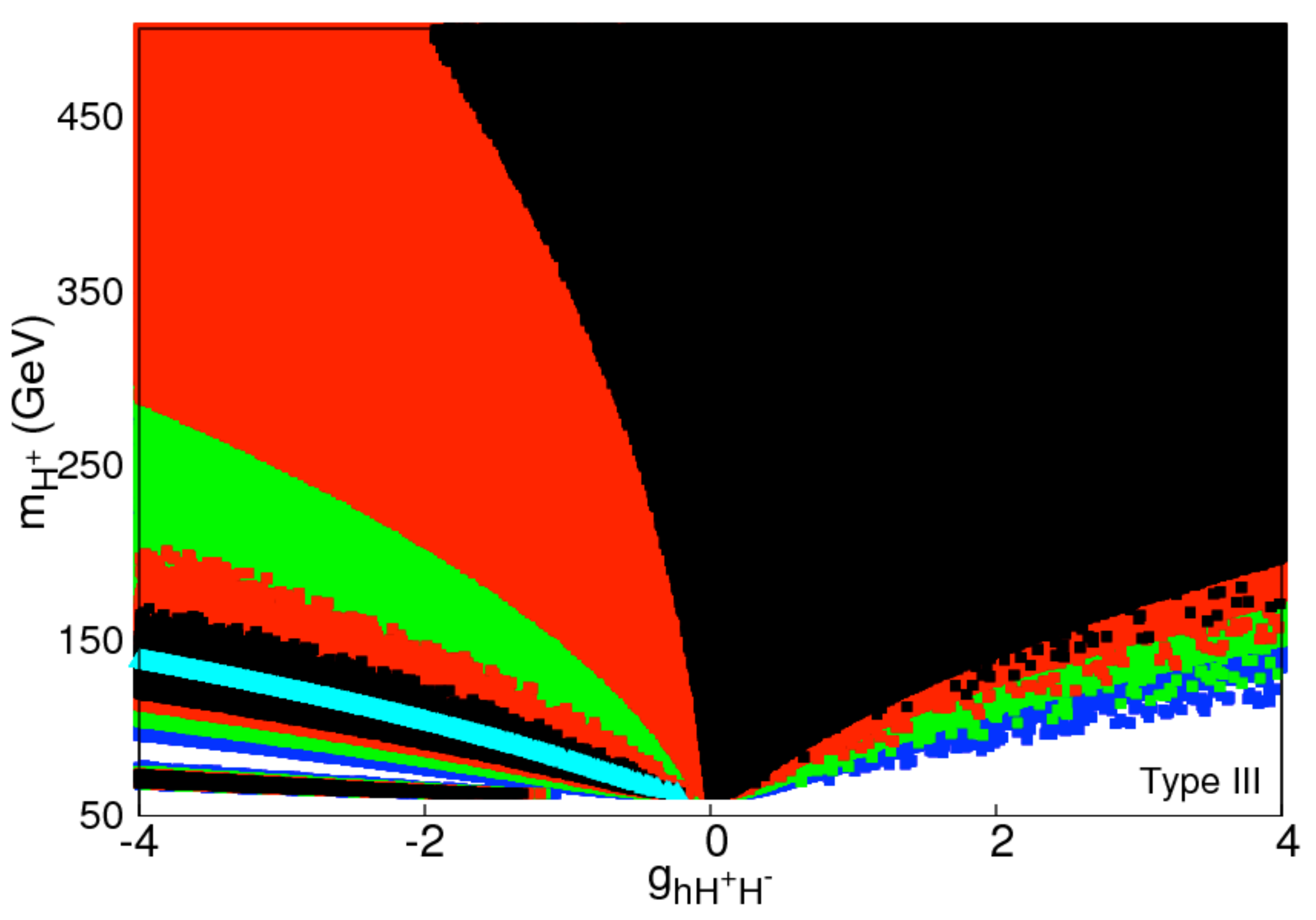}
\includegraphics[width=3.2in]{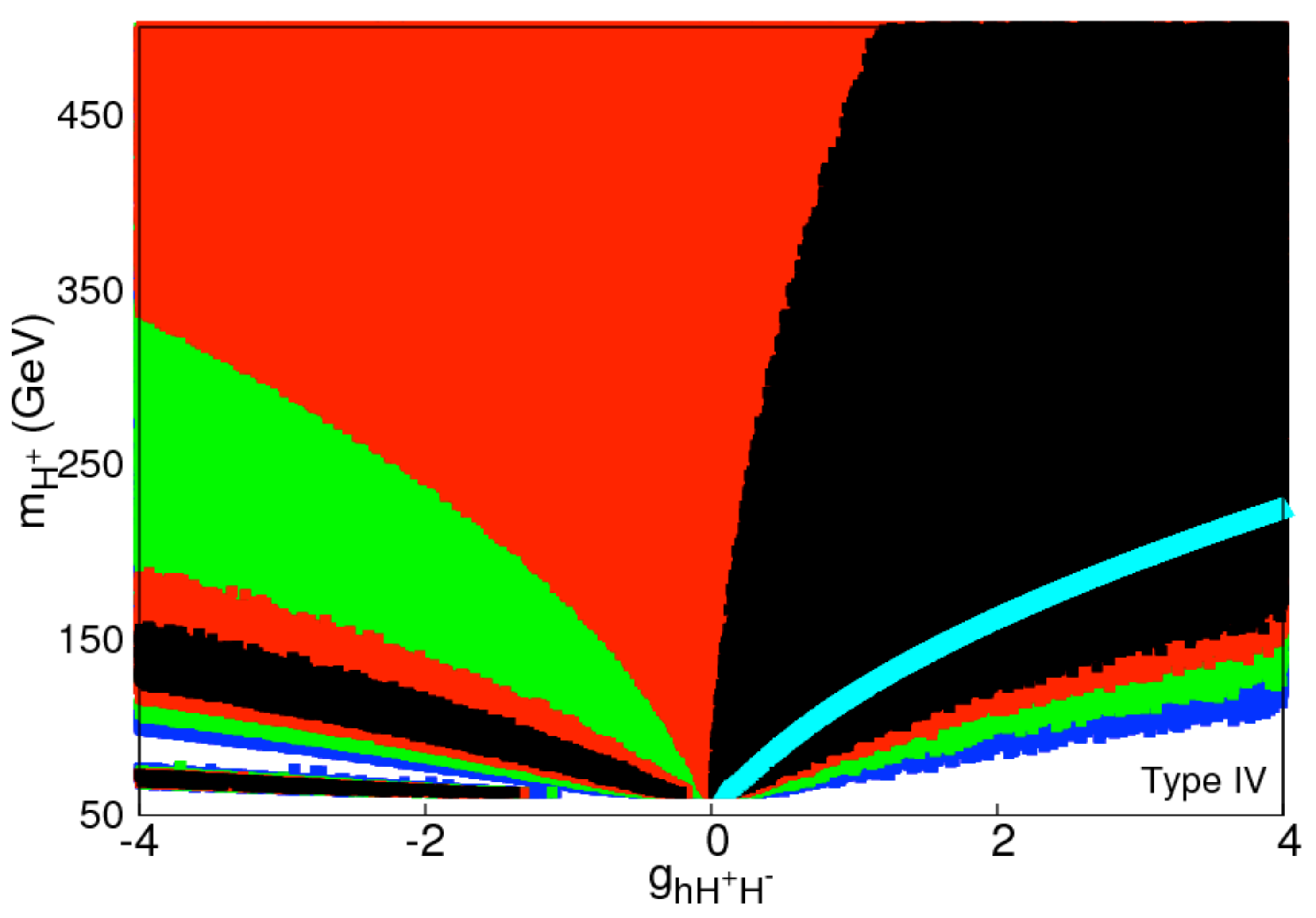}
\caption{\small \label{cpc4}
The same as Fig.~\ref{cpc3-cucv} but we used $g_{h H^+ H^-}$ and $m_{H^\pm}$ 
in place of $\left ( \Delta S^{\gamma} \right )^{H^\pm}$ ({\bf CPC3} case)
for Type I -- IV.
The contour regions shown are for 
$\Delta \chi^2 \le 1.0$ (black), 
$2.3$ (red), $5.99$ (green), and $11.83$ (blue) 
above the minimum, which 
correspond to confidence levels of $39.3\%$,
$68.3\%$, $95\%$, and $99.7\%$, respectively.
The best-fit points are denoted by a beam of cyan triangles.
}
\end{figure}

The single parameter $\left(\Delta S^\gamma\right)^{H^\pm}$
can be interpreted in terms of 
the charged Higgs mass $M_{H^\pm}$ and the neutral Higgs coupling to the
charged Higgses $g_{_{H_iH^+H^-}}$, as in Eq.~(\ref{eq:dsgamma}).
In Fig.~\ref{cpc4}, we show the CL regions in the plane of 
$M_{H^\pm}$ vs $g_{_{H_iH^+H^-}}$. 
Since the variation of $\chi^2$ is very mild, we add
one more region with $\Delta\chi^2\leq 1$ (black). The thick cyan 
lines denote the points giving the best-fit values of 
$\left(\Delta S^\gamma\right)^{H^\pm}$ in each type given by 
Eq.~(\ref{eq:dsgamma}). 
We see that a smaller charged Higgs mass is preferred 
when $g_{_{H_iH^+H^-}}<0$, because this corresponds to $C_u^S<0$ and so
a larger $\left(\Delta S^\gamma\right)^{H^\pm} \approx 2.3$ is required.
If the charged Higgs mass is larger than $\sim 300$ GeV as in
the type II model constrained by $B(b\to s \gamma)$, we can see 
that the positive $C_u^S$ case
with $\left(\Delta S^\gamma\right)^{H^\pm} \sim -0.8$ is somewhat preferred.
Nevertheless, the variation of $\chi^2$ is not large enough to have a 
conclusive statement based on the current Higgs data.

\subsection{CP violating fits}

In this subsection, we study the CP-violating case with a  
nonzero $C_u^P$ 
in addition to $C_u^S$, $C_v$ (or, equivalently, $\tan\beta$), and
$\left(\Delta S^\gamma\right)^{H^\pm}$.
In our numerical study, we again find that $\tan\beta$ is bounded 
from above when
$C_u^S$ deviates from its SM value $1$. 
So, as in the CP-conserving case,
the precise and independent future measurements of $C_u^S$ and $\tan\beta$ 
can tell us the phenomenological viability of 2HDMs, thus providing some
possible model discriminating power.

\subsubsection{\bf CPV3}
\begin{figure}[th!]
\centering
\includegraphics[width=3.2in]{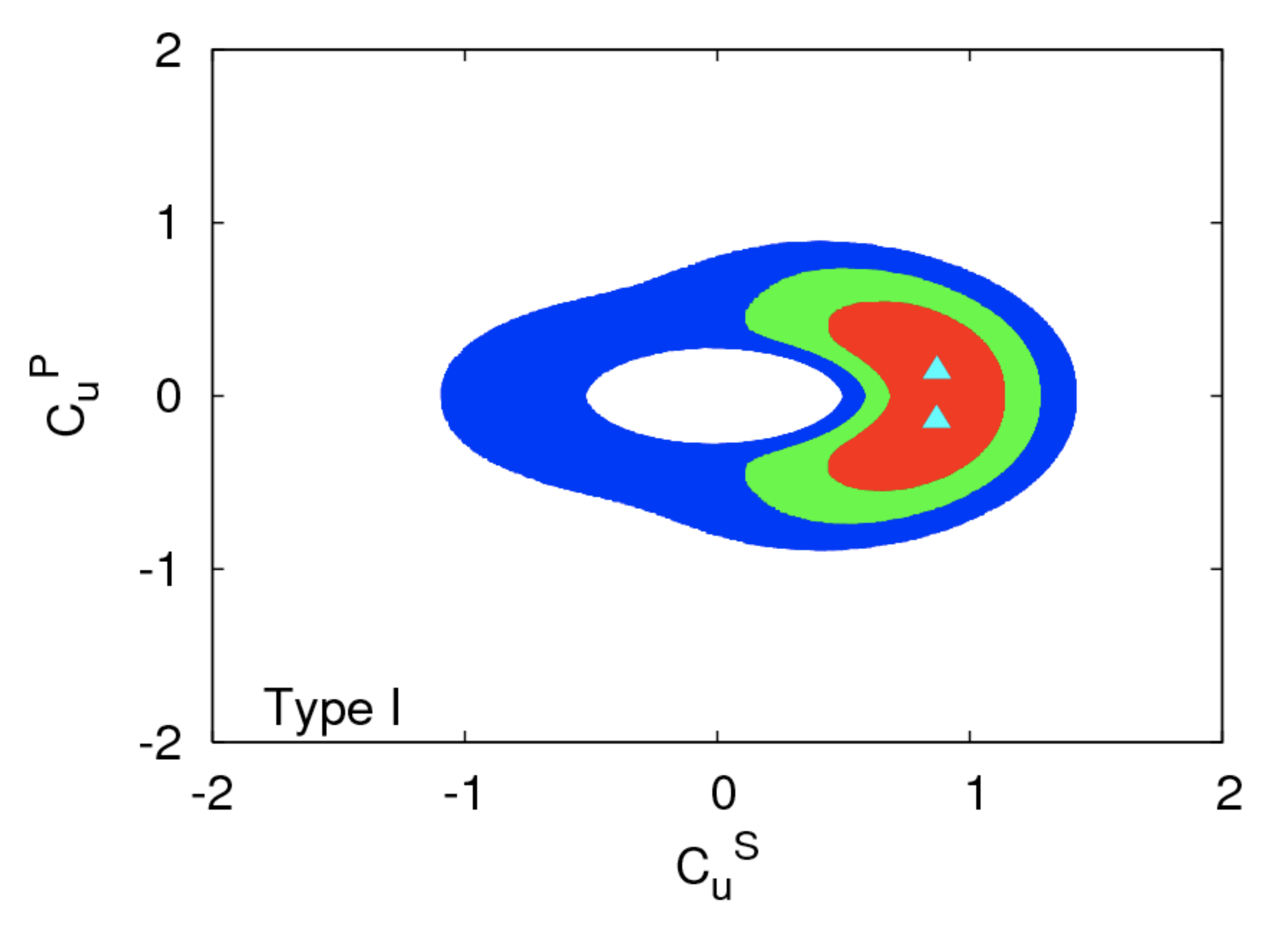}
\includegraphics[width=3.2in]{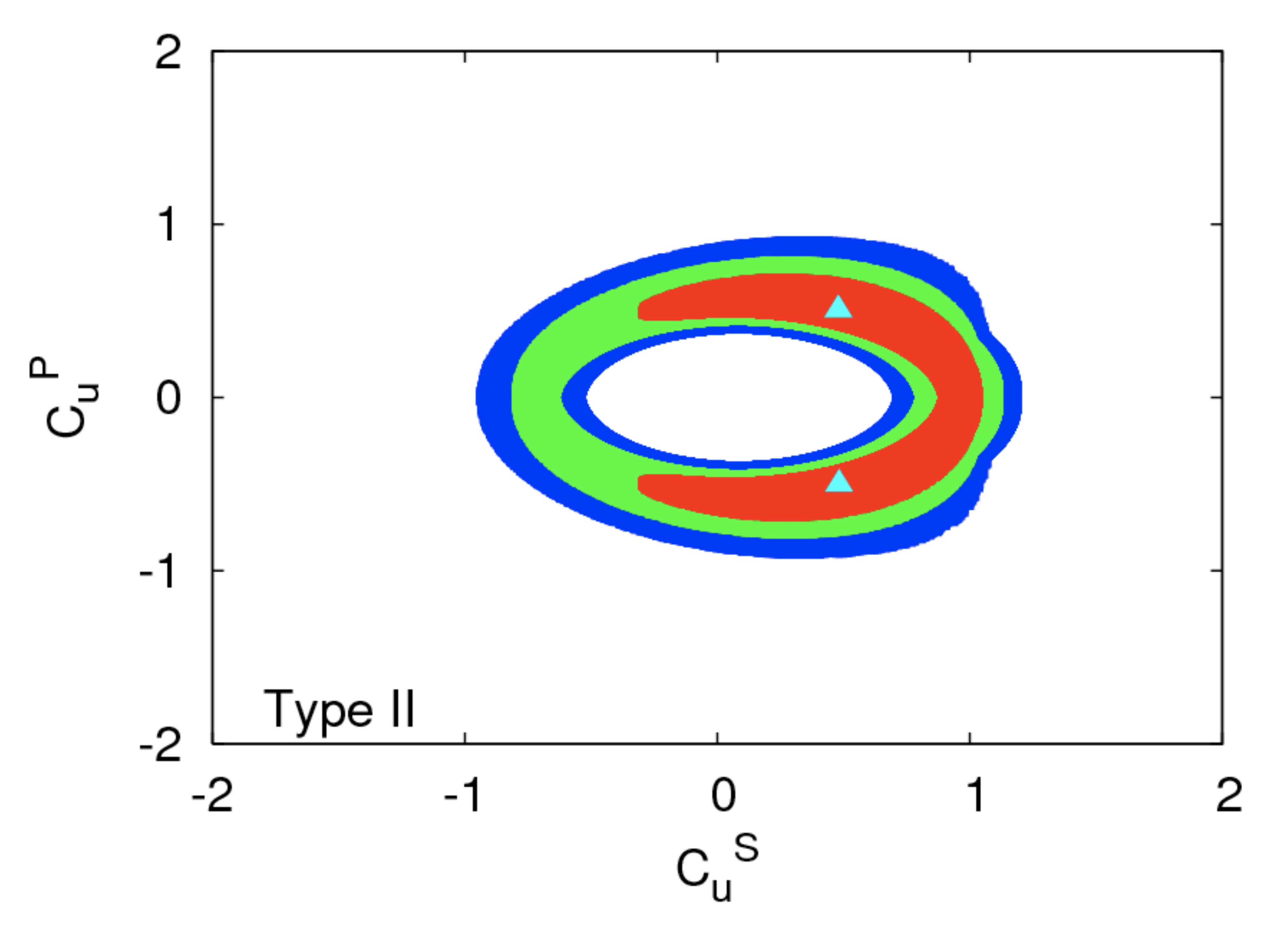}
\includegraphics[width=3.2in]{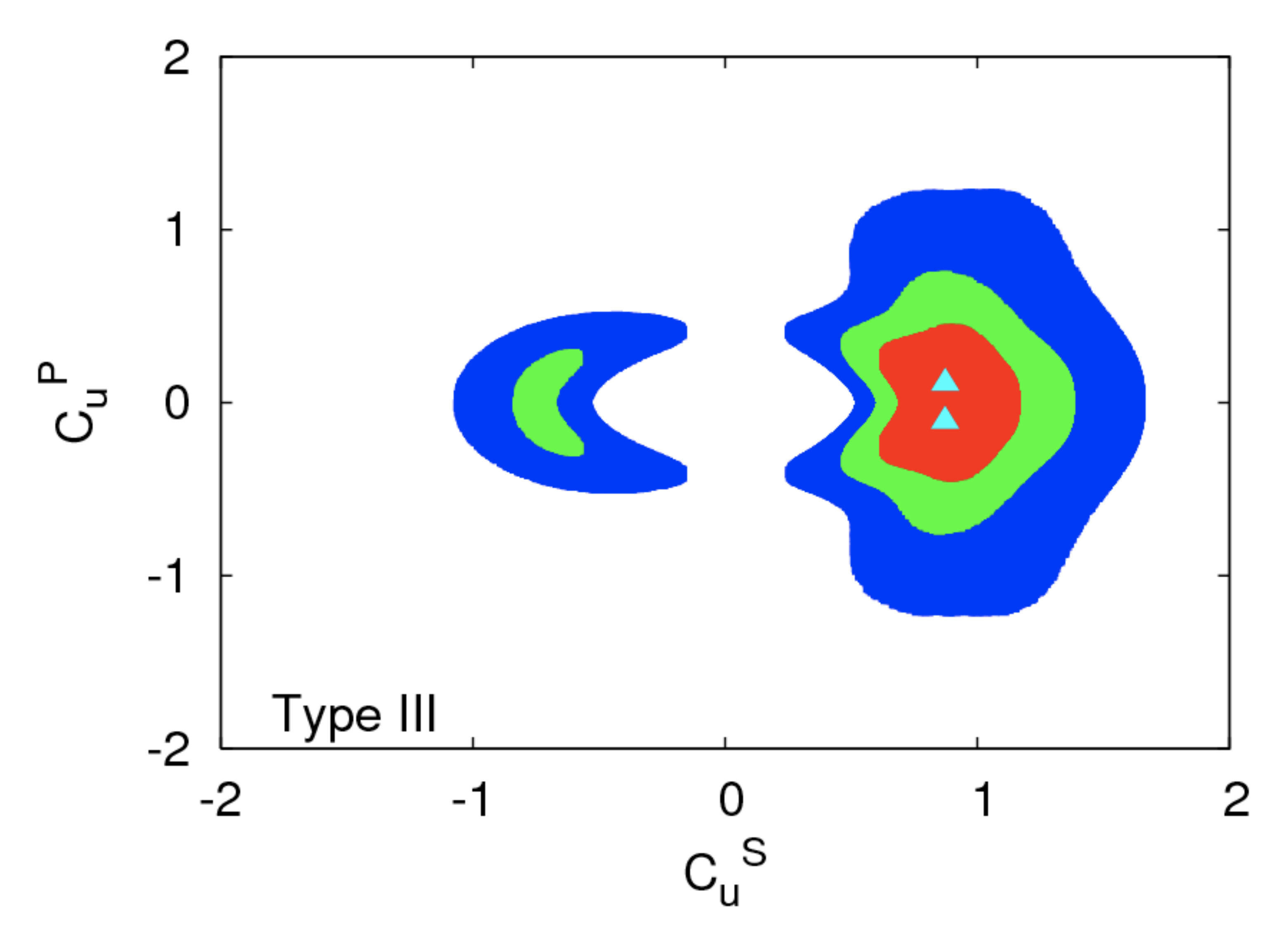}
\includegraphics[width=3.2in]{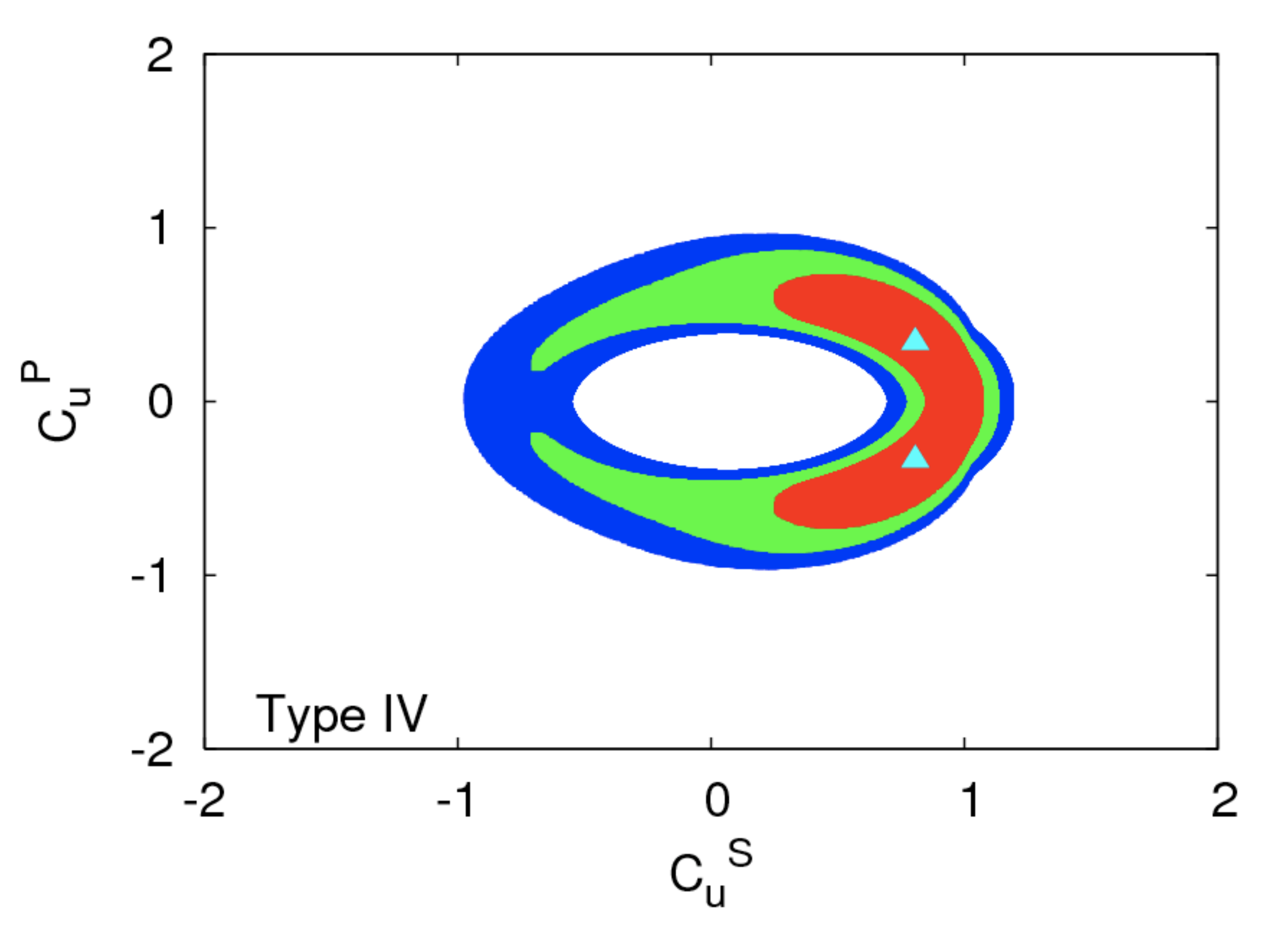}
\caption{\small \label{cpv1-cu-cup}
The confidence-level regions of the fit by varying $C_u^S$, $C_u^P$, and
$\log_{10} \tan\beta$ ({\bf CPV3} case) 
in the plane of $C_u^S$ vs $C_u^P$ for Type I -- IV.
The contour regions shown are for 
$\Delta \chi^2 \le 2.3$ (red), $5.99$ (green), and $11.83$ (blue) 
above the minimum, which 
correspond to confidence levels of
$68.3\%$, $95\%$, and $99.7\%$, respectively.
The best-fit points are denoted by the triangle.
}
\end{figure}

\begin{figure}[th!]
\centering
\includegraphics[width=3.2in]{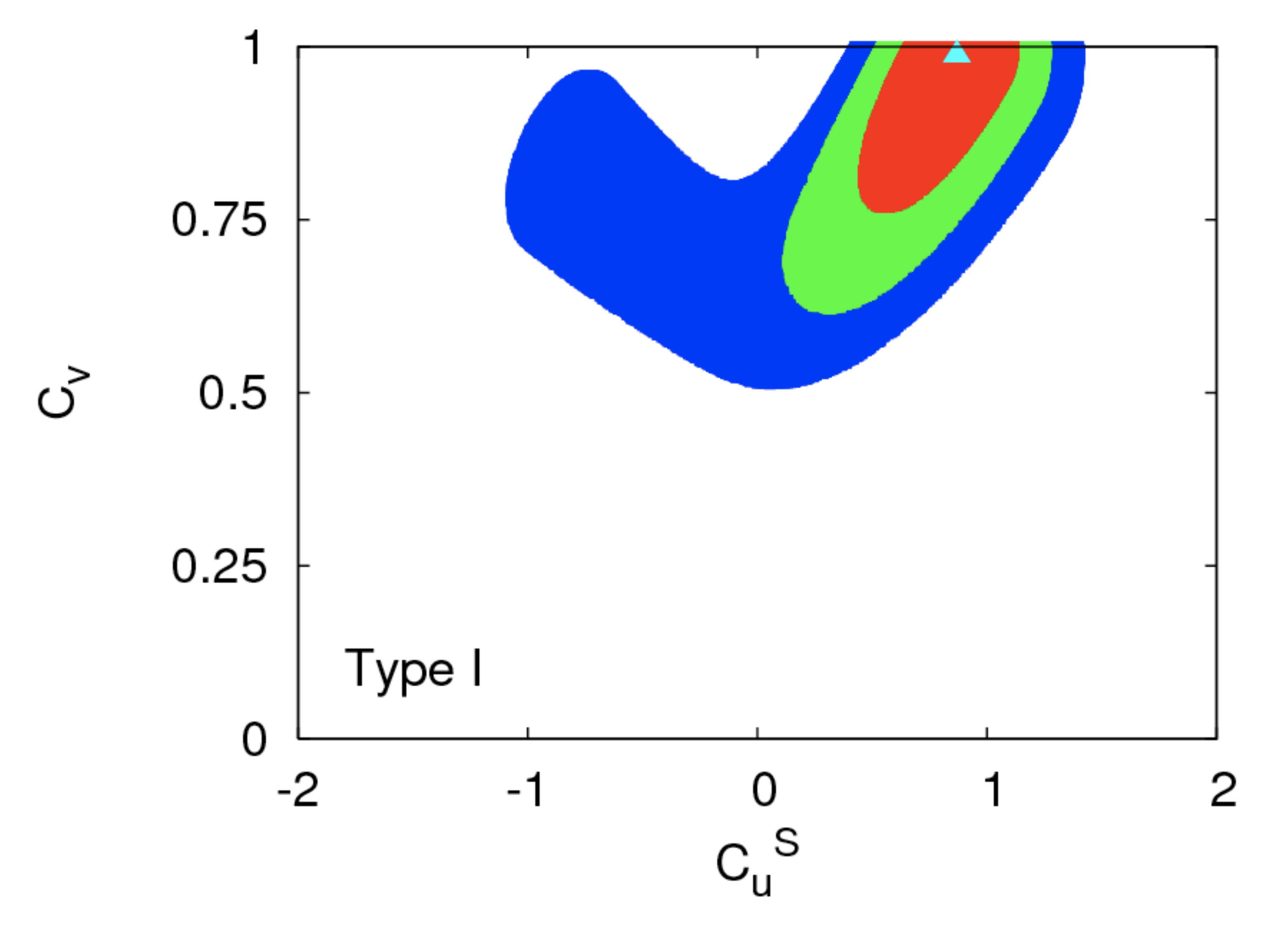}
\includegraphics[width=3.2in]{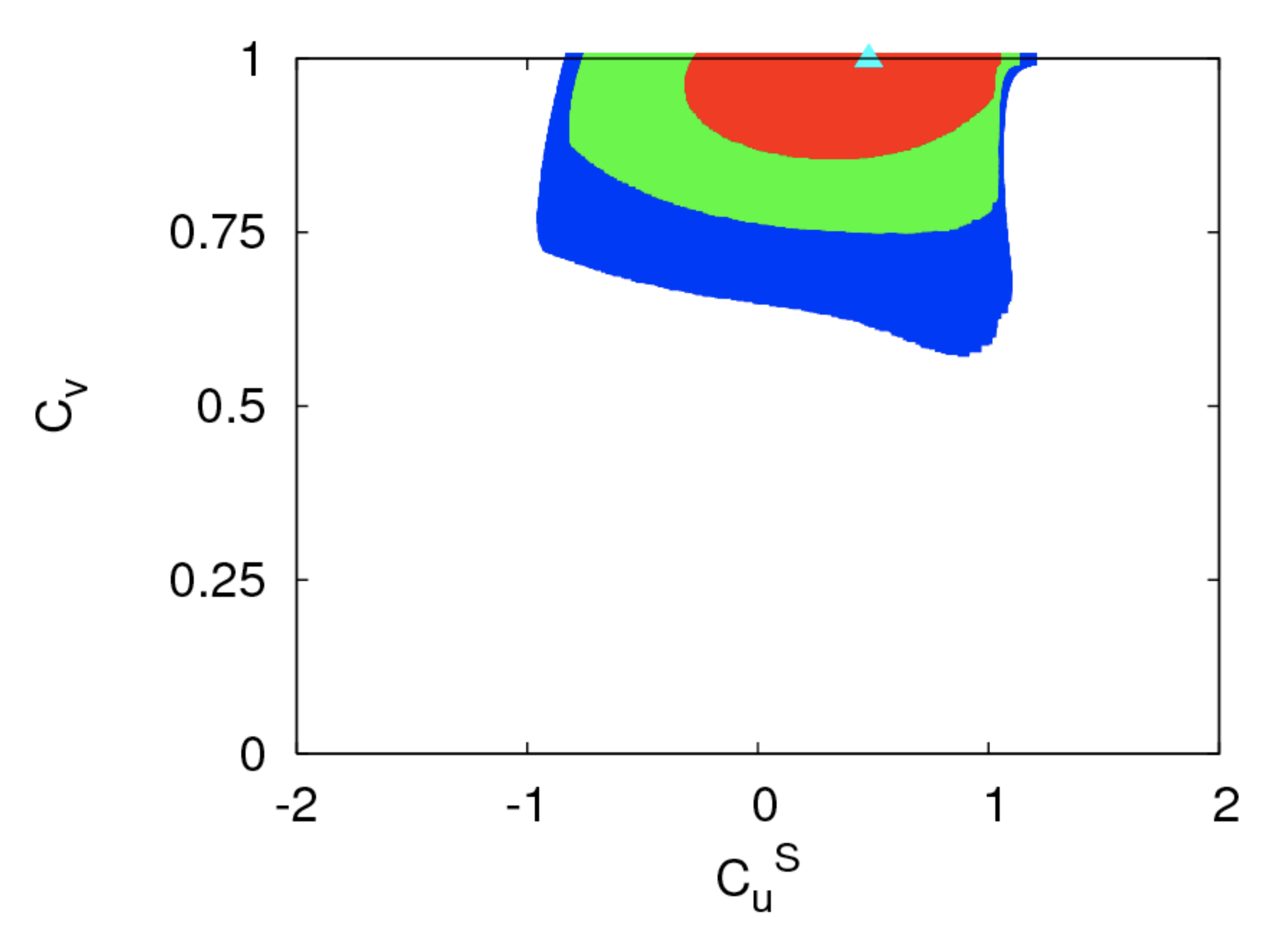}
\includegraphics[width=3.2in]{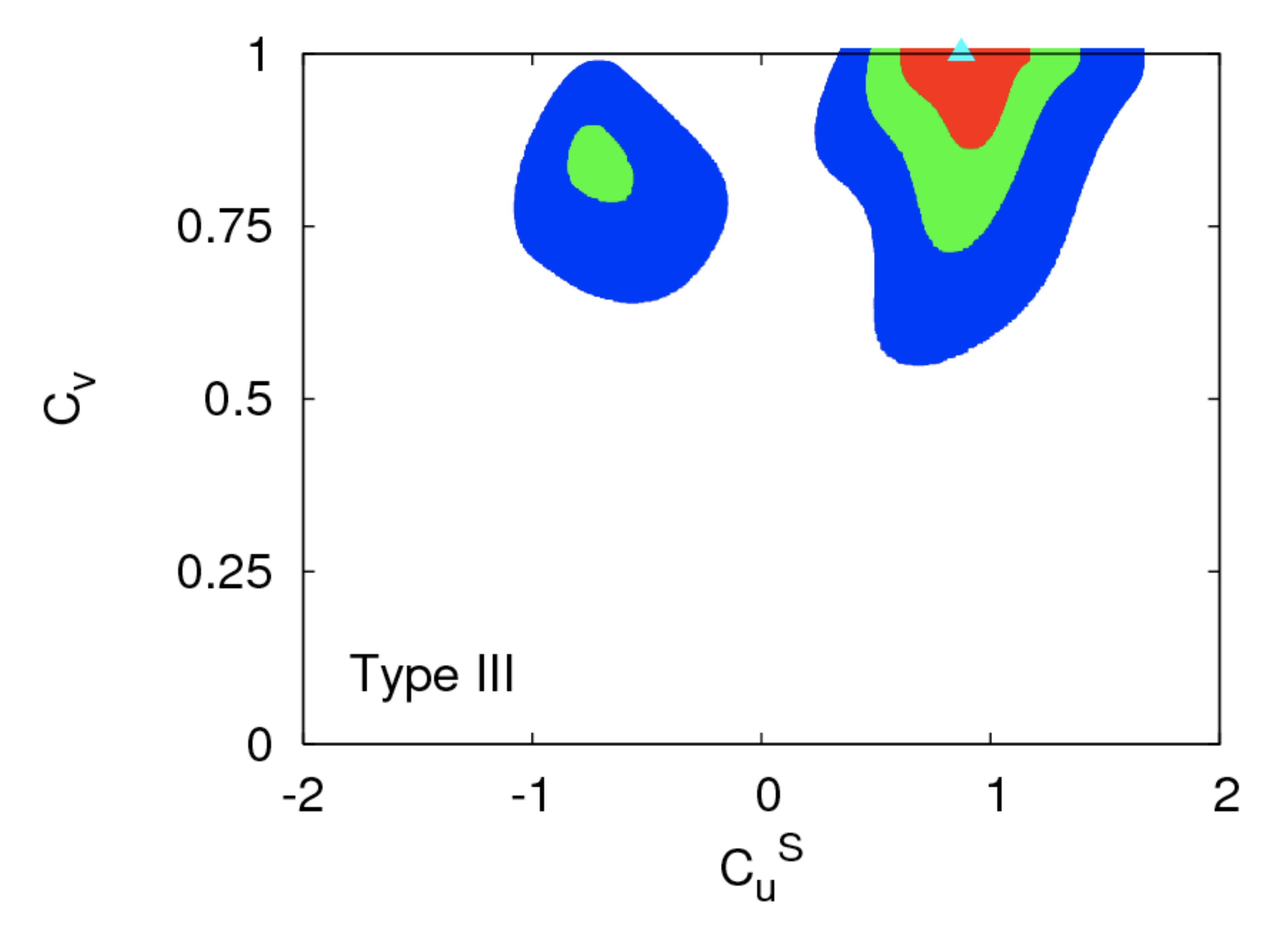}
\includegraphics[width=3.2in]{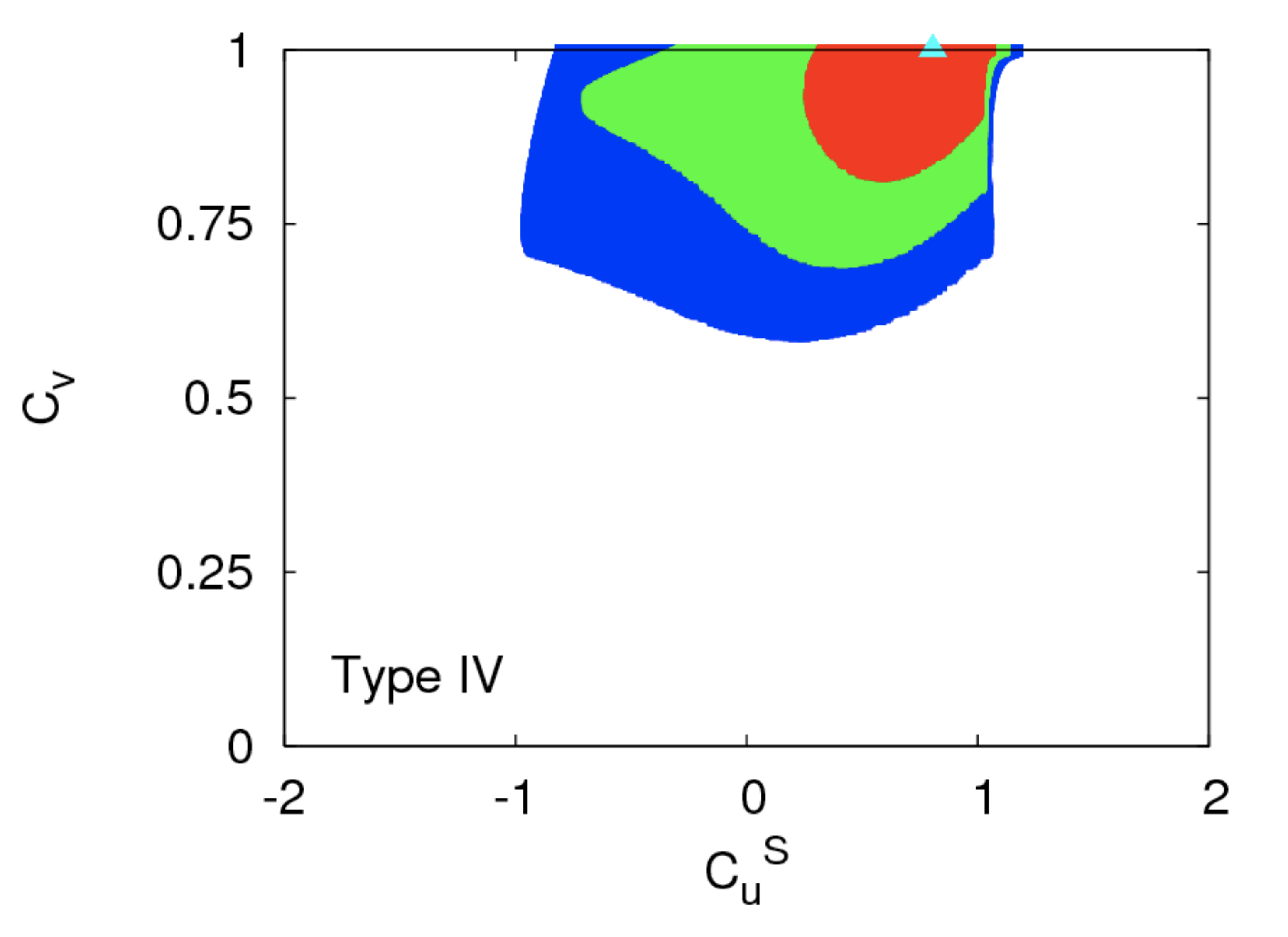}
\caption{\small \label{cpv1-cu-cv}
The same as in Fig.~\ref{cpv1-cu-cup} but 
in the plane of $C_u^S$ vs $C_v$ for Type I -- IV ({\bf CPV3}).
The description of the confidence regions is the same as Fig.~\ref{cpv1-cu-cup}.
}
\end{figure}

\begin{figure}[th!]
\centering
\includegraphics[width=3.2in]{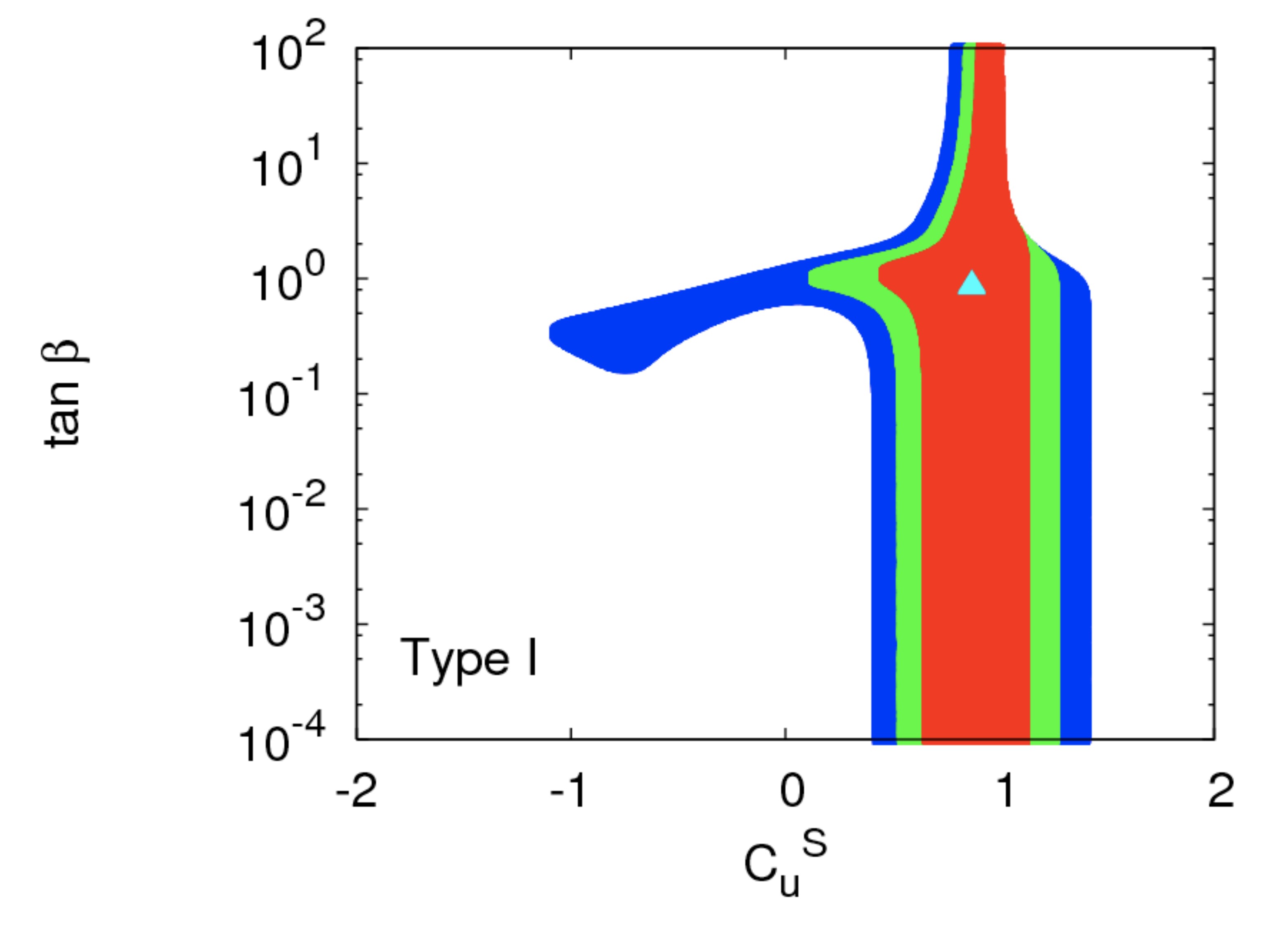}
\includegraphics[width=3.2in]{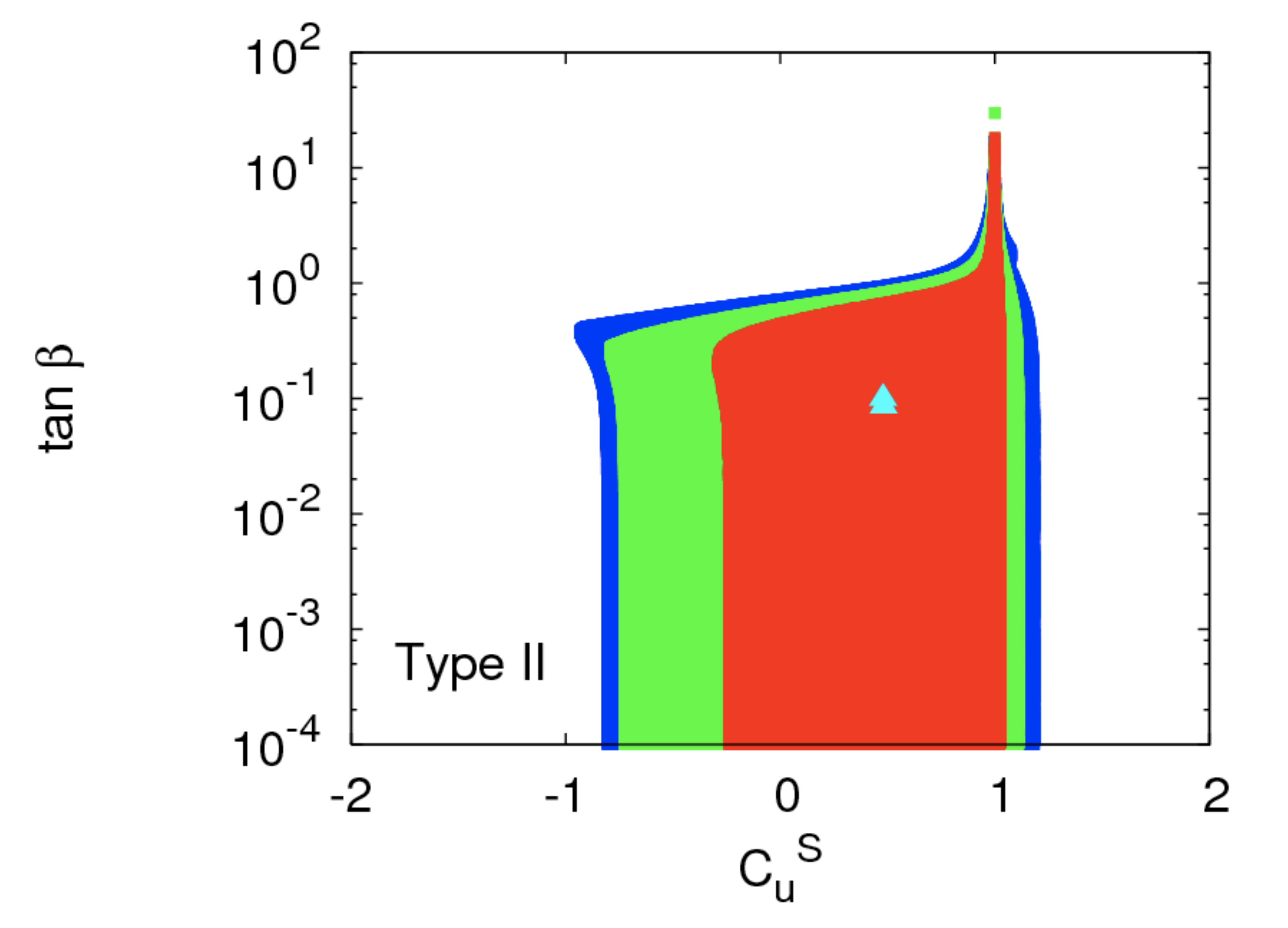}
\includegraphics[width=3.2in]{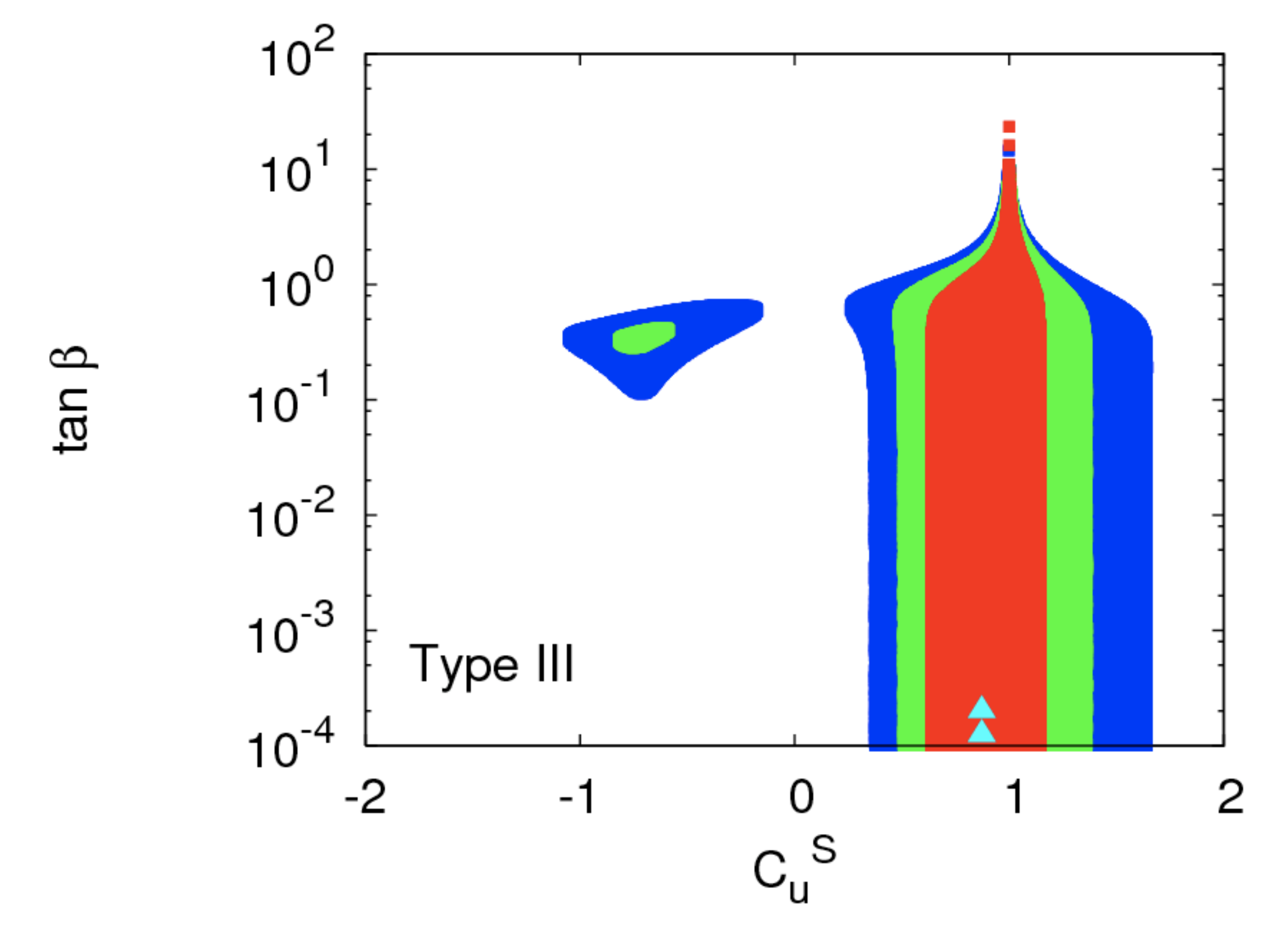}
\includegraphics[width=3.2in]{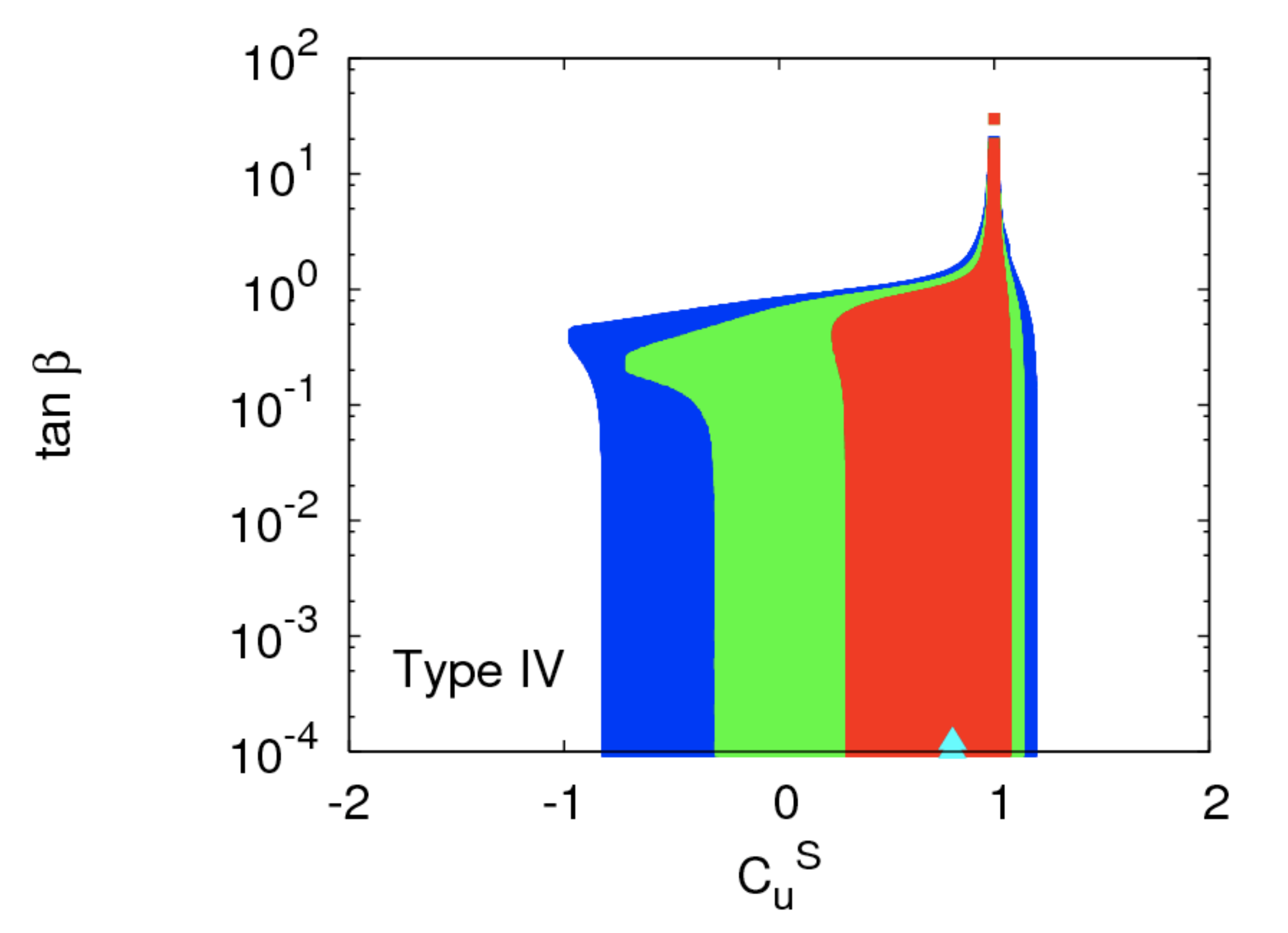}
\caption{\small \label{cpv1-cu-tanb}
The same as in Fig.~\ref{cpv1-cu-cup} but 
in the plane of $C_u^S$ vs $\tan\beta$ for Type I -- IV ({\bf CPV3}).
The description of the confidence regions is the same as Fig.~\ref{cpv1-cu-cup}.
}
\end{figure}

\begin{figure}[th!]
\centering
\includegraphics[width=3.2in]{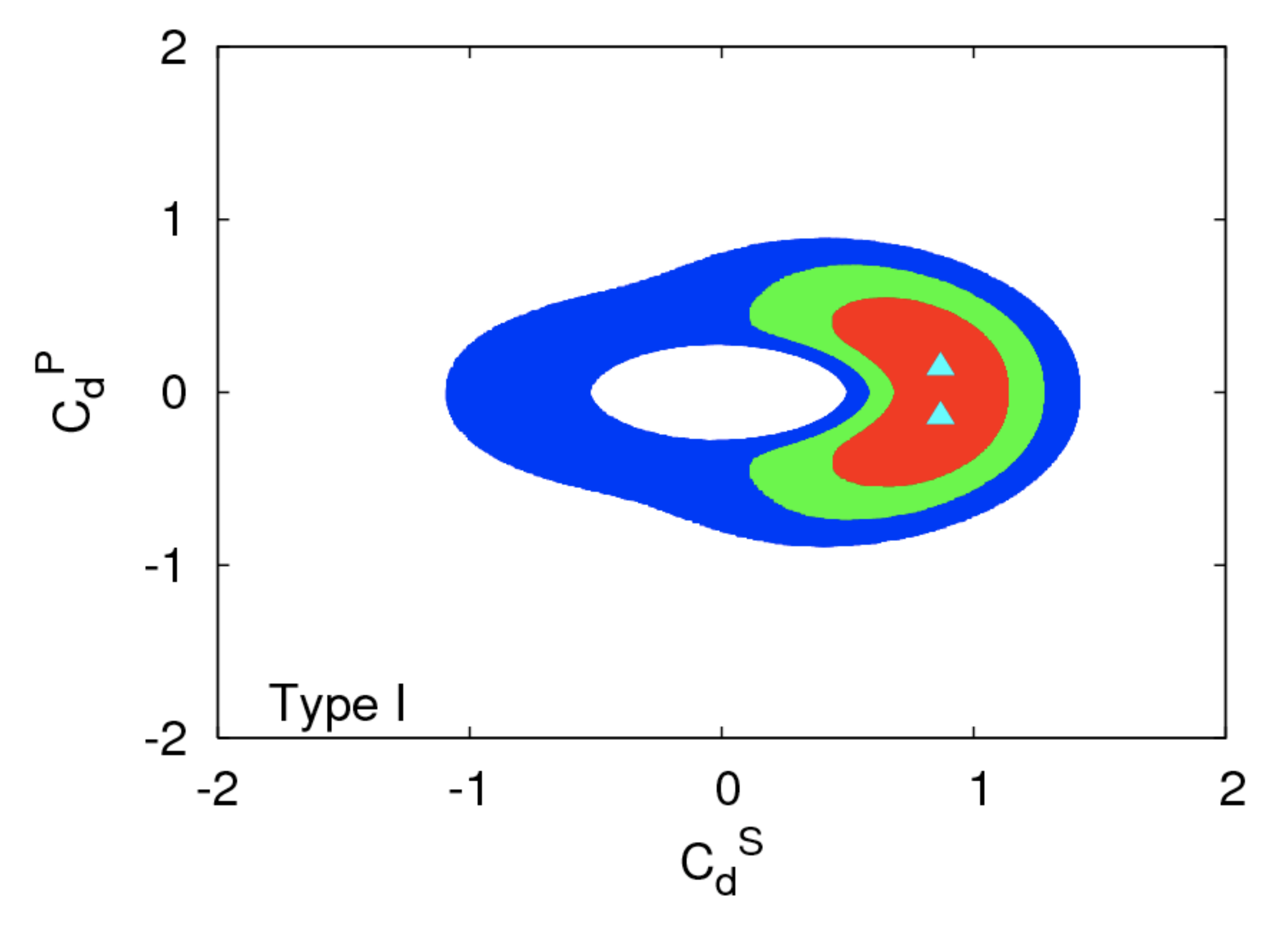}
\includegraphics[width=3.2in]{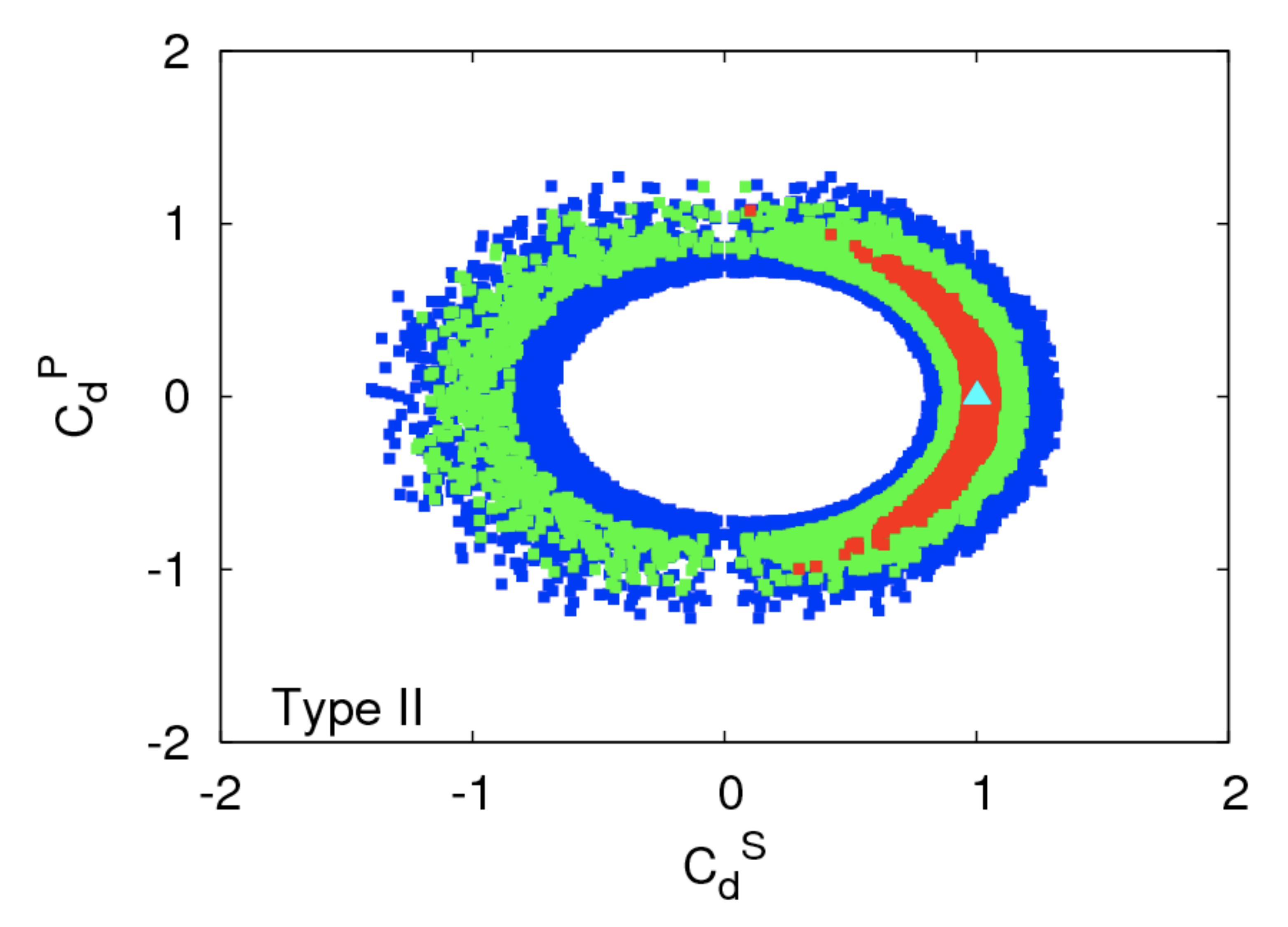}
\includegraphics[width=3.2in]{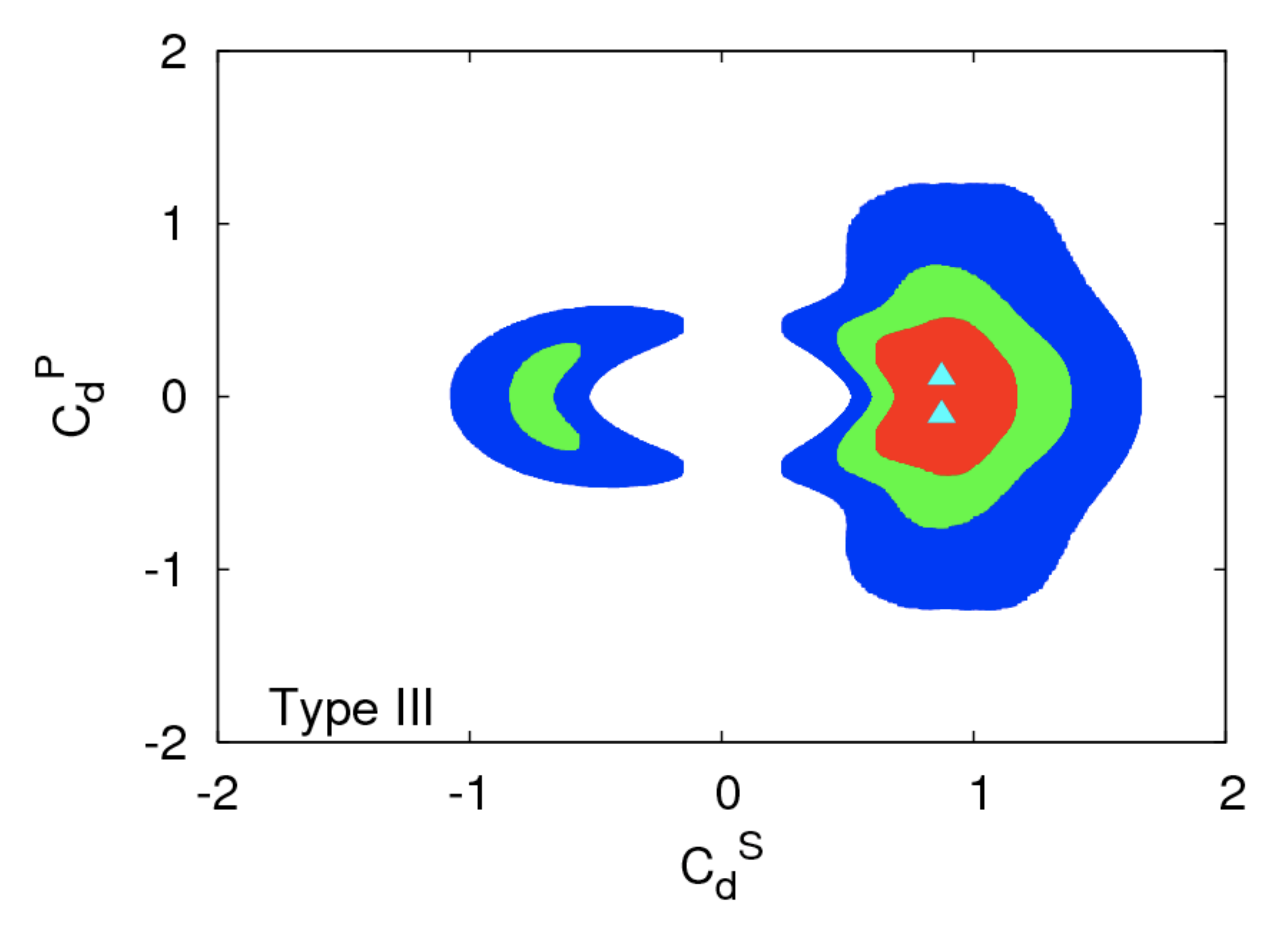}
\includegraphics[width=3.2in]{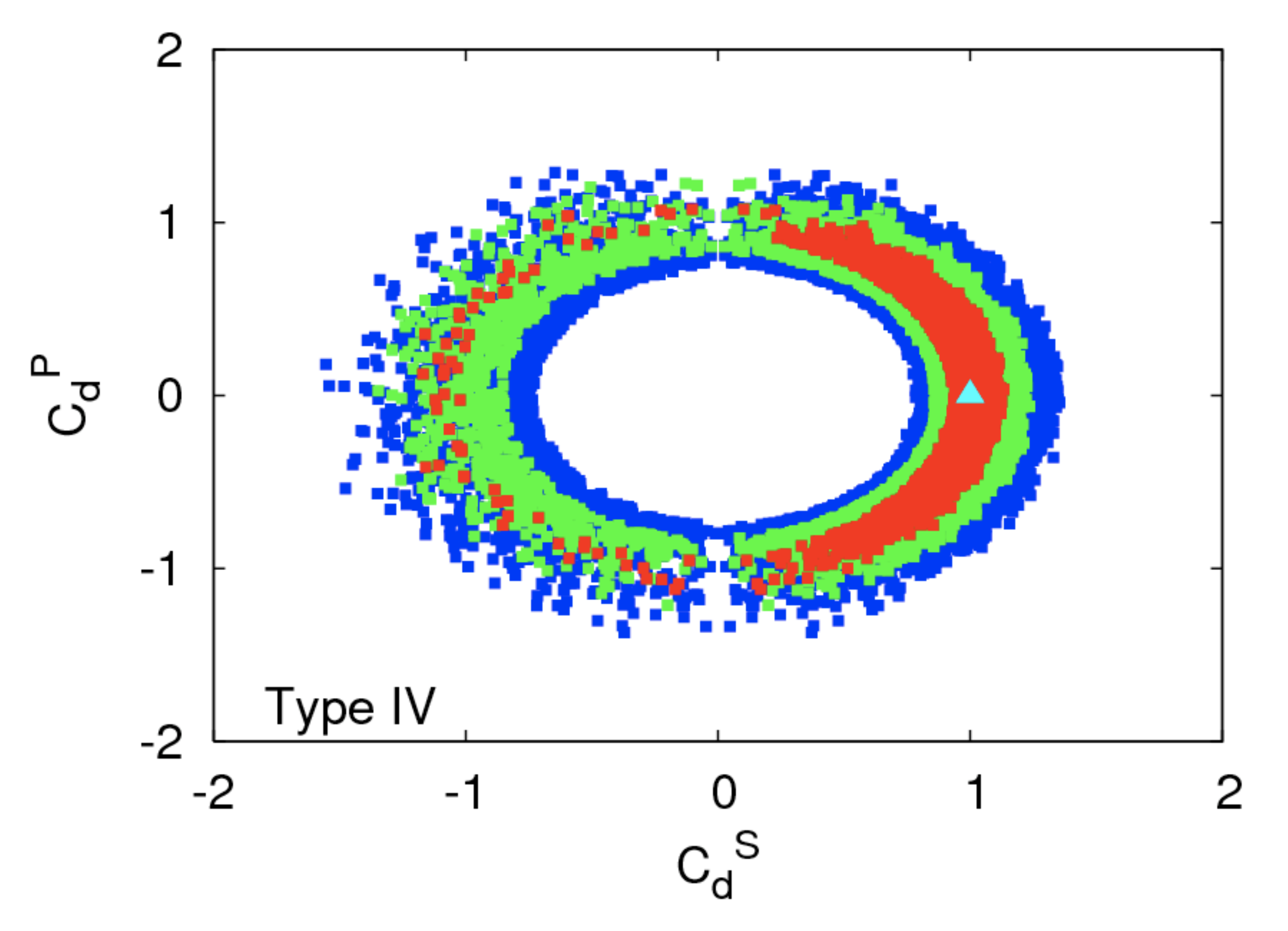}
\caption{\small \label{cpv1-cd-cdp}
The same as in Fig.~\ref{cpv1-cu-cup} but 
in the plane of $C_d^S$ vs $C_d^P$ for Type I -- IV ({\bf CPV3}).
The description of the confidence regions is the same as Fig.~\ref{cpv1-cu-cup}.
}
\end{figure}

\begin{figure}[th!]
\centering
\includegraphics[width=3.2in]{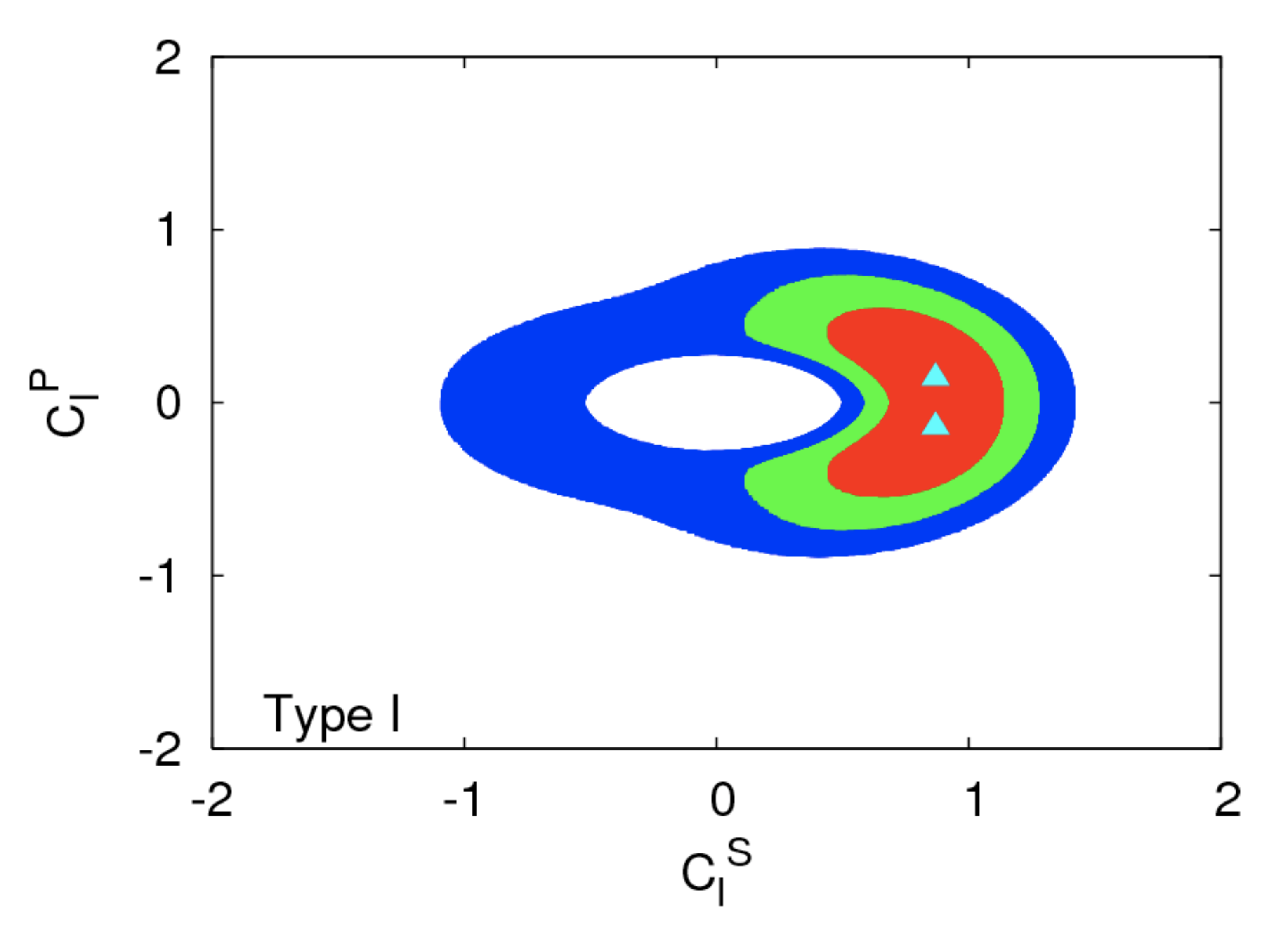}
\includegraphics[width=3.2in]{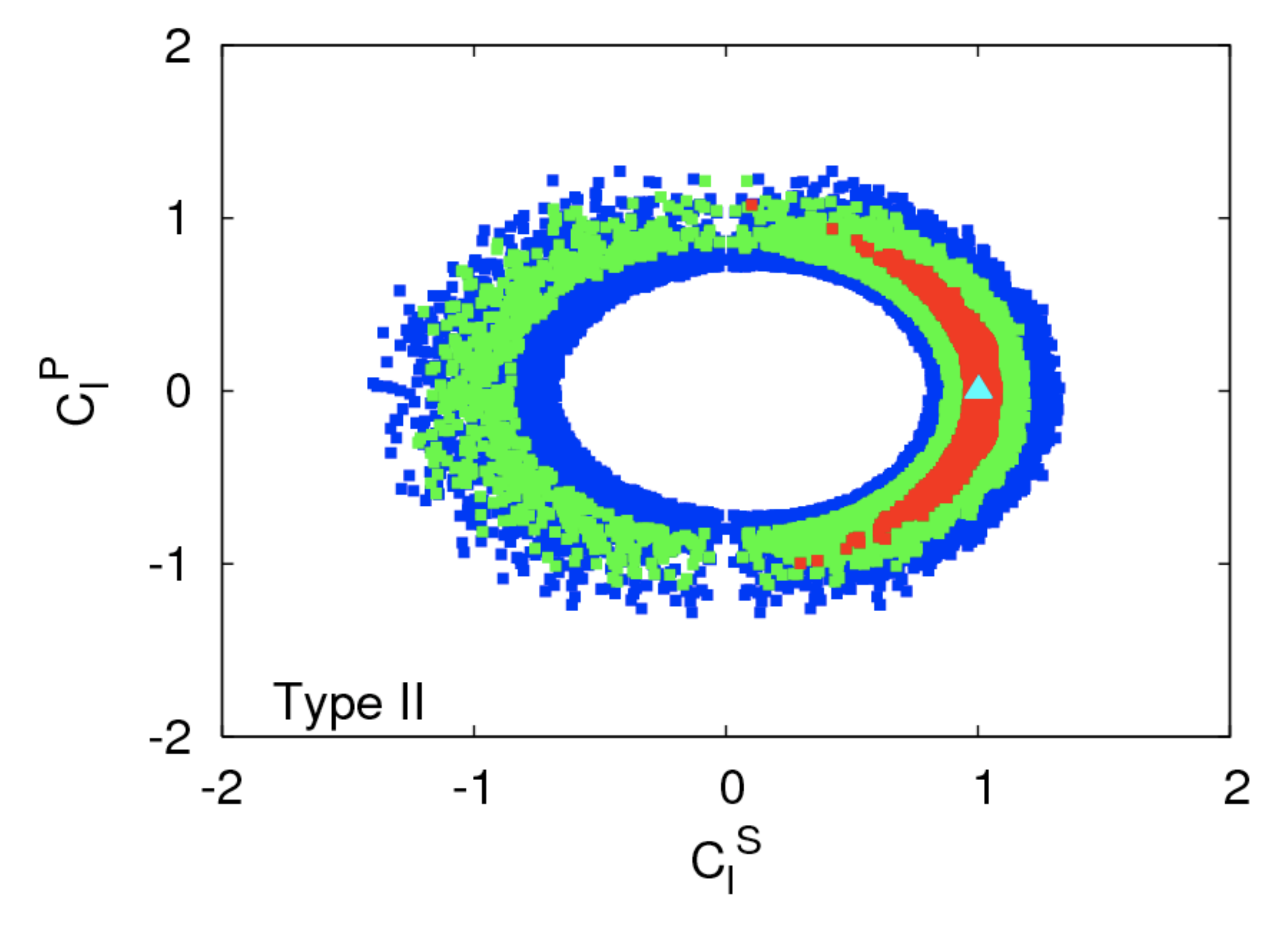}
\includegraphics[width=3.2in]{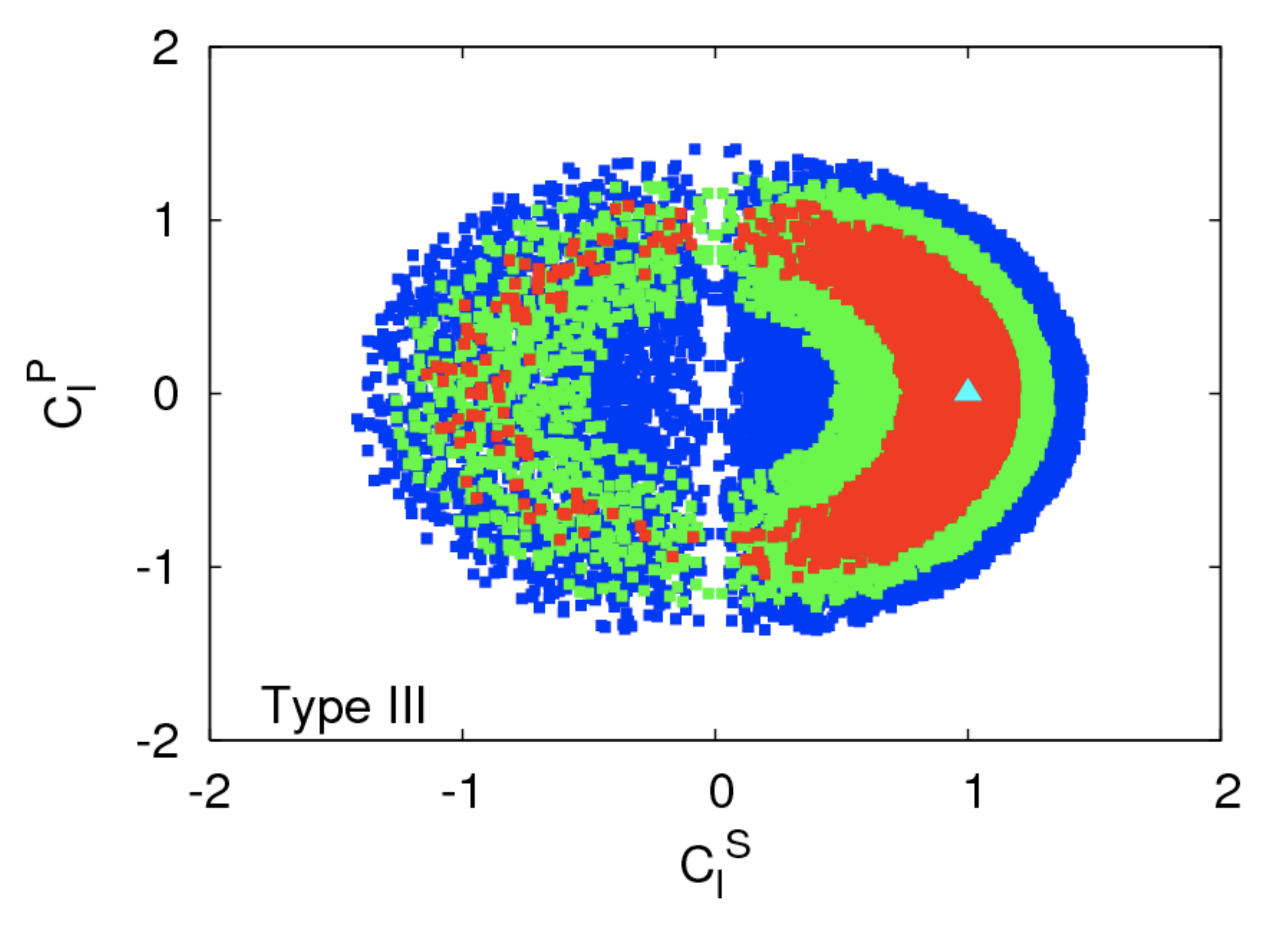}
\includegraphics[width=3.2in]{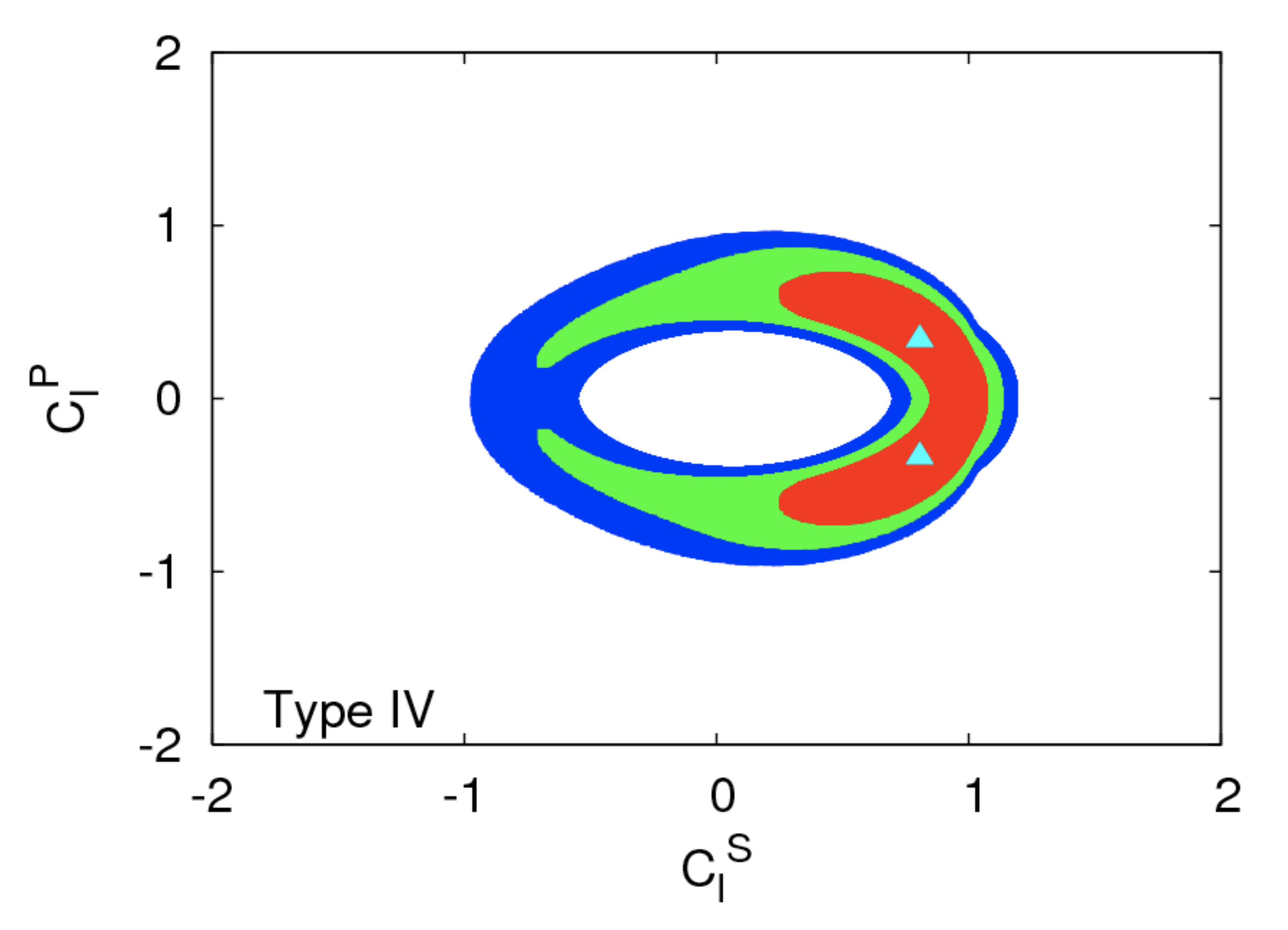}
\caption{\small \label{cpv1-cl-clp}
The same as in Fig.~\ref{cpv1-cu-cup} but 
in the plane of $C_l^S$ vs $C_l^P$ for Type I -- IV ({\bf CPV3}).
The description of the confidence regions is the same as Fig.~\ref{cpv1-cu-cup}.
}
\end{figure}

In the {\bf CPV3} fit, we vary $C_u^S$, $C_u^P$, and $C_v$ 
(or equivalently $\log_{10}\tan\beta$).
The other couplings $C_{d,l}^{S,P}$ are given by the 
relations shown in Table~\ref{tab:cdcl}.
The total $\chi^2$,
$\chi^2/{\rm dof}$, $p$-value, and the best-fit values for
$C_u^S$, $C_u^P$, and $C_v$ ($\tan\beta$) 
for the types I -- IV 2HDMs
are shown in the lower half of Table~\ref{tab:best}.
We show the contour plots for confidence-level regions
as functions $C_u^S$ vs $C_u^P$, 
$C_u^S$ vs $C_v$, $C_u^S$ vs $\tan\beta$, 
$C_d^S$ vs $C_d^P$, and 
$C_l^S$ vs $C_l^P$ in Figs.~\ref{cpv1-cu-cup} -- \ref{cpv1-cl-clp}, 
respectively.
We found that type II gives the smallest $\chi^2$
and the variation of
total $\chi^2$ among the 4 types is within $1.2$, which is 
about $4$ times larger compared to
the CP-conserving case. Yet, such small $\chi^2$ differences cannot
help us to preferentially select one of the types.
The best $p$-value for type II is $0.578$, which is the largest
among all the fits considered in this work, but it is still
smaller than the SM $p_{\rm SM} =0.65$.

The best-fit values for $C_u^S$ are 
all positive: $0.87$ (I), $0.48$ (II), $0.87$ (III) and $0.81$ (IV);
while we have both the positive and negative best-fit values for $C_u^P$:
$\pm 0.15$ (I), $\pm 0.51$ (II), $\pm 0.11$ (III) and $\pm 0.34$ (IV).
Note that the largest (almost maximal) CP violation can occur 
in type II with $C_u^S \sim |C_u^P| \sim 0.5$.
The best-fit values for $C_v$ are $0.99$ (I) and 1 (II, III and IV),
and those for $\tan\beta$ are $0.9$ (I), $0.1$ (II), 
and $\sim 10^{-4}$ (III and IV).

The CL regions in the $C_u^S$ and $C_u^P$ plane 
are shown in Fig.~\ref{cpv1-cu-cup}.
A positive $C_u^S$ is in general preferred and it takes a value between
$0.44$ and $1.1$ (I),
$-0.30$ and $1.1$ (II),
$0.64$ and $1.2$ (III), and
$0.26$ and $1.1$ (IV)
at 68.3 \% CL.
For $C_u^P$, the 68.3 \% CL regions are between:
$-0.55$ and $+0.55$ (I),
$-0.70$ and $+0.70$ (II),
$-0.45$ and $+0.45$ (III), and
$-0.73$ and $+0.73$ (IV).
We note that maximal CP violation with 
$C_u^S\sim |C_u^P|$ is possible even when
$C_v \simeq 1$. 
This can be understood by considering the relation Eq.~(\ref{eq:cv}), 
which takes on a form of
\begin{equation}
C_v=1-\frac{1}{2}\beta^2\left[\left(C_u^S-1\right)^2+\left(C_u^P\right)^2\right]
+{\cal O}(\beta^3)
\end{equation}
in the $\tan\beta=0$ limit.
Taking an example of $C_u^S=C_u^P=1/2$, one may have
\begin{equation}
O_{\phi_2 i}=\beta/2\,, \ \
O_{a i}=-\beta/2\,, \ \
O_{\phi_1 i}=1-\beta^2/4\,, \ \
C_v=1-\beta^2/4\,
\end{equation}
up to ${\cal O}(\beta^3)$.
Hence, although the 126-GeV observed state is mostly CP-even dominated by the
$\phi_1$ component,
it can have maximally CP-violating couplings to the up-type quarks
with $C_u^S=|C_u^P|=1/2$.

In Figs.~\ref{cpv1-cu-cv} and \ref{cpv1-cu-tanb}, we show the CL regions in the
$C_u^S$ vs $C_v$ and $C_u^S$ vs $\tan\beta$ planes, respectively.
Compared to the CPC case, we observe that 
the two islands are now merged together, except for type III.
We again find that $\tan\beta$ is bounded from above:
$\tan\beta\lsim 1$ (II),
$\tan\beta\lsim 3$ (III), and
$\tan\beta\lsim 2$ (IV).
As in the CPC case, considerable deviation of $C_u^S$ from 1 for
large $\tan\beta\gsim 10$ is not possible in the type II, III, IV models.

In Fig.~\ref{cpv1-cd-cdp}, we show the Higgs couplings to
the down-type quarks. The behavior can be understood by observing
the relations
$C_d^S=C_u^S$ and $C_d^P=-C_u^P$ (I and III) and
$C_d^S=\pm\left\{1+t_\beta^2[1-(C_u^S)^2]
-t_\beta^4/s_\beta^2\,(C_u^P)^2 \right\}^{1/2}$
and $C_d^P=t_\beta^2 C_u^P$ (II and IV),
see Table \ref{tab:cdcl}. 
Note that $|C_d^P| \lsim 1$ at 99.7 \% CL.
We observe large CP violation is possible in the Higgs
couplings to the down-type quarks.

In Fig.~\ref{cpv1-cl-clp}, we show the Higgs couplings to
the charged leptons. Now the couplings are given by
$C_l^S=C_u^S$ and $C_l^P=-C_u^P$ (I and IV) and
$C_l^S=\pm\left\{1+t_\beta^2[1-(C_u^S)^2]
-t_\beta^4/s_\beta^2\,(C_u^P)^2 \right\}^{1/2}$
and $C_l^P=t_\beta^2 C_u^P$ (II and III). 
Again we note that $|C_l^P| \lsim 1$ at 99.7 \% CL
and large CP violation is also possible in the Higgs
couplings to the charged leptons.

Before we close this sub-subsection, we make a comment on
the figures for the CL regions in the planes of 
$C_d^S$ vs $C_d^P$ and $C_l^S$ vs $C_l^P$.
Unless $(C_{d,l}^S,C_{d,l}^P)=(C_u^S,-C_u^P)$,
the boundaries of the  CL regions are somewhat fuzzy
as shown 
in the frames for type II and IV of Fig.~\ref{cpv1-cd-cdp} 
and in those for type II and III of Fig.~\ref{cpv1-cl-clp}.
We figure out that this is because one has
$(C_{d,l}^S, C_{d,l}^P) \sim (1,0)$ in most of the parameters space
due to the coupling relations shown in Table~\ref{tab:cdcl}.
Furthermore, we have the fewer points
on the negative side of $C_d^S$ or $C_l^S$.
For the couplings $C_{d,l}^S$ to be negative,
the negative sign needs to be 
chosen for $O_{\phi_1 i}$ in Eq.~(\ref{eq:oai}).
But we note that the other positive
sign is chosen mostly for $O_{\phi_1 i}$
due to the choice of $C_v >0$ made in the analysis.
Similar behavior happens in Fig.~\ref{cpv3-cd-cdp}
and Fig.~\ref{cpv3-cl-clp}.

\subsubsection{\bf CPV4}

\begin{figure}[th!]
\centering
\includegraphics[width=3.2in]{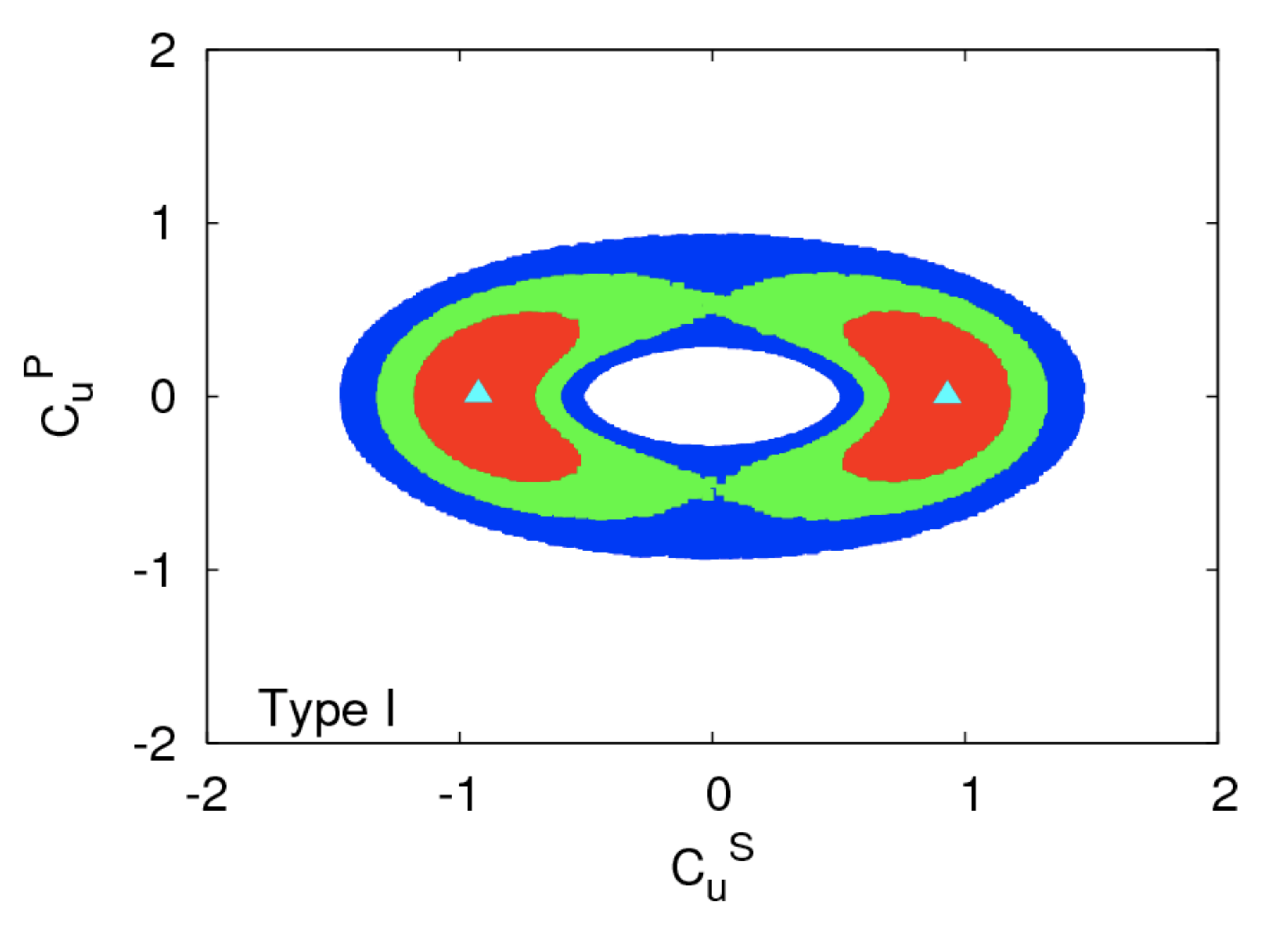}
\includegraphics[width=3.2in]{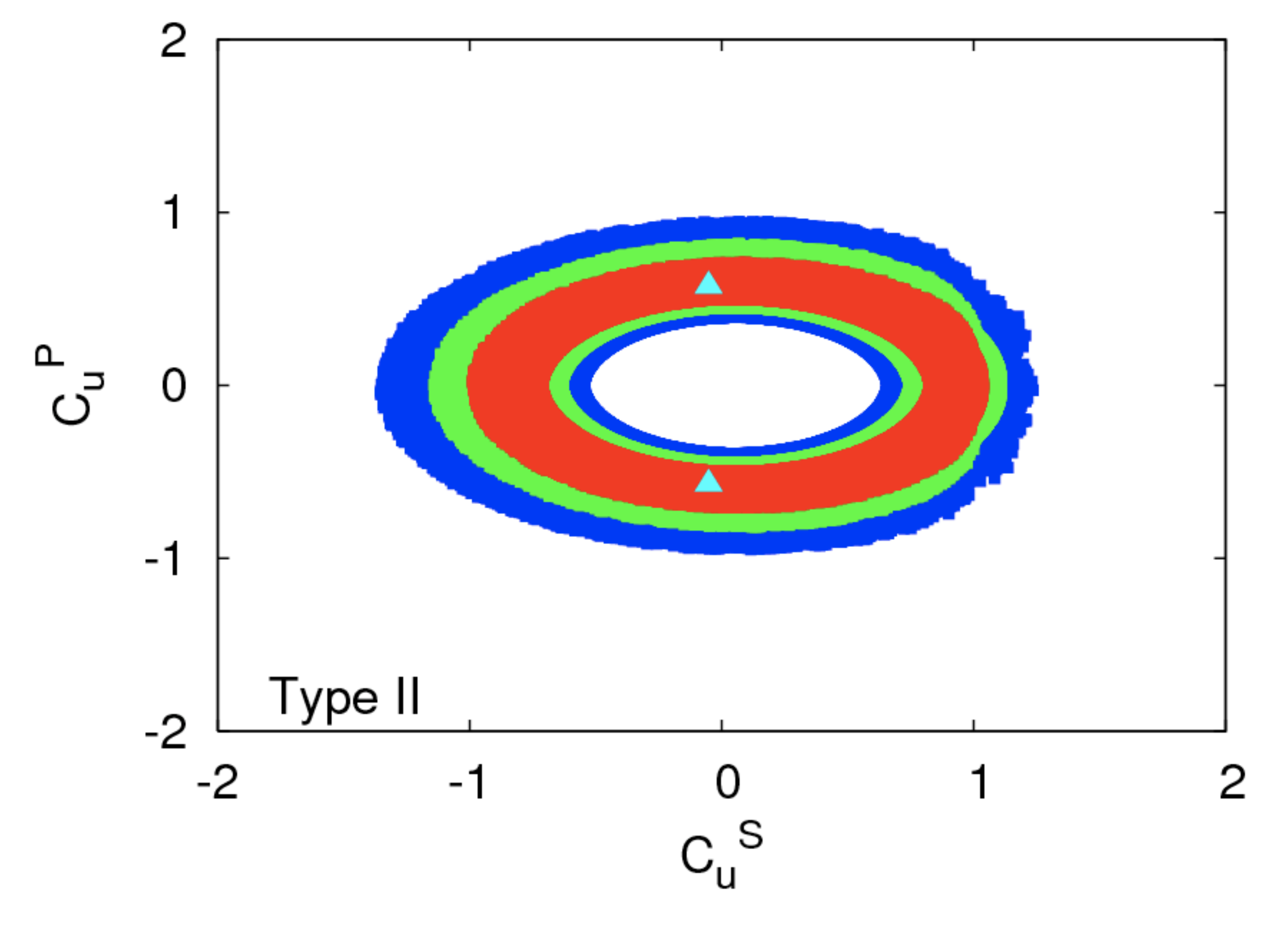}
\includegraphics[width=3.2in]{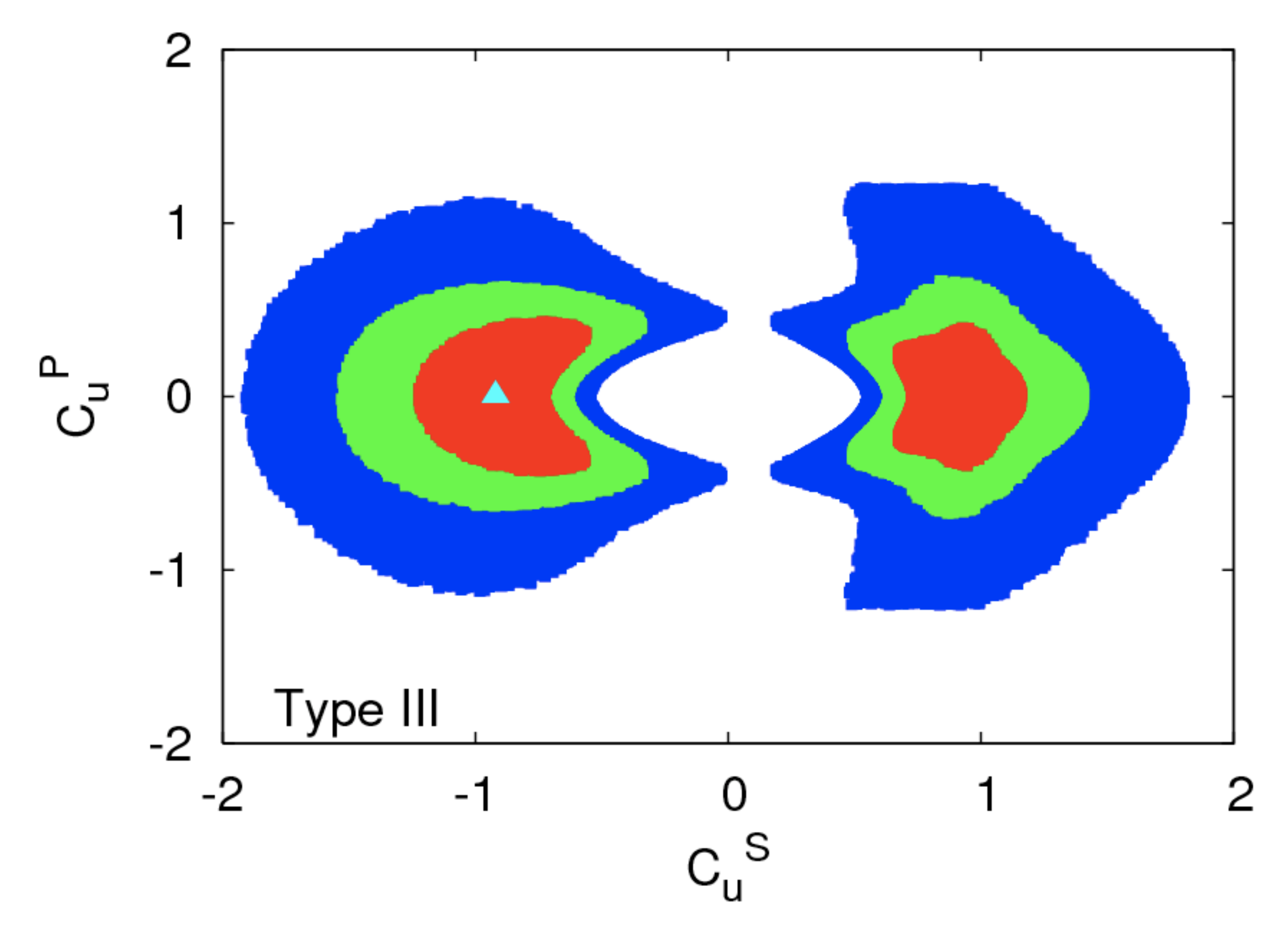}
\includegraphics[width=3.2in]{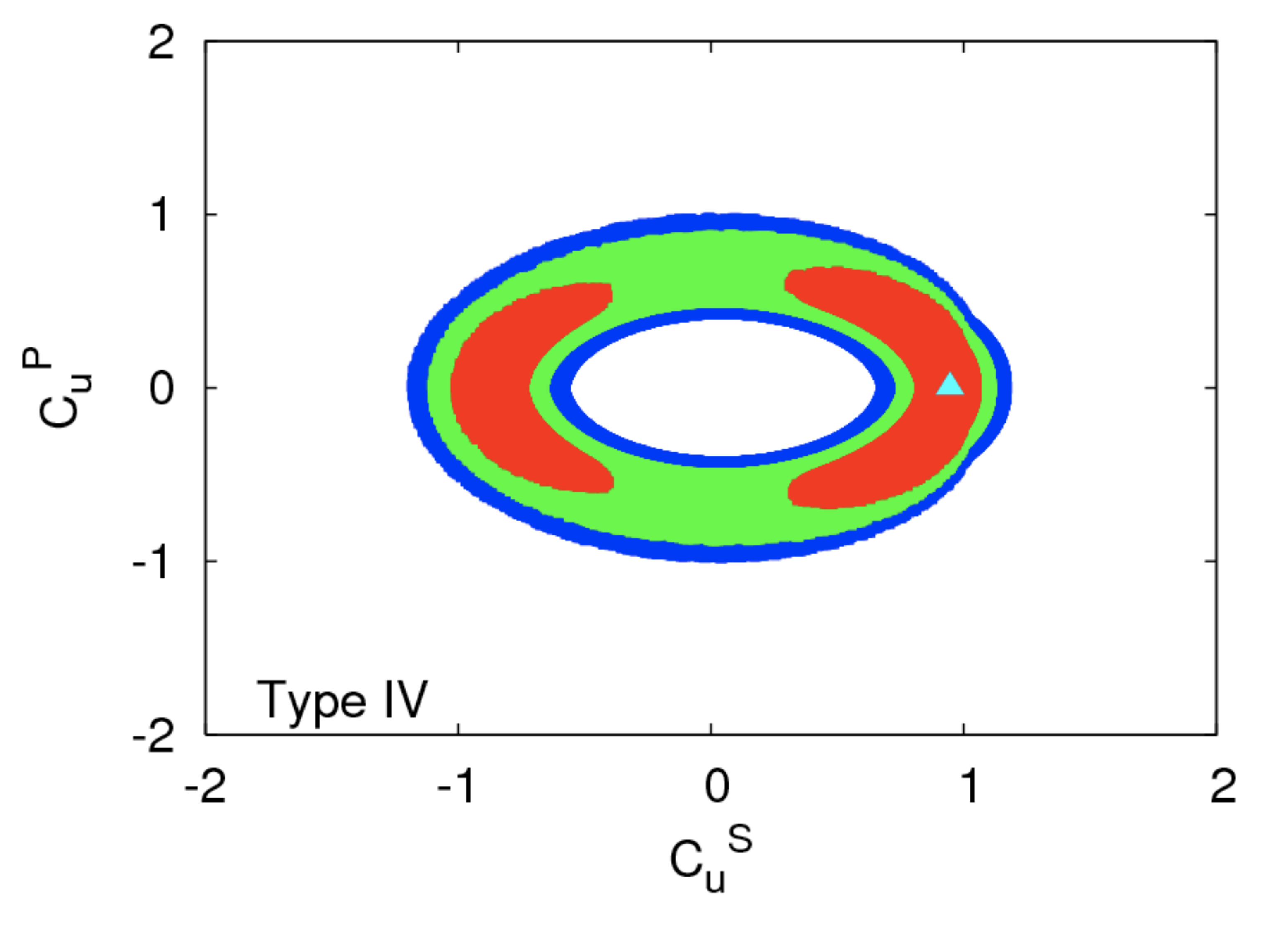}
\caption{\small \label{cpv3-cu-cup}
The confidence-level regions of the fit by varying $C_u^S$, $C_u^P$, 
$\log_{10} \tan\beta$, and $\left(\Delta S^\gamma\right)^{H^\pm}$  ({\bf CPV4} case) 
in the plane of $C_u^S$ vs $C_u^P$ for Type I -- IV.
The contour regions shown are for 
$\Delta \chi^2 \le 2.3$ (red), $5.99$ (green), and $11.83$ (blue) 
above the minimum, which 
correspond to confidence levels of
$68.3\%$, $95\%$, and $99.7\%$, respectively.
The best-fit points are denoted by the triangle.
}
\end{figure}

\begin{figure}[th!]
\centering
\includegraphics[width=3.2in]{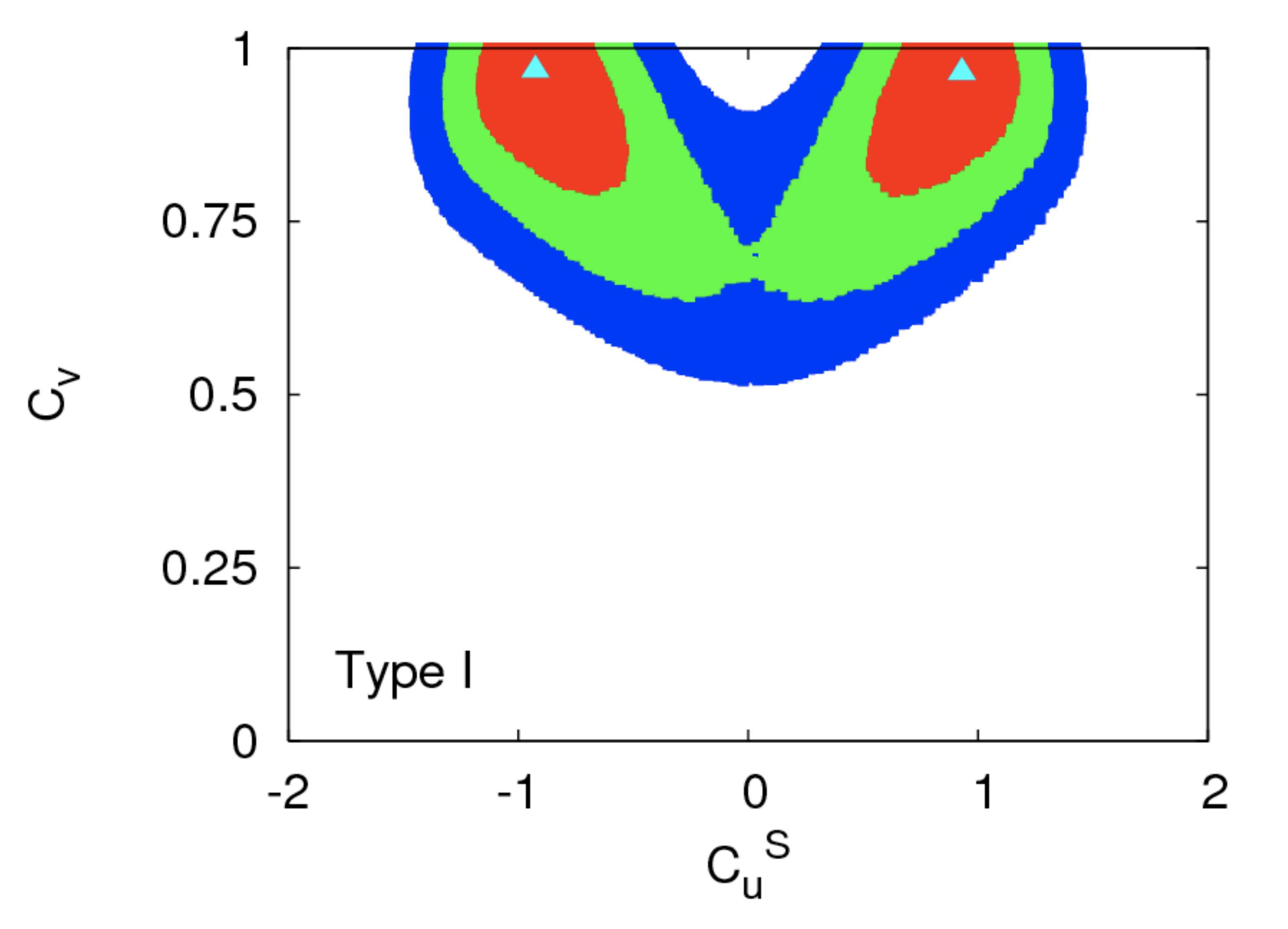}
\includegraphics[width=3.2in]{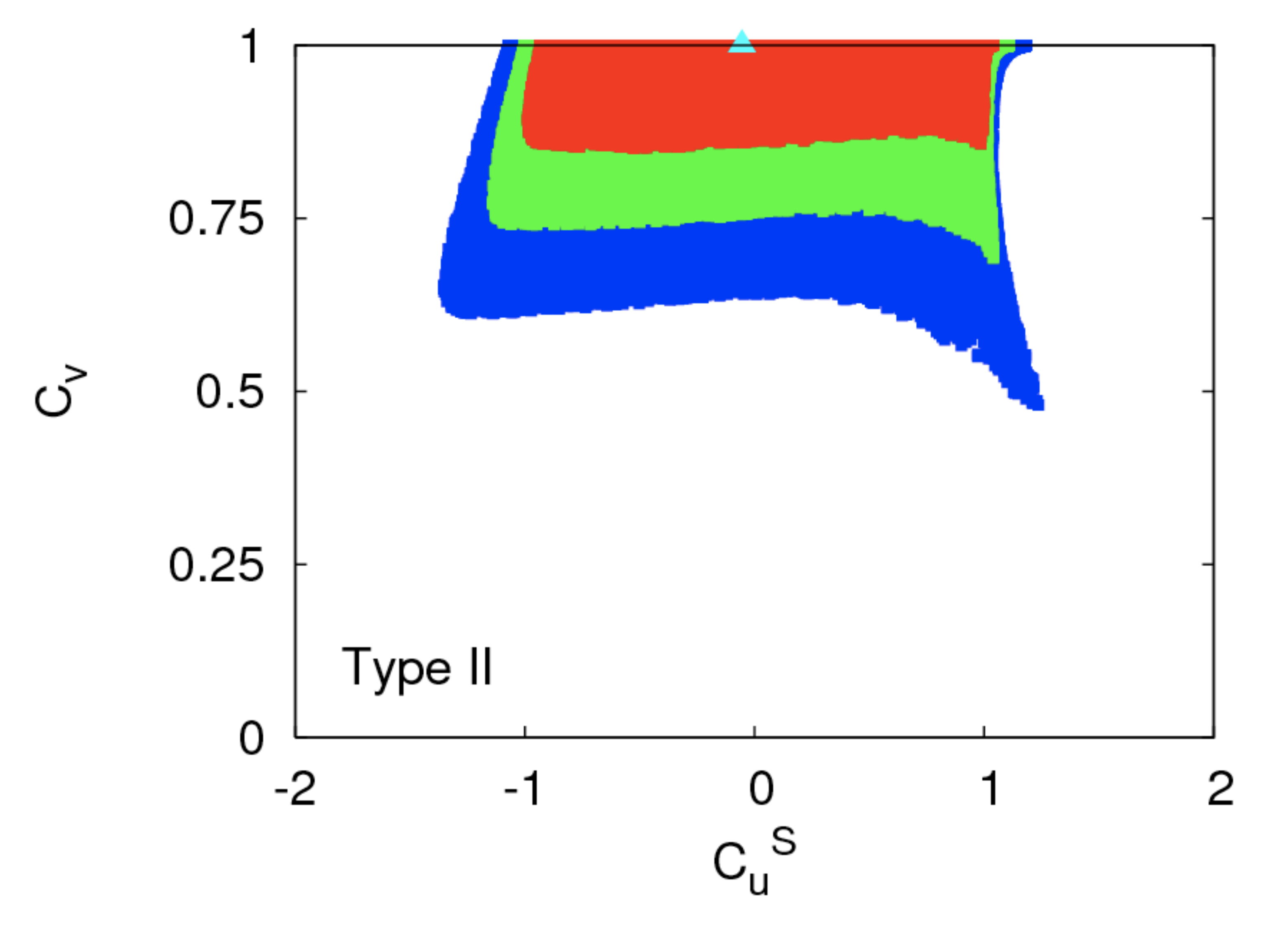}
\includegraphics[width=3.2in]{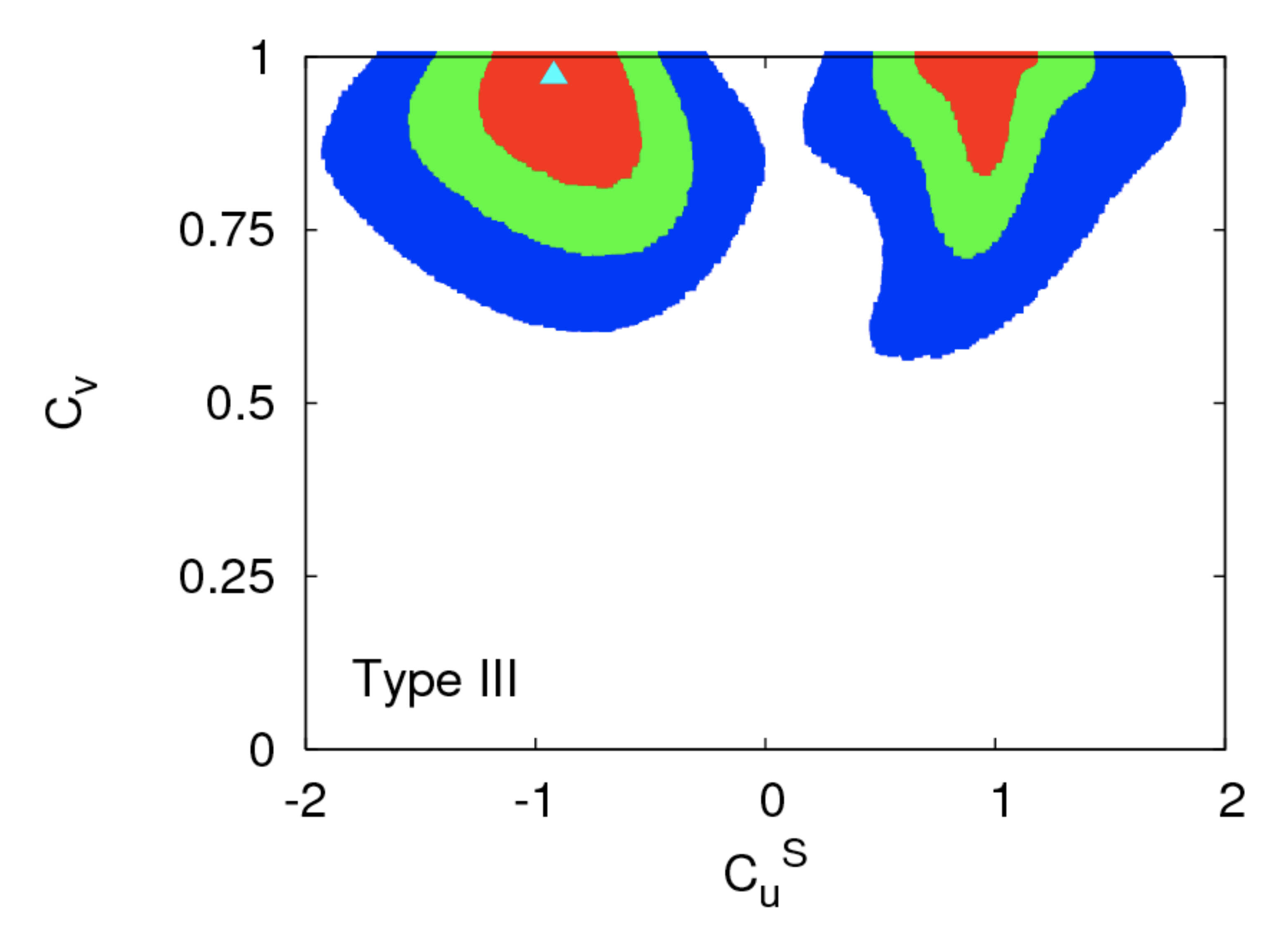}
\includegraphics[width=3.2in]{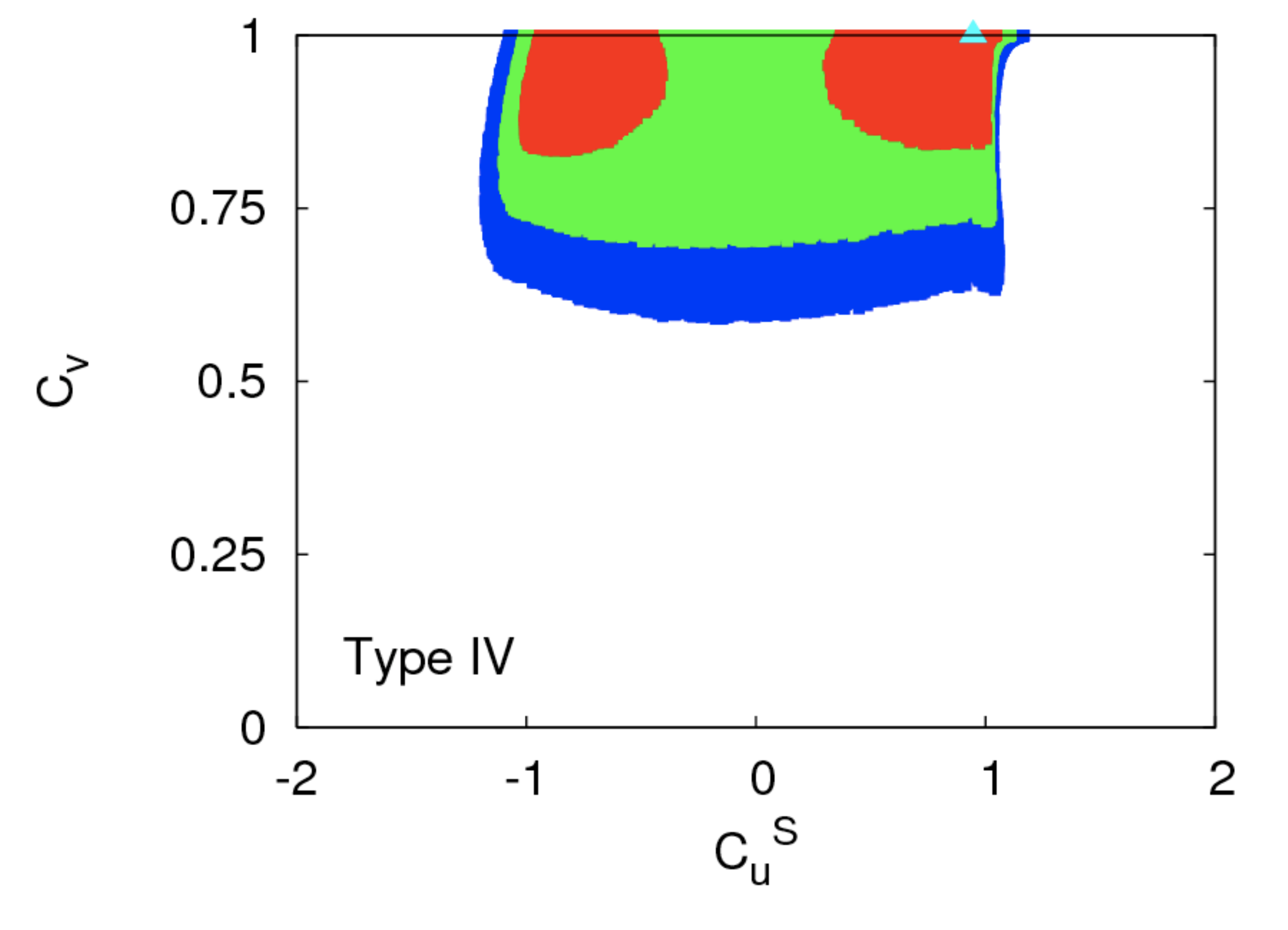}
\caption{\small \label{cpv3-cu-cv}
The same as Fig.~\ref{cpv3-cu-cup} but in the plane of $C_u^S$ vs $C_v$ 
for Type I -- IV ({\bf CPV4}).
The description of the confidence regions is the same as Fig.~\ref{cpv3-cu-cup}.
}
\end{figure}

\begin{figure}[th!]
\centering
\includegraphics[width=3.2in]{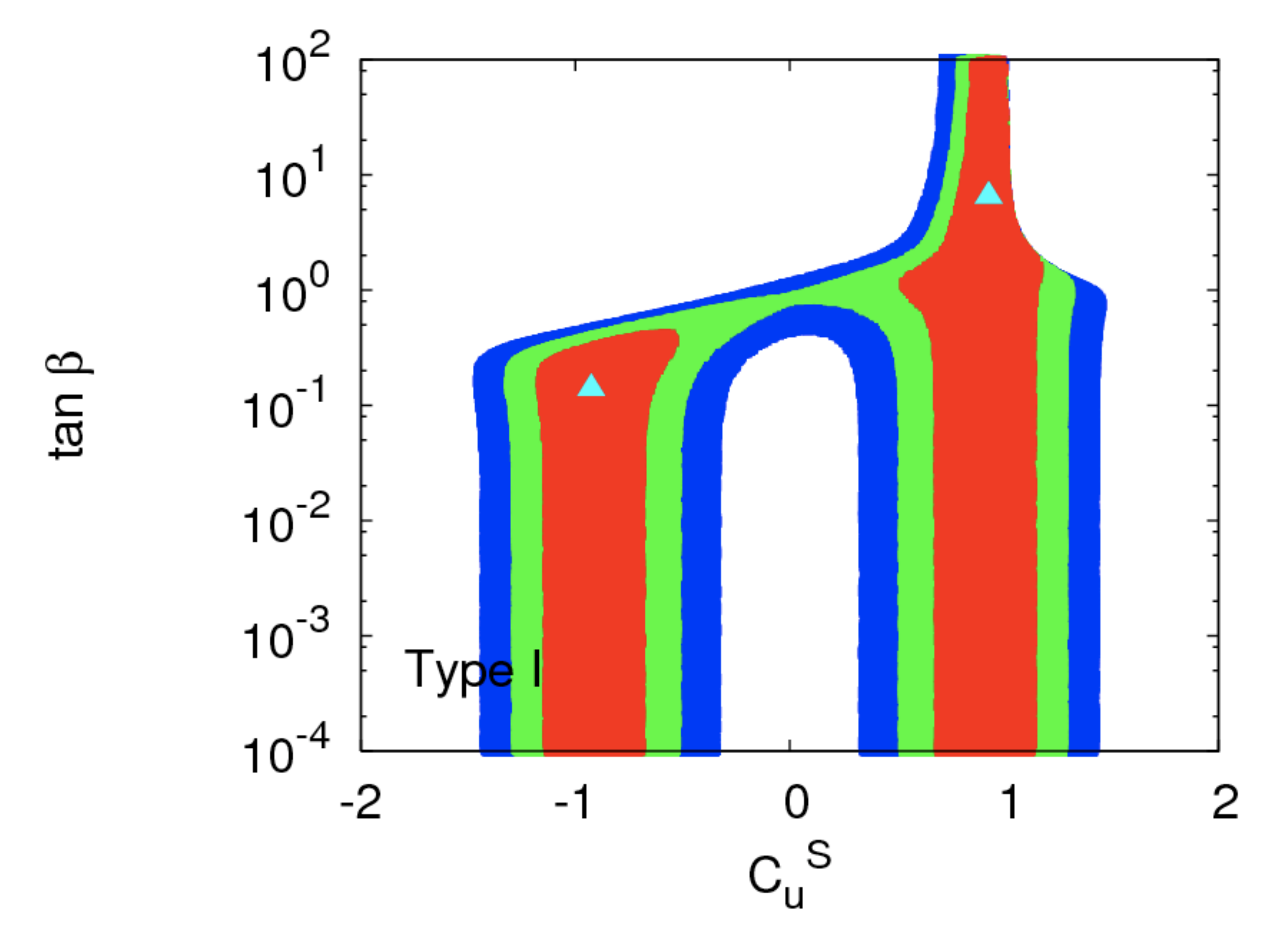}
\includegraphics[width=3.2in]{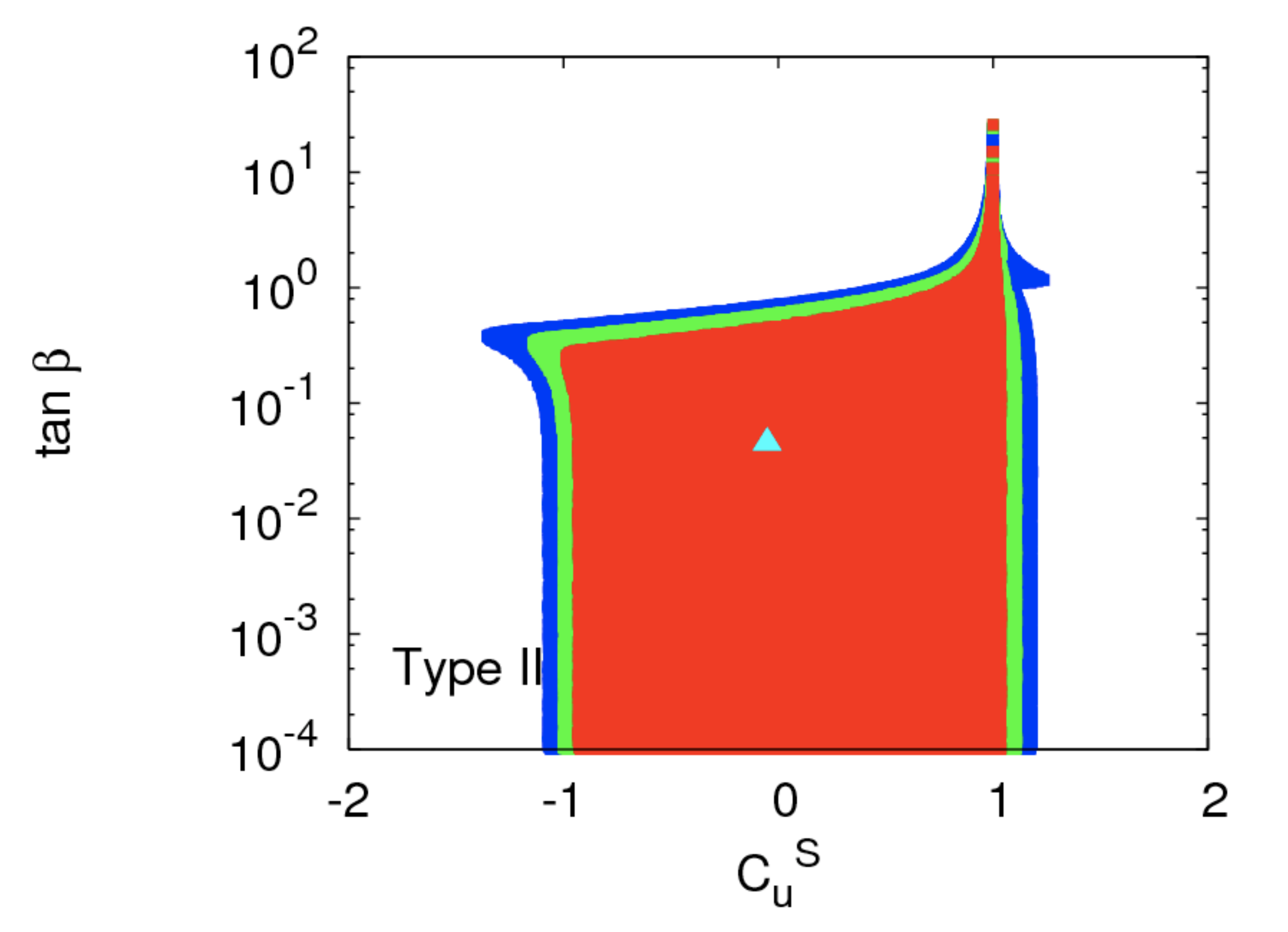}
\includegraphics[width=3.2in]{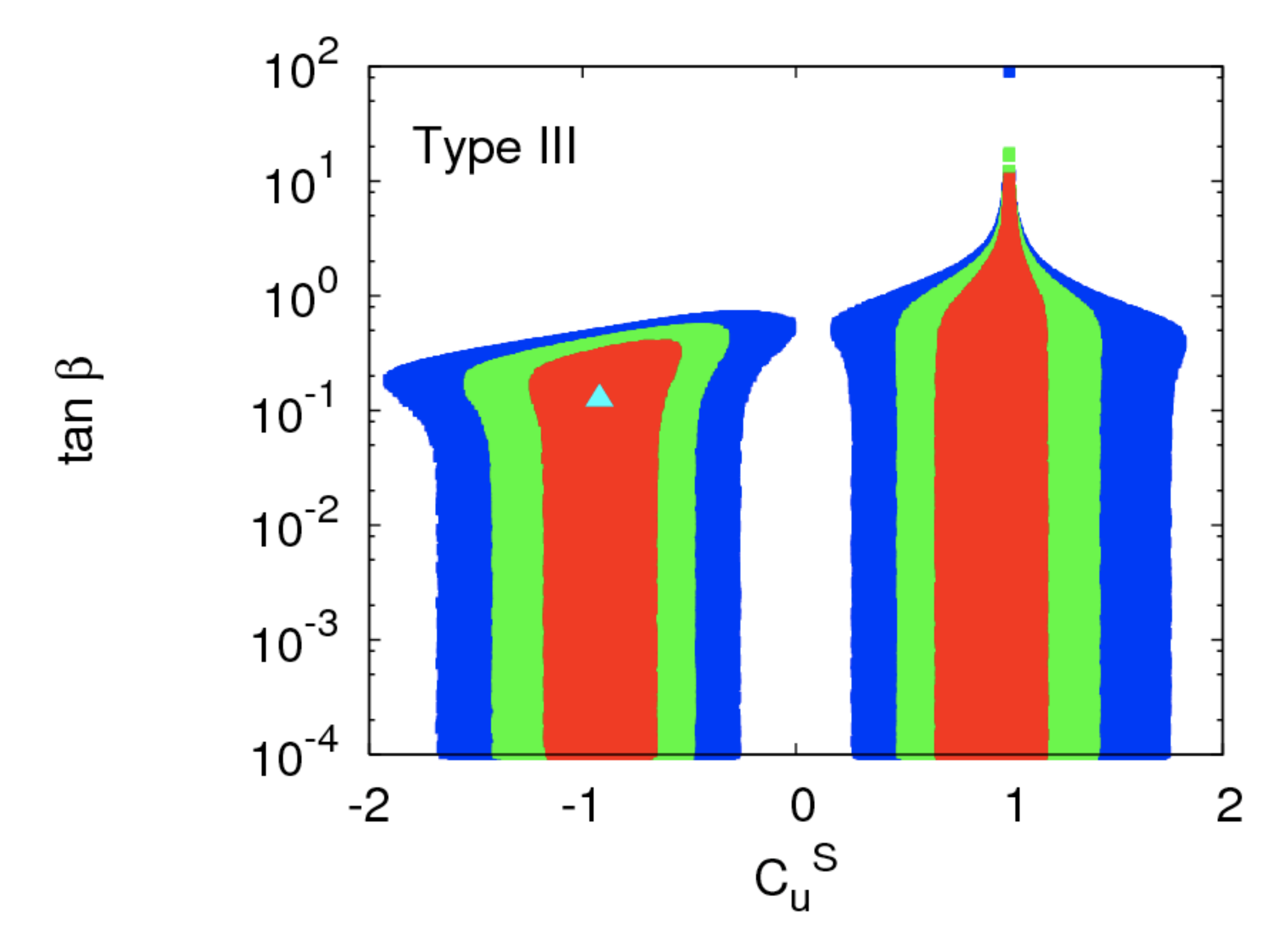}
\includegraphics[width=3.2in]{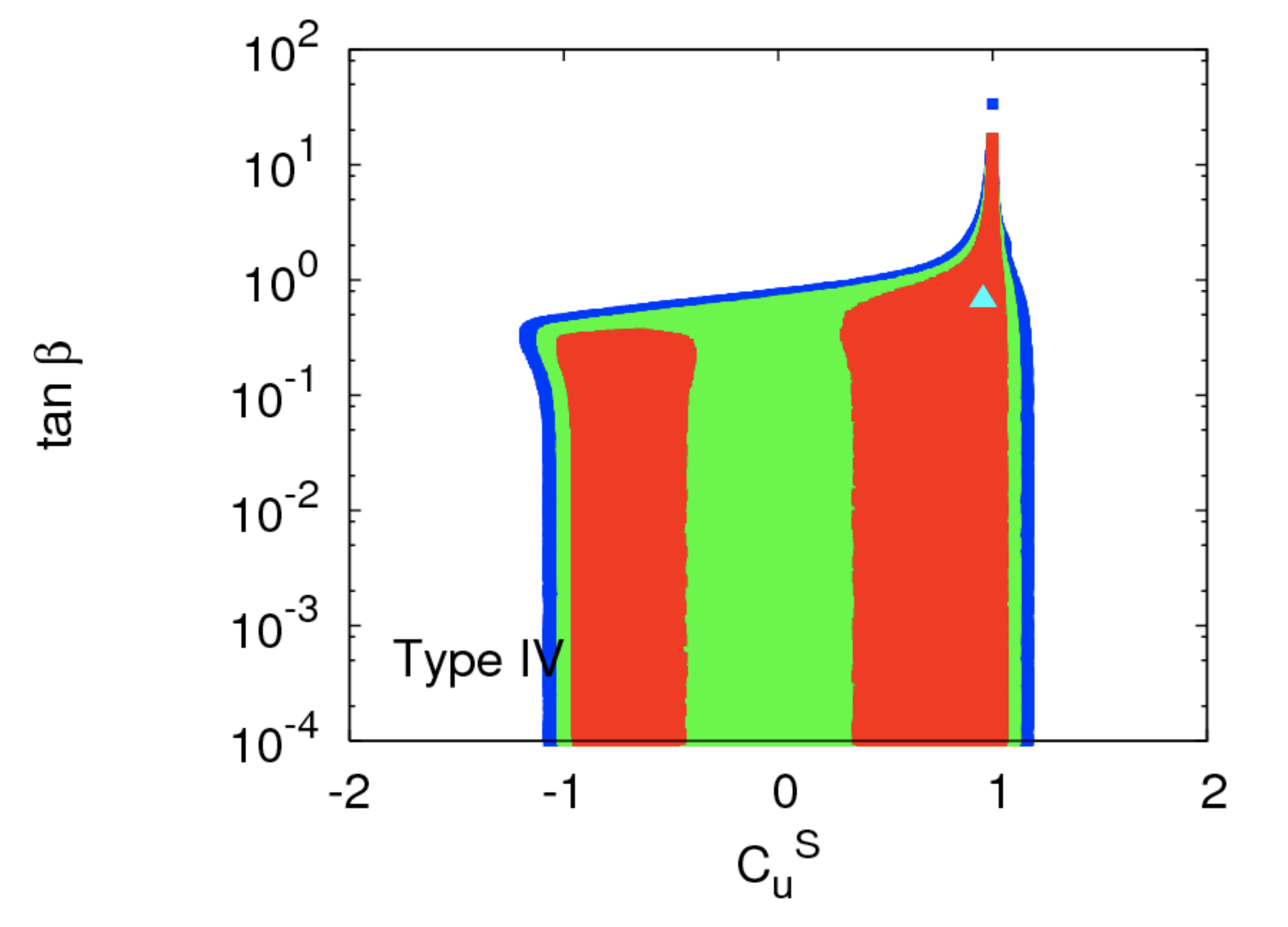}
\caption{\small \label{cpv3-cu-tanb}
The same as Fig.~\ref{cpv3-cu-cup} but 
in the plane of $C_u^S$ vs $\tan\beta$ for Type I -- IV ({\bf CPV4} case).
The description of the confidence regions is the same as Fig.~\ref{cpv3-cu-cup}.
}
\end{figure}

\begin{figure}[th!]
\centering
\includegraphics[width=3.2in]{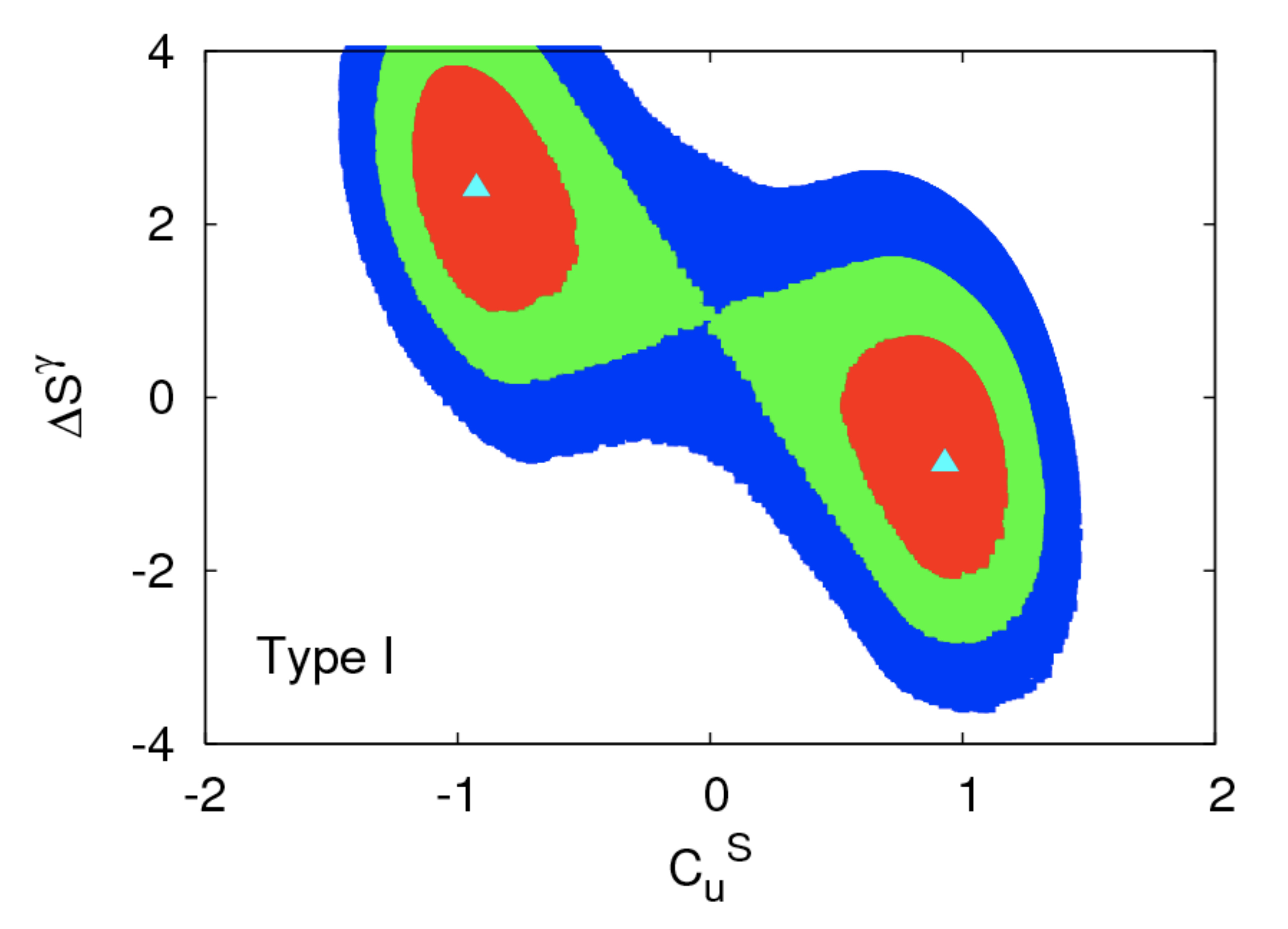}
\includegraphics[width=3.2in]{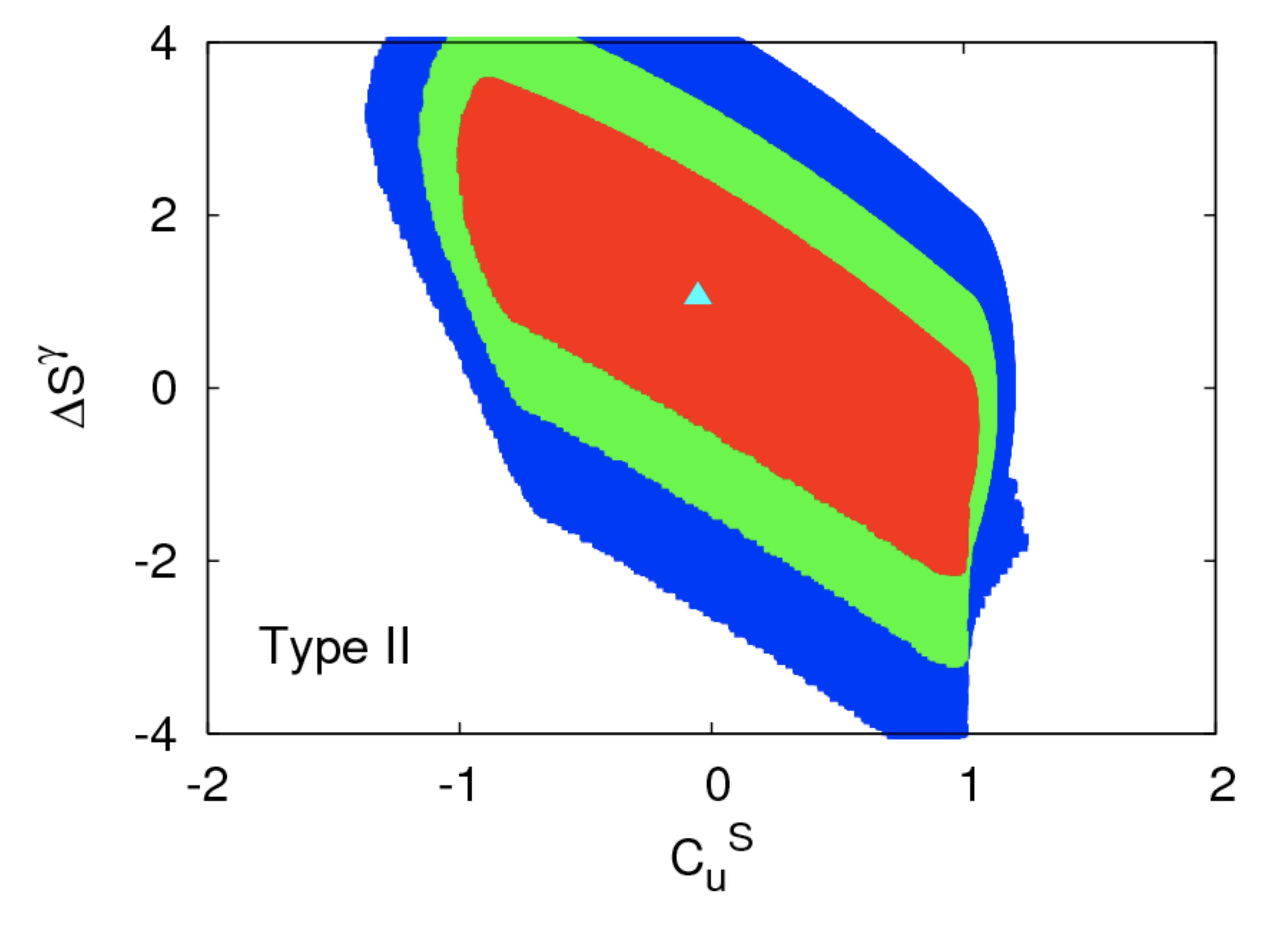}
\includegraphics[width=3.2in]{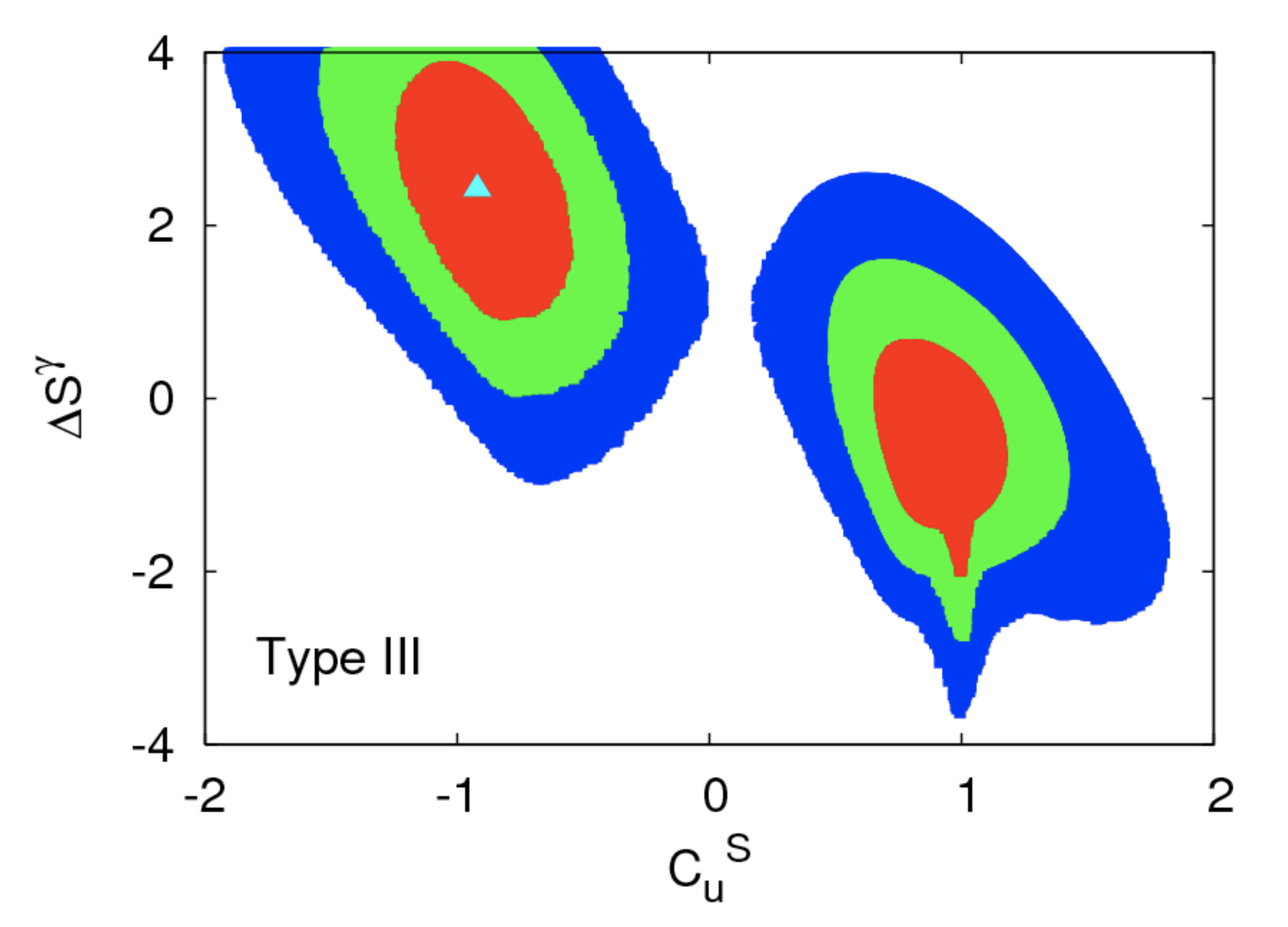}
\includegraphics[width=3.2in]{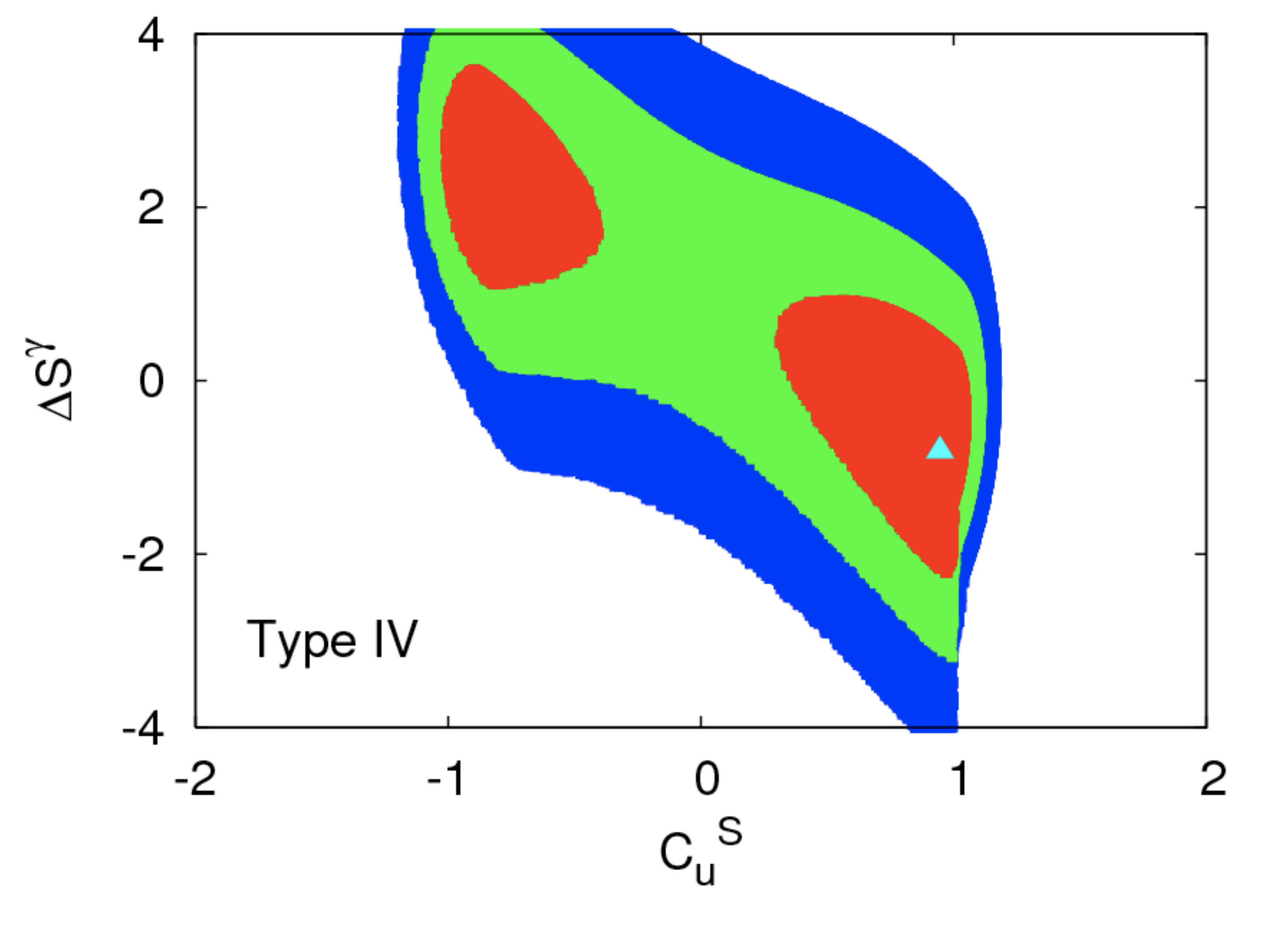}
\caption{\small \label{cpv3-cu-dsp}
The same as Fig.~\ref{cpv3-cu-cup} but 
in the plane of $C_u^S$ vs $\left(\Delta S^\gamma\right)^{H^\pm}$ 
for Type I -- IV ({\bf CPV4}).
The description of the confidence regions is the same as Fig.~\ref{cpv3-cu-cup}.
}
\end{figure}

\begin{figure}[th!]
\centering
\includegraphics[width=3.2in]{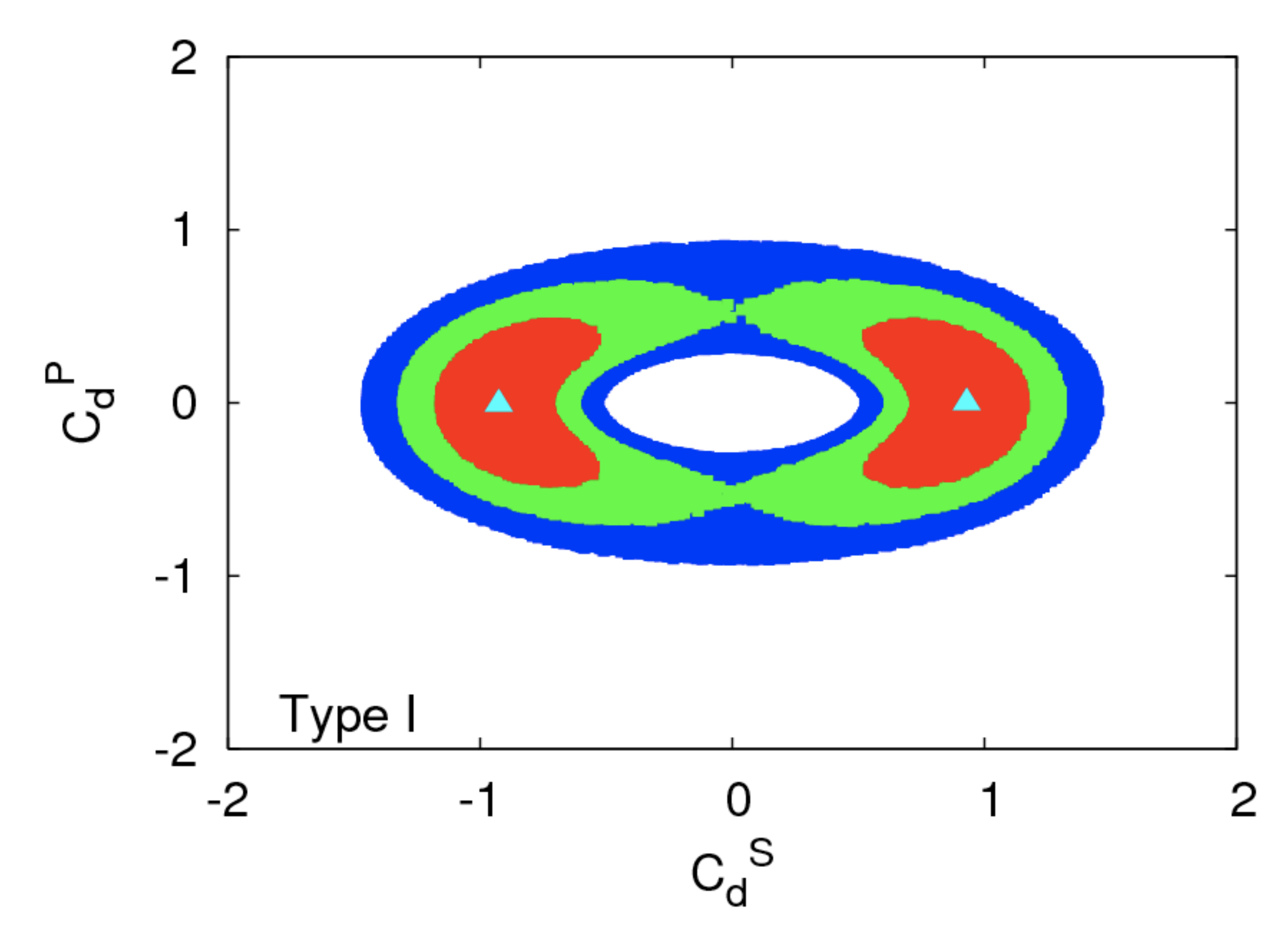}
\includegraphics[width=3.2in]{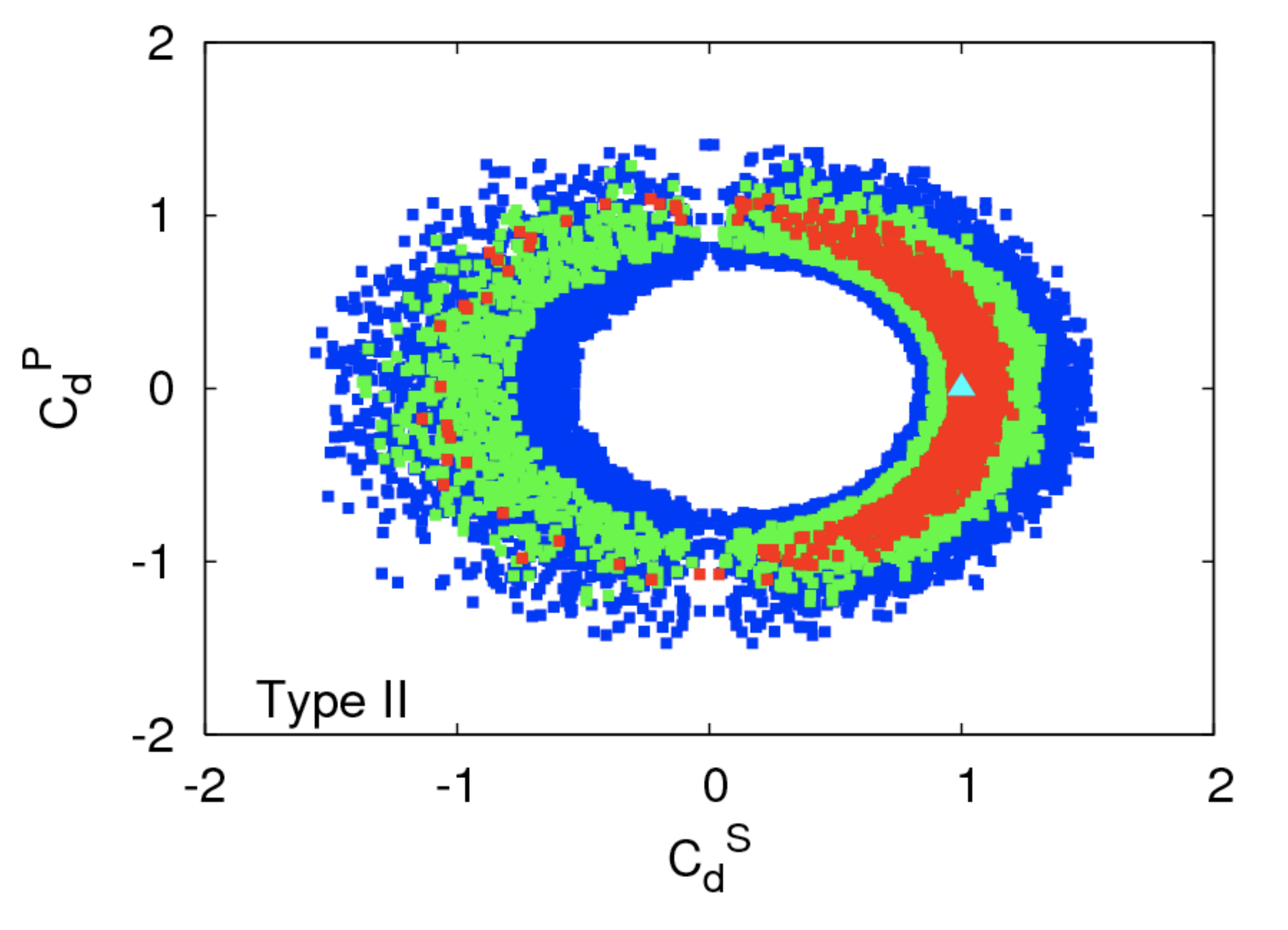}
\includegraphics[width=3.2in]{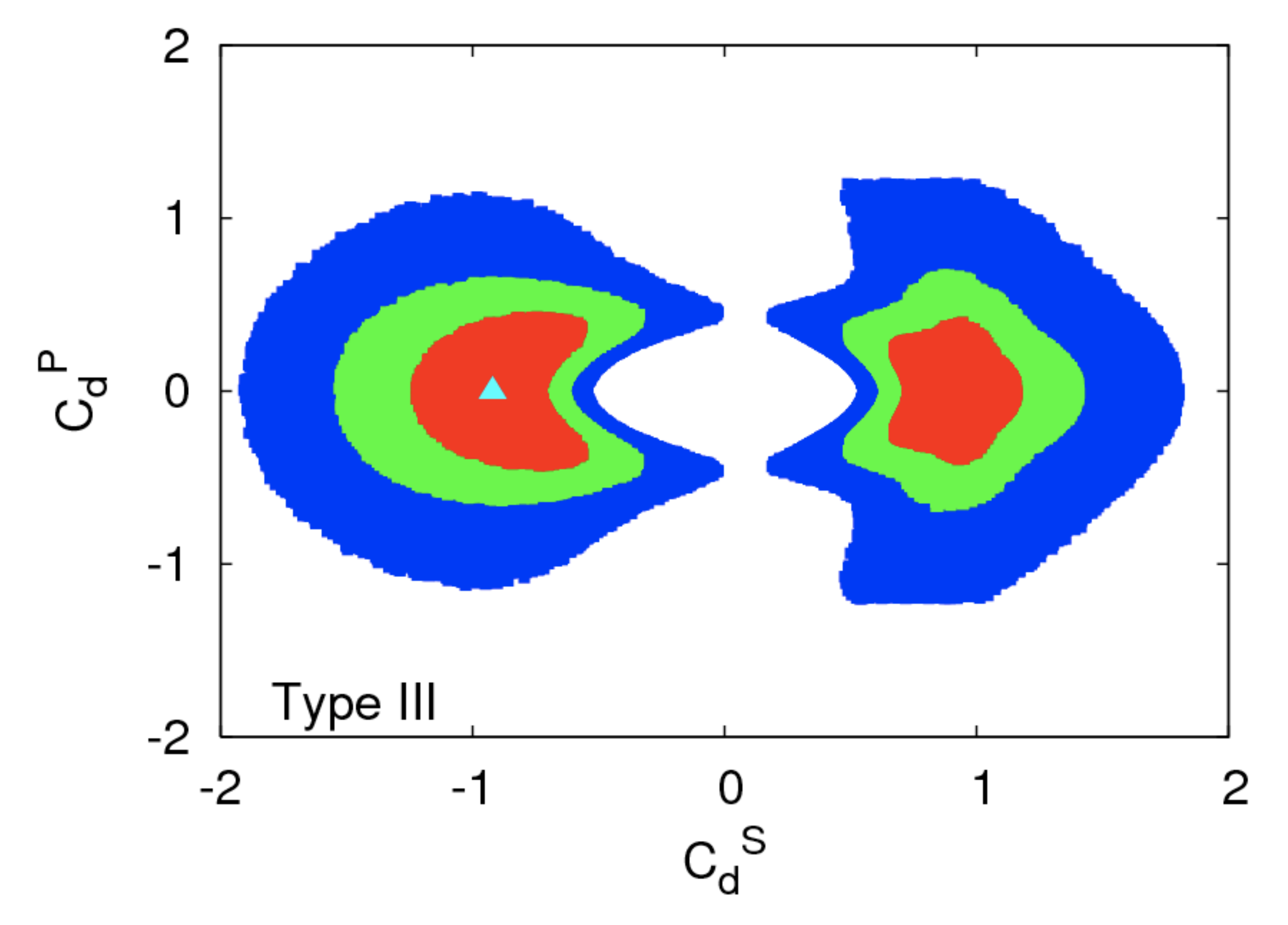}
\includegraphics[width=3.2in]{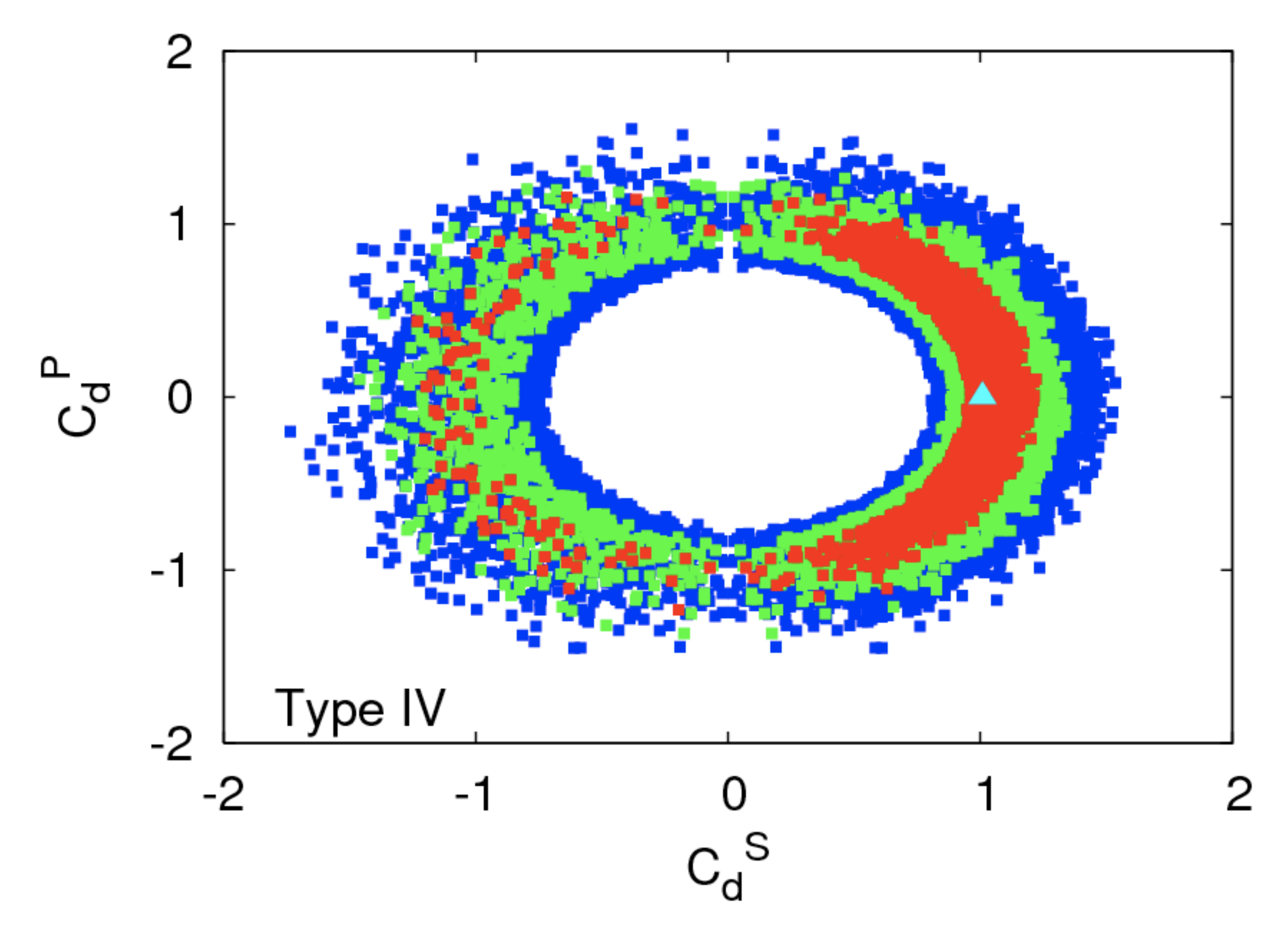}
\caption{\small \label{cpv3-cd-cdp}
The same as Fig.~\ref{cpv3-cu-cup} but 
in the plane of $C_d^S$ vs $C_d^P$ for Type I -- IV ({\bf CPV4}).
The description of the confidence regions is the same as Fig.~\ref{cpv3-cu-cup}.
}
\end{figure}

\begin{figure}[th!]
\centering
\includegraphics[width=3.2in]{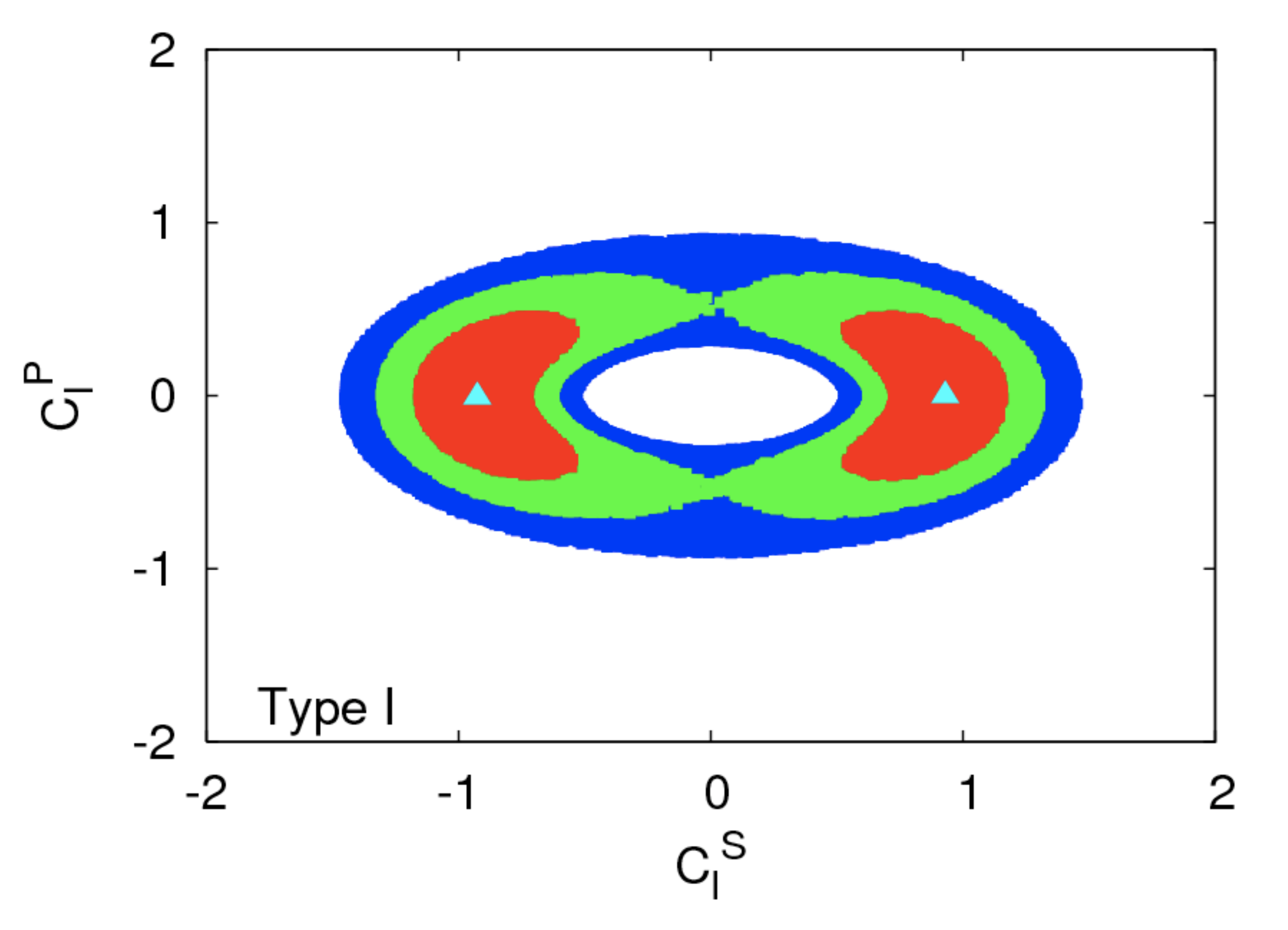}
\includegraphics[width=3.2in]{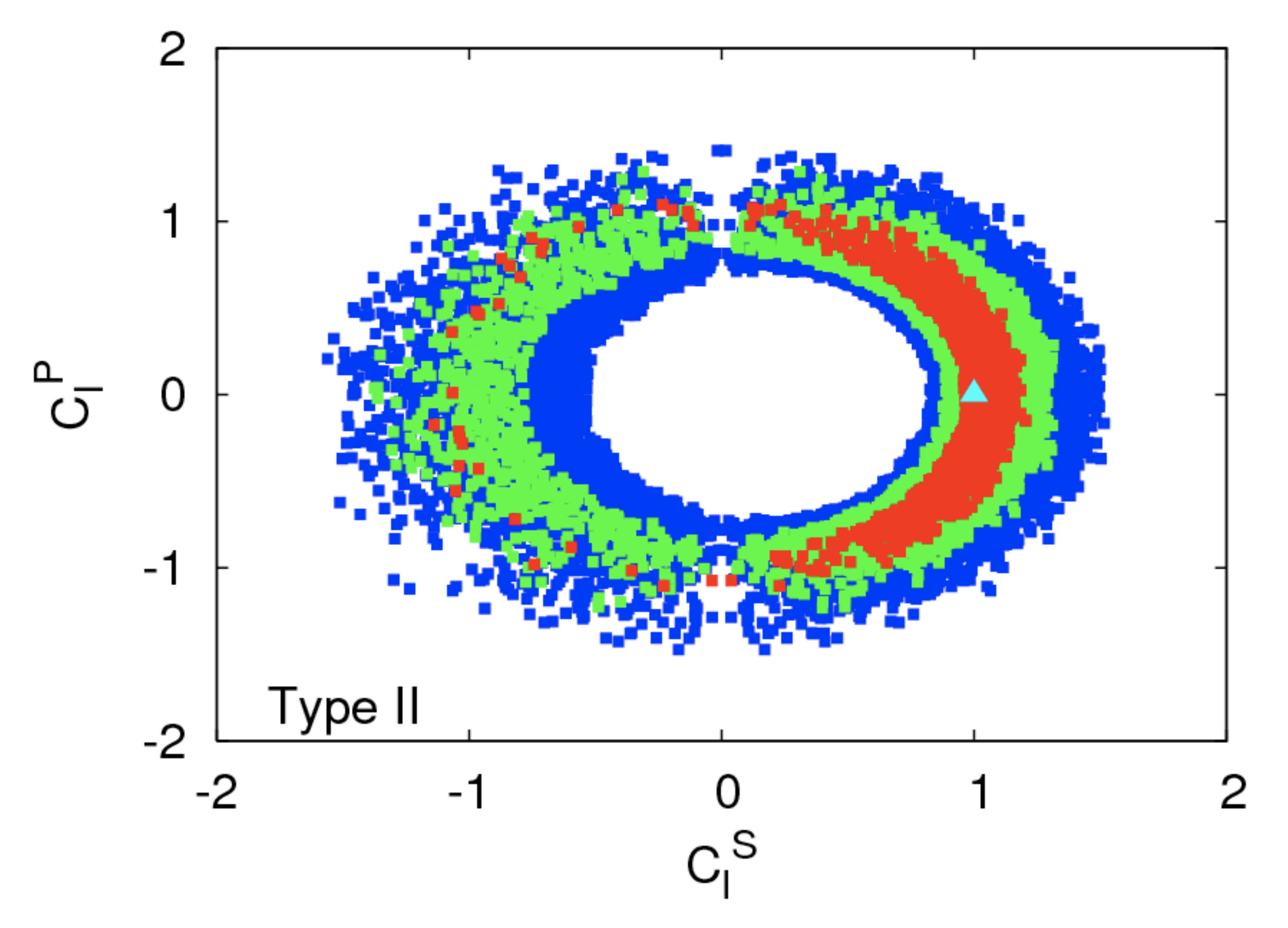}
\includegraphics[width=3.2in]{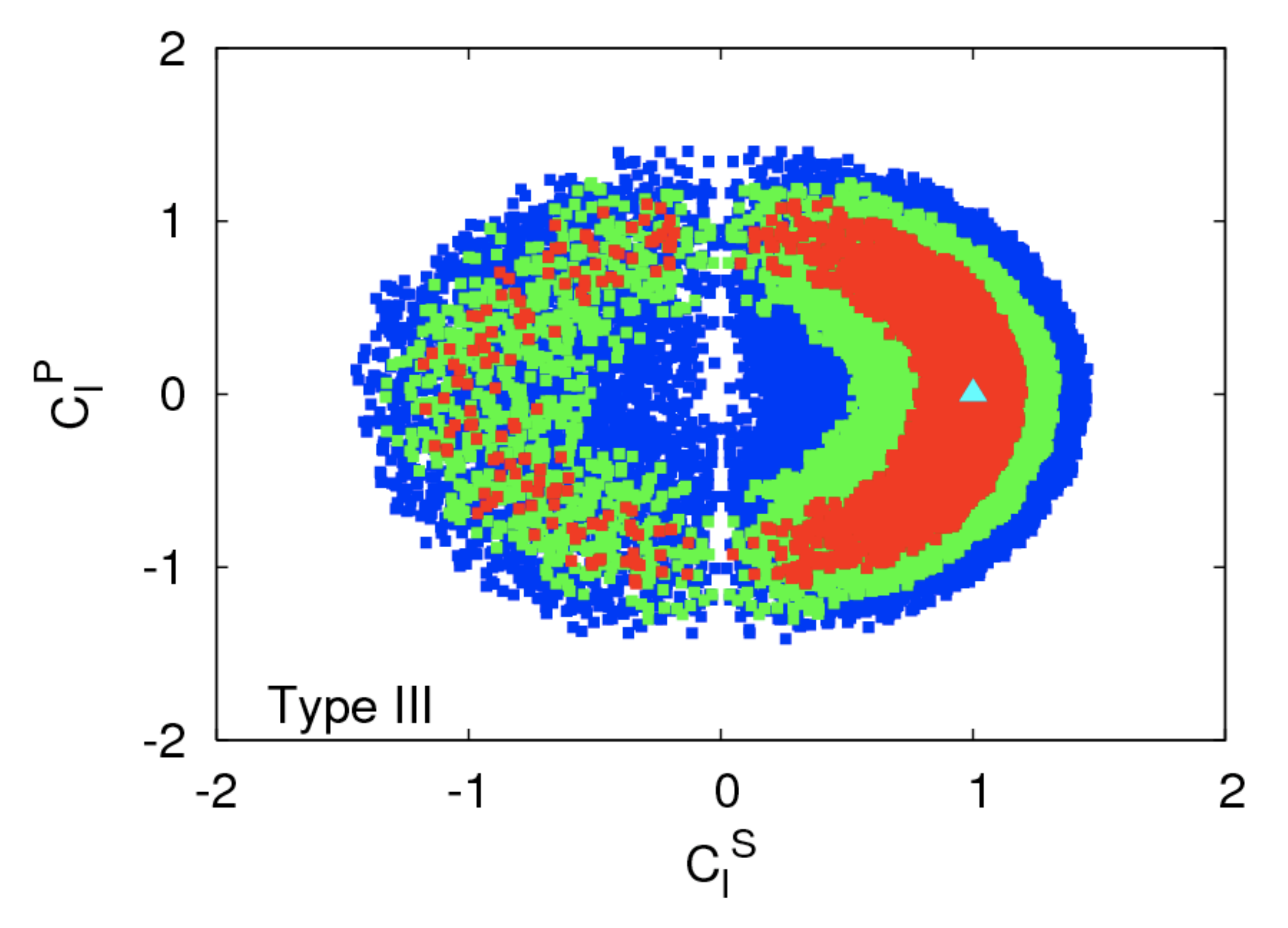}
\includegraphics[width=3.2in]{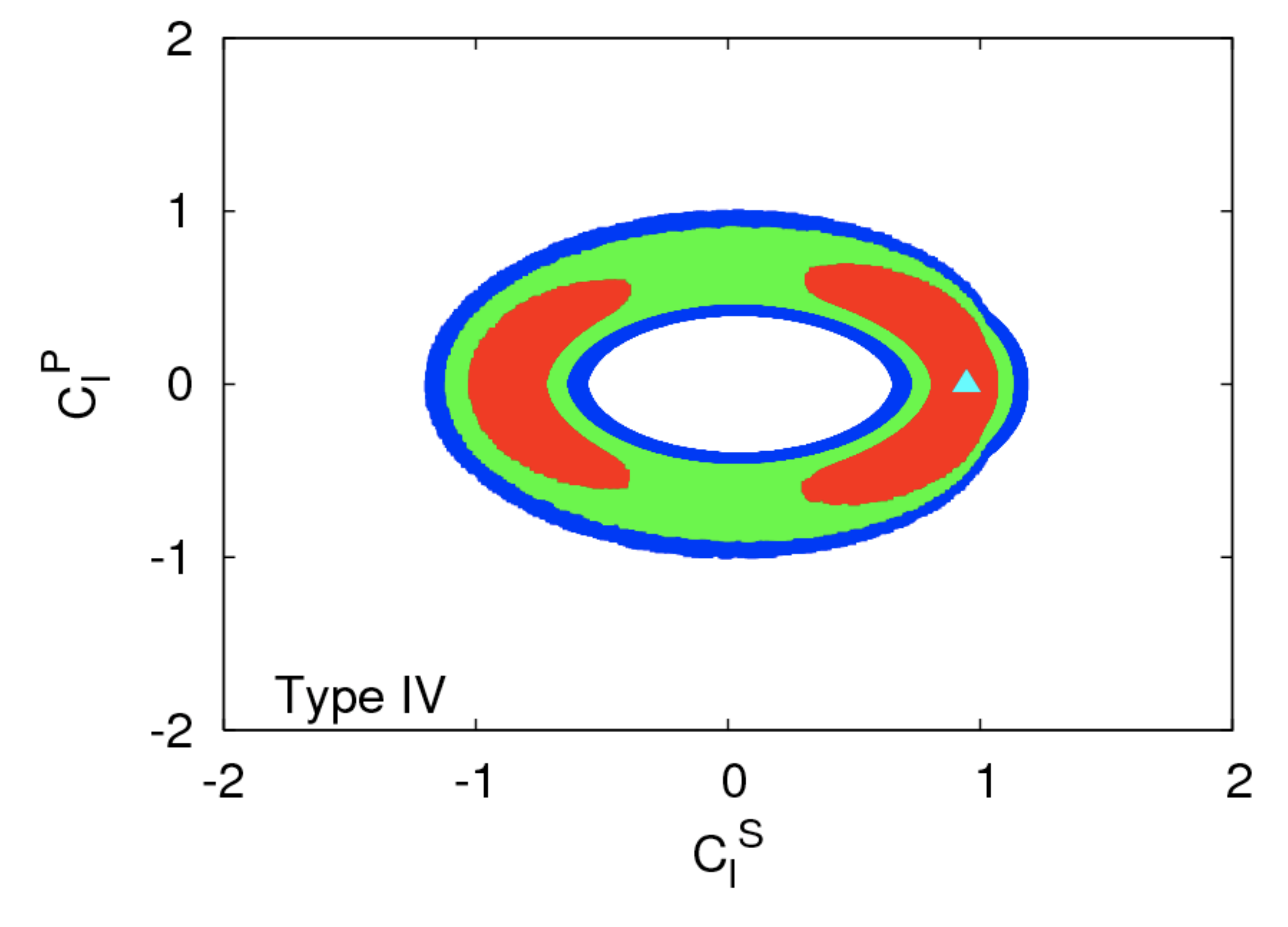}
\caption{\small \label{cpv3-cl-clp}
The same as Fig.~\ref{cpv3-cu-cup} but 
in the plane of $C_\ell^S$ vs $C_\ell^P$ for Type I -- IV ({\bf CPV4}).
The description of the confidence regions is the same as Fig.~\ref{cpv3-cu-cup}.
}
\end{figure}

In the {\bf CPV4} fit, we  vary $C_u^S$, $C_u^P$,
$C_v$ (or $\log_{10}\tan\beta$ equivalently), and
$\left(\Delta S^\gamma\right)^{H^\pm}$. 
The other couplings $C_{d,l}^{S,P}$ are given by the 
relations shown in Table~\ref{tab:cdcl}.
The total $\chi^2$,
$\chi^2/{\rm dof}$, $p$-value and the best-fit values for
$C_u^S$, $C_u^P$, $C_v$ ($\tan\beta$), and $\left(\Delta S^\gamma\right)^{H^\pm}$
for the four types of 2HDMs 
can be found in the lower half of Table~\ref{tab:best}.
We show the contour plots for confidence-level regions
as functions $C_u^S$ vs $C_u^P$, 
$C_u^S$ vs $C_v$, $C_u^S$ vs $\tan\beta$, 
$C_u^S$ vs $\left( \Delta S^\gamma \right )^{H^\pm} $, 
$C_d^S$ vs $C_d^P$, and 
$C_l^S$ vs $C_l^P$ in Figs.~\ref{cpv3-cu-cup} -- \ref{cpv3-cl-clp}, 
respectively.
We find that type II gives the smallest $\chi^2$
and its variation among the 4 types is within $0.57$, which is smaller than 
that of the {\bf CPV3} fits.  The $p$-values of the 
{\bf CPV4} fits are also worse than the {\bf CPV3} fits.

The best-fit values for $C_u^S$ are about 
$\pm 0.92$ (I), $-0.05$ (II), $-0.91$ (III) and $0.96$ (IV), while 
those of $C_u^P$ are 
about $0$ (I), $\pm 0.57$ (II), $0.03$ (III), and $-0.02$ (IV).
%
In type II, we note the best-fit value for $C_u^S$
is almost $0$ and those of $C_u^P$ are very small except for type II.
Therefore, in terms of the best-fit values the measure of CP-violating effect 
$2 C_u^S C_u^P/[(C_u^S)^2+(C_u^P)^2]$ is not significant in all 4 types
of 2HDMs.  
Nevertheless, the CP violation could be significant taking account
of the errors.
For the Higgs couplings to the down-type quarks and charged leptons,
we find that all the
couplings $C_d^P$ and $C_l^P$ are almost vanishing: see Table \ref{tab:best2}.
The best-fit values for $C_v$ are about $0.97$ (I and III) and 1 (II and IV)
and those for $\tan\beta$ are ${\cal O}(0.1)$,
except for type I with positive $C_u^S$, 
where the best-fit value is $6.5$.
As will be shown below in the figures, variation of $\chi^2$ 
vs of $\tan\beta$ is small in a large region of parameter
space.
For $\left(\Delta S^\gamma\right)^{H^\pm}$, the best-fit values are
$-0.78$ and $2.4$ (I),
$1.0$ (II), $2.4$ (III), and $-0.83$ (IV).
This also can be understood from Eq.~(\ref{eq:sgamma}).

In Fig.~\ref{cpv3-cu-cup}, we show the CL regions in the 
$C_u^S$ and $C_u^P$ plane, and note that the positive and 
negative $C_u^S$ regions are providing equally good fits.
The 68 \% CL regions of $C_u^S$ are:
$-1.1 \sim -0.5$  and $0.5 \sim 1.1$ (I),
$-1 \sim  1$ (II),
$-1.2 \sim -0.5$  and $0.6 \sim 1.2$ (III), and
$-1 \sim -0.4$  and $0.3 \sim 1.1$ (IV).  Also, $C_u^P$ varies between 
$\pm 0.5$ (I, III) and
$\pm 0.7$ (II, IV) in the 68 \% CL regions. Therefore, the maximal CP violation
with $|C_u^S| = |C_u^P|$ is still possible.

We show the CL regions in the $C_u^S$--$C_v$ and $C_u^S$--$\tan\beta$ planes
in Figs.~\ref{cpv3-cu-cv} and \ref{cpv3-cu-tanb}, respectively.
$C_v \gsim 0.8$ at 68 \% CL 
and $\tan\beta$ are bounded from above for the type II, III, and IV,
except for a narrow region around $C_u^S=1$.
The CL regions in the plane of $C_u^S$ and $\left(\Delta S^\gamma\right)^{H^\pm}$
are shown in Fig.~\ref{cpv3-cu-dsp}. Roughly speaking, 
$-2 \lsim \left(\Delta S^\gamma\right)^{H^\pm} \lsim 3.5$ (68 \% CL).
In Figs.~\ref{cpv3-cd-cdp} and \ref{cpv3-cl-clp}, 
the Higgs couplings to the down-type
quarks and charged leptons are shown.

\begin{figure}[th!]
\centering
\includegraphics[width=3.2in]{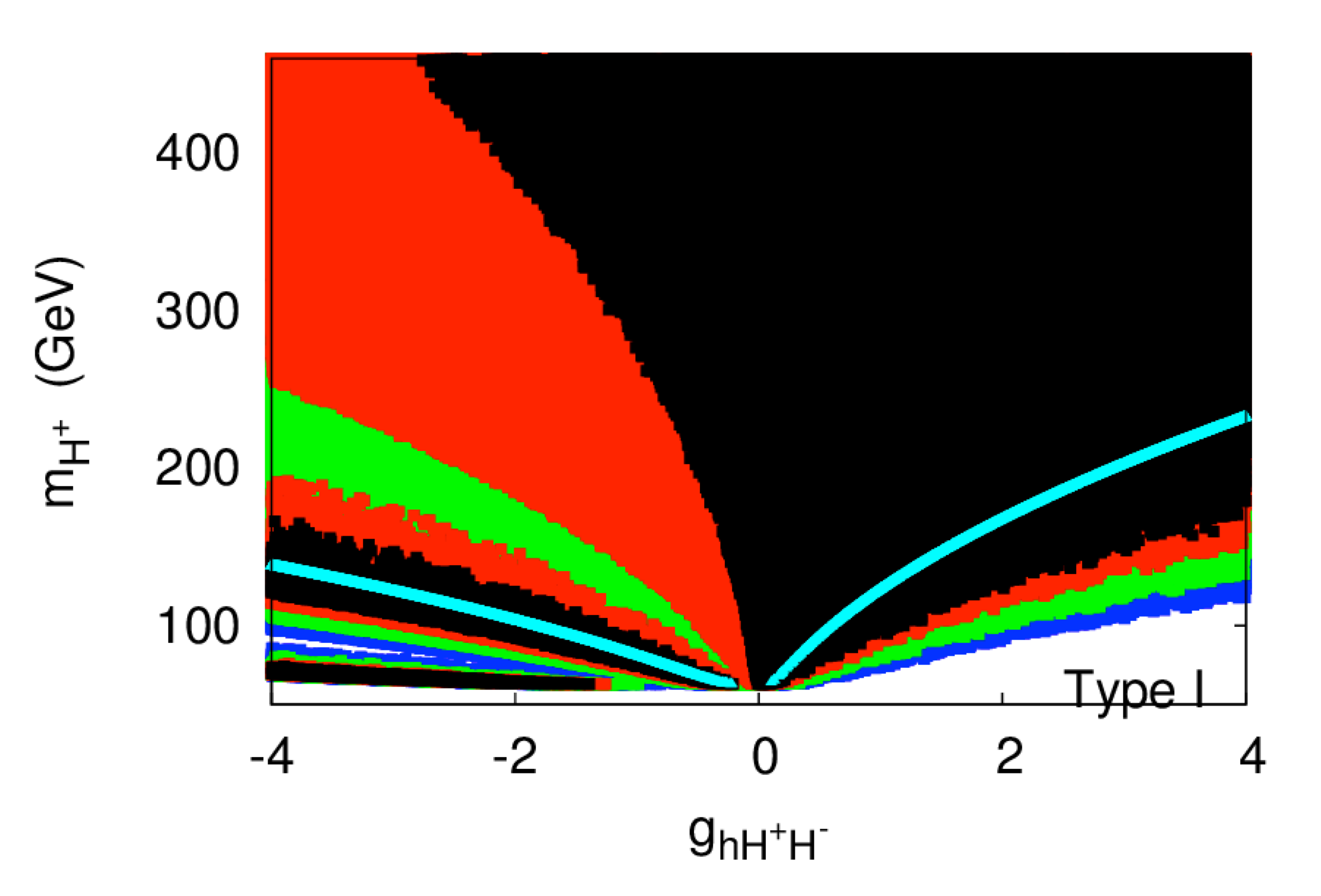}
\includegraphics[width=3.2in]{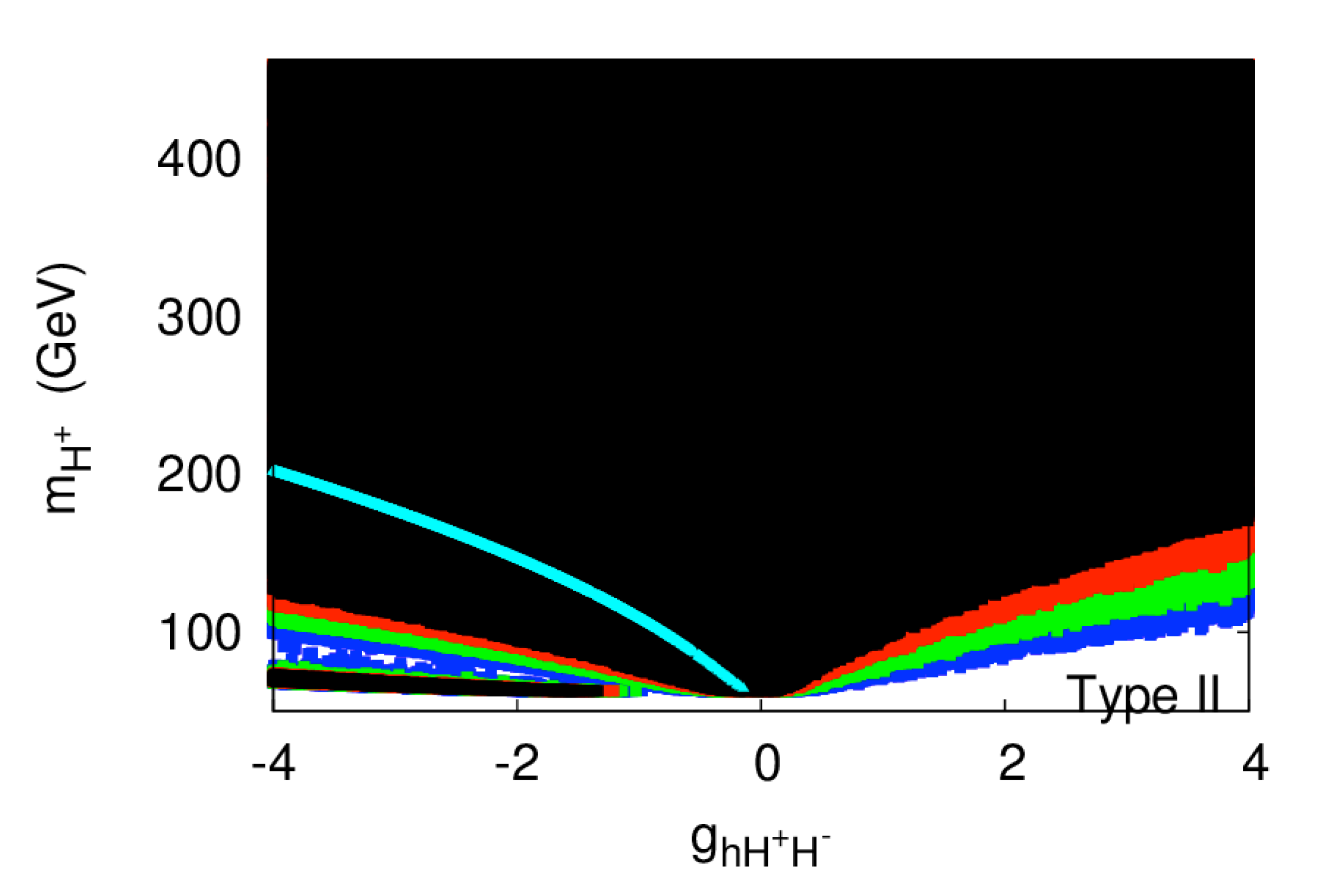}
\includegraphics[width=3.2in]{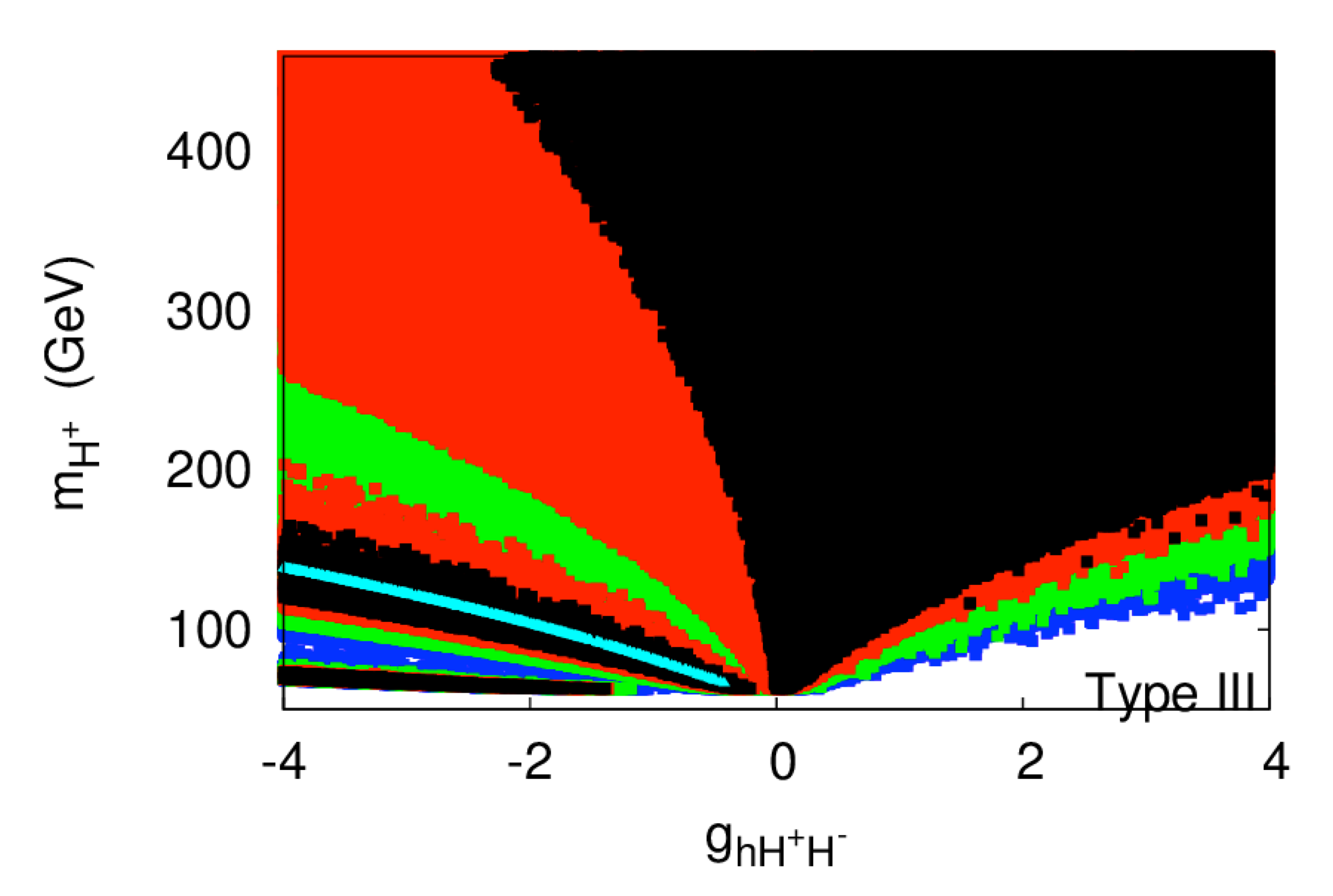}
\includegraphics[width=3.2in]{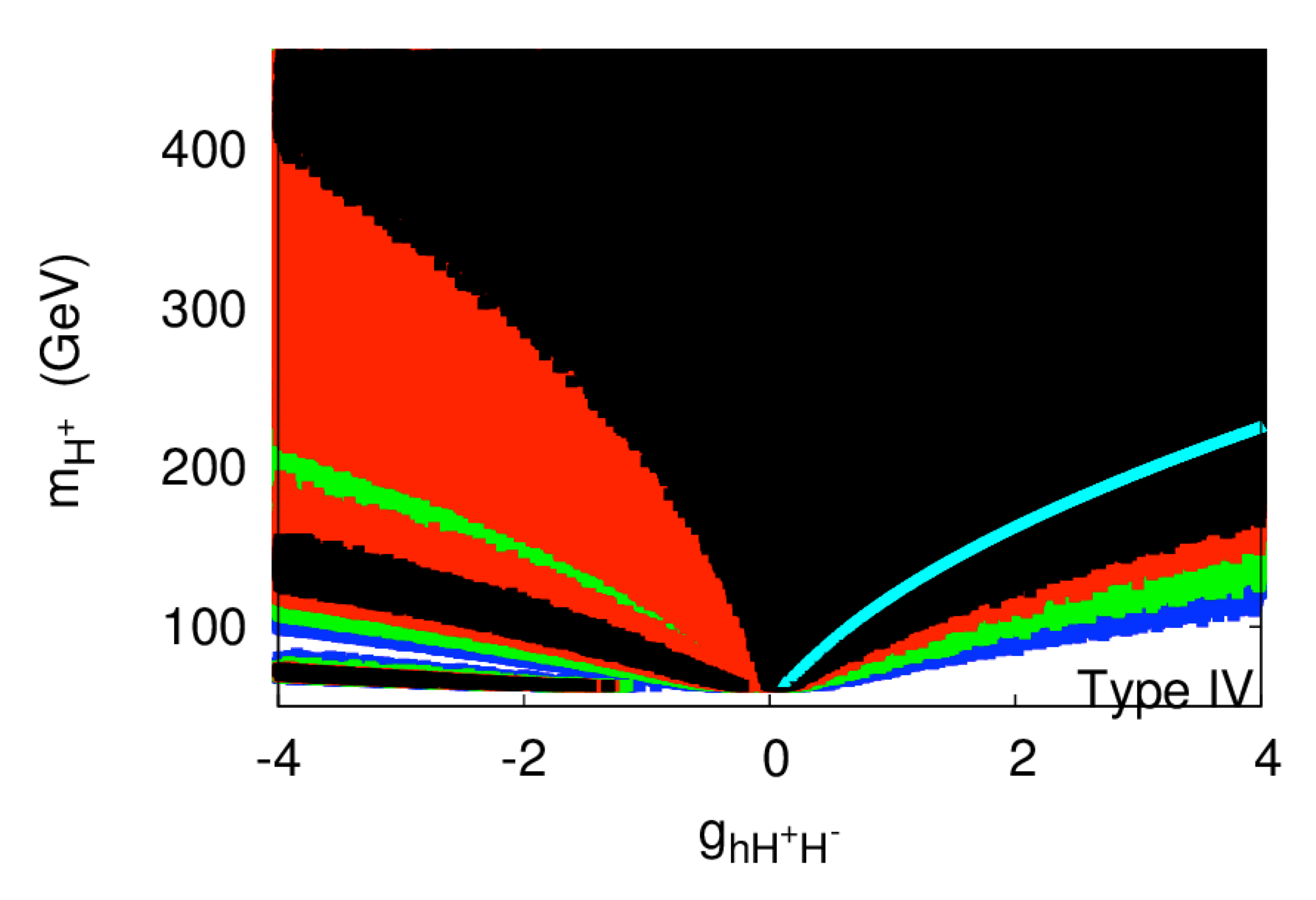}
\caption{\small \label{cpv4}
The same as Fig.~\ref{cpv3-cu-cup} but we used $g_{h H^+ H^-}$ and $m_{H^\pm}$ 
in place of $\left( \Delta S^{\gamma} \right )^{H^\pm}$ 
for Type I -- IV ({\bf CPV4}).
The contour regions shown are for 
$\Delta \chi^2 \le 1.0$ (black), 
$2.3$ (red), $5.99$ (green), and $11.83$ (blue) 
above the minimum, which 
correspond to confidence levels of $39.3\%$,
$68.3\%$, $95\%$, and $99.7\%$, respectively.
The best-fit points are denoted by a beam of cyan triangles.
}
\end{figure}

Finally, 
the single parameter $\left(\Delta S^\gamma\right)^{H^\pm}$
can be interpreted in terms of 
the charged Higgs mass $M_{H^\pm}$ and the neutral Higgs coupling to the
charged Higgses $g_{_{H_iH^+H^-}}$, as in Eq.~(\ref{eq:dsgamma}).
In Fig.~\ref{cpv4}, we show the CL regions in the plane of 
$M_{H^\pm}$ vs $g_{_{H_iH^+H^-}}$. 
Compared to the {\bf CPC3} case, we have $\Delta\chi^2\leq 1$ in the wider range.

\section{Discussion}

In this work, we have applied our previous model-independent approach
\cite{Cheung:2013kla},
which analyzes all the observed Higgs boson signal strengths 
and fits to all the Higgs boson couplings, to the 2HDMs.
In 2HDMs, the Higgs couplings to up-type and down-type quarks, and 
charged leptons are related by a set of relations shown in 
Table~\ref{tab:cdcl}.
We have shown that the whole analysis can be performed with at most $3$
independent parameters: $C_u^S$, $C_v$ (or $\tan\beta$), and
$\left(\Delta S^{\gamma} \right)^{H^\pm}$ for CP-conserving scenarios,
and only one more parameter $C_u^P$ for the CP-violating scenarios.
A number of relationships among the couplings of the
up- and down-type quarks and
charged leptons have been derived such that we need only $C_u^S$ and
$C_u^P$.

A set of discrete symmetries are often imposed in literature in order
to eliminate flavor-changing neutral currents, denoted by the
parameters $\eta^{d,l}_{1,2}$, which take up the values either 0 or 1.
The four combinations of $(\eta^{d}_{1},\eta^{l}_1)
=(0,0),\;(1,1),\;(0,1),\;(1,0)$ 
correspond to type I, II, III, and IV, respectively. 
We have demonstrated that the
current Higgs boson data have no preference for any of the four types
of 2HDMs, because
statistically the $\chi^2$ difference among type I--IV is only
$0.3$ for CPC cases and $1.2$ for CPV cases: see Table~\ref{tab:best}.
We also relaxed the discrete symmetries to allow continuous values
for $\eta^{d,l}_{1,2}$ subject to normalization $(\eta^{d,l}_{1})^2
+ (\eta^{d,l}_{2})^2 = 1$, and we found that in the whole plane of 
$ 0 \le \eta^{d}_{1}, \eta^l_1 \le 1$ the $\chi^2$ differences among the
best-fits are all within $\chi^2 < 1.2$. It is one of the main findings 
in this work -- no particular preference among type I to IV as long as
the current Higgs boson data are concerned.  

The Higgs data used are almost the final set out of the 7 TeV and 8 TeV
runs at the LHC. Further improvement to the fits will only be possible
when more data are pouring in the next run of 2015. So far,
the data have pointed to the SM Higgs boson with a large $p$-value,
while all other extensions to the SM,
such as the 2HDMs studied in this work or more model-independently
in Ref.~\cite{Cheung:2013kla}, provide fits with smaller $p$-values
than the SM. It means that the SM Higgs boson is currently the best
explanation to all the Higgs boson data. 

We offer a few more comments before we conclude.
\begin{enumerate}
\item
The up-type and down-type (charged lepton) Yukawa couplings are related
by quark masses and $\tan\beta$. Therefore, one set of parameters
$C_u^S$, $C_u^P$, and $\tan\beta$ 
is sufficient to define all the fermionic couplings.

\item 
When we relax the discrete symmetries by varying $\eta^{d}_1$ and 
$\eta^l_1$, we found the best-fit values for them are neither 0 nor 1. 
However, the $\chi^2$ differences in the whole plane of
$\eta^{d}_1$ vs $\eta^l_1$ are too small to claim any preference statistically.

\item 
The charged Higgs boson contributes to the one-loop vertex $H\gamma\gamma$.
In the studies, we first treated $\left( \Delta S^\gamma \right )^{H^\pm}$ as 
an independent parameter. Then we broke it down into the charged Higgs mass
$M_{H^\pm}$ and the coupling $g_{H_i H^+ H^-}$. When the $b\to s\gamma$ constraint
(roughly $M_{H^\pm} > 300$ GeV) is taken into account, positive
$g_{H_i H^+ H^-}$ is preferred.

\item 
The Higgs coupling to gauge bosons $C_v$ is constrained to be very close
to 1.  It means that the observed Higgs boson is entirely responsible
for breaking the electroweak symmetry.

\item
Future precision
measurements of $C_u^S$ and $\tan\beta$ can provide us with the
discriminating power among various types of 2HDMs especially when
$C_u^S$ deviates from its SM value $1$.

\item 
The parameters $C_u^S$ and $C_u^P$ are constrained in the form of some
ellipses. The current Higgs observables are not sensitive to 
CP-violating effects, and so only combinations of scalar and 
pseudoscalar Yukawa couplings are constrained, as shown in 
Figs.~\ref{cpv1-cu-cup} and \ref{cpv3-cu-cup}.

\item
Among the 2HDM fits considered in this work, the type-II CP-violating case
with $\left( \Delta S^\gamma \right )^{H^\pm}=0$ (the {\bf CPV3} type-II fit)
gives the best fit with $\chi^2=17.17$ and $p$-value$=0.578$
when $C_u^S \sim |C_u^P| = 1/2$.

\end{enumerate}

\section*{Acknowledgment}  
This work was supported the National Science
Council of Taiwan under Grants No. 99-2112-M-007-005-MY3 and
102-2112-M-007-015-MY3.
J.S.L. was supported by 
the National Research Foundation of Korea (NRF) grant 
(No. 2013R1A2A2A01015406).
This study was also
financially supported by Chonnam National University, 2012.


\begin{table}[thb!]
  \caption{\small \label{tab:best}
The best-fit values for various CPC and CPV fits.
The SM values are: $\chi^2=18.94$, $\chi^2/{\rm dof}=0.86$,
and $p$-value$=0.65$.
  }
\begin{ruledtabular}
\begin{tabular}{ cc|ccc|rrccr}
Fits & Type & $\chi^2$ & $\chi^2$/dof & $p$-value &
\multicolumn{5}{c}{Best-fit values} \\
& & & & & $C_u^S$ & $C_u^P$ & $C_v$ & $\tan\beta$ &
$\left(\Delta S^\gamma\right)^{H^\pm}$ \\
\hline\hline
& I   & $18.39$ & $0.920$ & $0.562$ & $0.895$ & $0$ &
$1.000$ & $limit$ & $0$ \\
{\bf CPC2}
& II  & $18.68$ & $0.934$ & $0.543$ & $0.963$ & $0$ &
$1.000$ & $limit$ & $0$ \\
& III & $18.44$ & $0.922$ & $0.558$ & $0.892$ & $0$ &
$1.000$ & $limit$ & $0$ \\
& IV  & $18.66$ & $0.933$ & $0.544$ & $0.965$ & $0$ &
$1.000$ & $limit$ & $0$ \\
\hline
& I   & $17.64$ & $0.928$ & $0.547$ & $ 0.924$ & $0$ &
$0.965$ & $6.308$ & $-0.756$ \\
& I   & $17.64$ & $0.928$ & $0.547$ & $-0.921$ & $0$ &
$0.965$ & $0.144$ & $2.377$ \\
{\bf CPC3}
& II  & $17.30$ & $0.910$ & $0.570$ & $-0.822$ & $0$ &
$1.000$ & $2\times 10^{-4}$ & $2.218$ \\
& III & $17.63$ & $0.928$ & $0.547$ & $-0.912$ & $0$ &
$0.967$ & $0.137$ & $2.365$ \\
& IV  & $17.54$ & $0.923$ & $0.553$ & $0.955$ & $0$ &
$1.000$ & $0.662$ & $-0.835$ \\
\hline\hline
& I   & $18.37$ & $0.967$ & $0.498$ & $0.867$ & $0.142$ &
$0.988$ & $0.840$ & $0$ \\
& I   & $18.37$ & $0.967$ & $0.498$ &
$0.867$ & $-0.142$ &
$0.988$ & $0.840$ & $0$ \\
{\bf CPV3}
& II  & $17.17$ & $0.904$ & $0.578$ & $0.476$ & $-0.505$ &
$0.998$ & $0.082$ & $0$ \\
& II  & $17.17$ & $0.904$ & $0.578$ &
$0.475$ & $0.505$ &
$0.998$ & $0.095$ & $0$ \\
& III & $18.41$ & $0.969$ & $0.495$ & $0.873$ & $-0.110$ &
$1.000$ & $2\times 10^{-4}$ & $0$ \\
& III & $18.41$ & $0.969$ & $0.495$ & $0.873$ &
$0.109$ &
$1.000$ & $1.2\times 10^{-4}$ & $0$ \\
& IV  & $18.16$ & $0.956$ & $0.512$ & $0.806$ & $0.339$ &
$1.000$ & $limit$ & $0$ \\
& IV  & $18.16$ & $0.956$ & $0.512$ & $0.806$ &
$-0.339$ &
$1.000$ & $1.2\times 10^{-4}$ & $0$ \\
\hline
& I   & $17.64$ & $0.980$ & $0.480$ & $0.924$ & $-1.5\times 10^{-3}$ &
$0.964$ & $6.488$ & $-0.777$ \\
& I   & $17.64$ & $0.980$ & $0.480$ &
$-0.924$ & $2\times 10^{-4}$ &
$0.965$ & $0.139$ & $2.389$ \\
{\bf CPV4}
& II  & $17.07$ & $0.948$ & $0.518$ & $-0.052$ & $0.572$ &
$0.999$ & $0.045$ & $1.042$ \\
& II  & $17.07$ & $0.948$ &
$0.518$ & $-0.052$ & $-0.572$ &
$0.999$ & $0.045$ & $1.042$ \\
& III & $17.64$ & $0.980$ & $0.480$ & $-0.909$ & $0.032$ &
$0.972$ & $0.126$ & $2.370$ \\
& IV  & $17.54$ & $0.975$ & $0.486$ & $0.956$ & $-0.016$ &
$1.000$ & $0.670$ & $-0.831$ \\
\end{tabular}
\end{ruledtabular}
\end{table}
\begin{table}[thb!]
  \caption{\small \label{tab:best2}
Table showing the corresponding best-fit values for
$C_{d,l}^{S,P}$.
  }
\begin{ruledtabular}
\begin{tabular}{ cc|ccc|rrrr}
Fits & Type & $\chi^2$ & $\chi^2$/dof & $p$-value &
\multicolumn{4}{c}{Best-fit values} \\
& & & & & $C_d^S$ & $C_l^S$ & $C_d^P$ & $C_l^P$  \\
\hline\hline
& I   & $18.39$ & $0.920$ & $0.562$ & $0.896$ & $0.896$ &
$0$ & $0$  \\
{\bf CPC2}
& II  & $18.68$ & $0.934$ & $0.543$ & $1.000$ & $1.000$ &
$0$ & $0$  \\
& III & $18.44$ & $0.922$ & $0.558$ & $0.892$ & $1.000$ &
$0$ & $0$  \\
& IV  & $18.66$ & $0.933$ & $0.544$ & $1.000$ & $0.965$ &
$0$ & $0$  \\
\hline
& I   & $17.64$ & $0.928$ & $0.547$ & $ 0.923$ & $0.923$ &
$0$ & $0$ \\
& I   & $17.64$ & $0.928$ & $0.547$ & $-0.923$ & $-0.923$ &
$0$ & $0$ \\
{\bf CPC3}
& II  & $17.30$ & $0.910$ & $0.570$ & $1.000$ & $1.000$ &
$0$ & $0$  \\
& III & $17.63$ & $0.928$ & $0.547$ & $-0.914$ & $1.002$ &
$0$ & $0$  \\
& IV  & $17.54$ & $0.923$ & $0.553$ & $1.015$ & $0.951$ &
$0$ & $0$  \\
\hline\hline
& I   & $18.37$ & $0.967$ & $0.498$ & $0.867$ & $0.867$ &
$-0.142$ & $-0.142$  \\
& I   & $18.37$ & $0.967$ & $0.498$ &
$0.867$ & $0.867$ &
$0.142$ & $0.142$  \\
{\bf CPV3}
& II  & $17.17$ & $0.904$ & $0.578$ & $1.002$ & $1.002$ &
$-4.6\times 10^{-3}$ & $-4.6\times 10^{-3}$  \\
& II  & $17.17$ & $0.904$ & $0.578$ &
$1.002$ & $1.002$ &
$4.6\times 10^{-3}$ & $4.6\times 10^{-3}$  \\
& III & $18.41$ & $0.969$ & $0.495$ & $0.873$ & $1.000$ &
$0.109$ & $0$  \\
& III & $18.41$ & $0.969$ & $0.495$ & $0.873$ & $1.000$ &
$-0.109$ & $0$  \\
& IV  & $18.16$ & $0.956$ & $0.512$ & $1.000$ & $0.806$ &
$0$ & $-0.339$  \\
& IV  & $18.16$ & $0.956$ & $0.512$ & $1.000$ & $0.806$ &
$0$ & $0.339$  \\
\hline
& I   & $17.64$ & $0.980$ & $0.480$ & $0.924$ & $0.924$ &
$1.5\times 10^{-3}$ & $1.5\times 10^{-3}$  \\
& I   & $17.64$ & $0.980$ & $0.480$ &
$-0.924$ & $-0.924$ &
$-2\times 10^{-4}$ & $-2\times 10^{-4}$  \\
{\bf CPV4}
& II  & $17.07$ & $0.948$ & $0.518$ & $1.001$ & $1.001$ &
$1.2\times 10^{-3}$ & $1.2\times 10^{-3}$  \\
& II  & $17.07$ & $0.948$ &
$0.518$ & $1.001$ & $1.001$ &
$-1.2\times 10^{-3}$ & $-1.2\times 10^{-3}$  \\
& III & $17.64$ & $0.980$ & $0.480$ & $-0.914$ & $1.002$ &
$-3\times 10^{-5}$ & $0$  \\
& IV  & $17.54$ & $0.975$ & $0.486$ & $1.015$ & $0.951$ &
$-1\times 10^{-3}$ & $3\times 10^{-3}$  \\
\end{tabular}
\end{ruledtabular}
\end{table}

 \clearpage

 \clearpage

\end{document}